\PassOptionsToPackage{pdfpagelabels=false}{hyperref} 
\documentclass[a4paper,fleqn,usenatbib]{mnras}

\usepackage{newtxtext,newtxmath}
\usepackage[flushleft]{threeparttable}
\usepackage[T1]{fontenc}
\usepackage{ae,aecompl}

\usepackage{graphicx}	% Including figure files
\usepackage{amsmath}	% Advanced maths commands
\usepackage{nccmath}
\usepackage{amssymb}	% Extra maths symbols
\usepackage{pdflscape}	% table landscape
\usepackage{multicol}
\usepackage{afterpage}

\title[Observed colours of Type II supernovae]{Observed Type II supernova colours from the Carnegie Supernova Project-I}

\author[de Jaeger et al.]
{T. de Jaeger$^{1}$\thanks{e-mail: tdejaeger@berkeley.edu}, J. P. Anderson$^{2}$, L. Galbany$^{3}$, S. Gonz\'alez-Gait\'an$^{4}$, M. Hamuy$^{5,6}$,
\newauthor 
M. M. Phillips$^{7}$, 
M. D. Stritzinger$^{8}$, 
C. Contreras$^{7}$,
G. Folatelli$^{9,10}$,
C. P. Guti\'errez$^{11}$,
\newauthor
E. Y. Hsiao$^{12}$, 
N. Morrell$^{7}$,
N. B. Suntzeff$^{13}$,
L. Dessart$^{14}$, 
A.V. Filippenko$^{1,15}$\\
$^{1}$Department of Astronomy, University of California, Berkeley, CA 94720-3411, USA.\\
$^{2}$European Southern Observatory, Alonso de C\'ordova 3107, Casilla 19, Santiago, Chile.\\
$^{3}$PITT PACC, Department of Physics and Astronomy, University of Pittsburgh, Pittsburgh, PA 15260, USA.\\
$^{4}$Centro Multidisciplinar de Astrofisica, Instituto Superior Tecnico, Av. Rovisco Pais 1, 1049--001, Lisbon, Portugal.\\
$^{5}$Departamento de Astronom\'ia -- Universidad de Chile, Camino el Observatorio 1515, Santiago, Chile.\\
$^{6}$Millennium Institute of Astrophysics, Santiago, Chile.\\
$^{7}$Las Campanas Observatory, Carnegie Observatories, Casilla 601, La Serena, Chile.\\
$^{8}$Department of Physics and Astronomy, Aarhus University, Ny Munkegade 120, DK-8000 Aarhus C, Denmark.\\
$^{9}$Facultad de Ciencias Astron\'{o}micas y Geof\'{i}sicas, UNLP, IALP, CONICET, Paseo del Bosque S/N, B1900FWA La Plata, Argentina.\\
$^{10}$Kavli Institute for the Physics and Mathematics of the Universe (WPI), The University of Tokyo,\ 
5-1-5 Kashiwanoha, Kashiwa, Chiba 277-8583, Japan.\\
$^{11}$Department of Physics and Astronomy, University of Southampton, Southampton SO17 1BJ, UK.\\
$^{12}$Department of Physics, Florida State University, Tallahassee, FL 32306, USA.\\
$^{13}$George P. and Cynthia Woods Mitchell Institute for Fundamental Physics and Astronomy, Department of Physics and Astronomy, Texas A\&M University, College Station, TX 77843, USA.\\
$^{14}$Unidad Mixta Internacional Franco-Chilena de Astronom\'ia (CNRS UMI 3386), Departamento de Astronom\'ia,\
Universidad de Chile, Camino El Observatorio 1515, Las Condes, Santiago, Chile.\\
$^{15}$Miller Senior Fellow, Miller Institute for Basic Research in Science, University of California, Berkeley, CA 94720, USA.
}

\date{Accepted XXX. Received YYY; in original form ZZZ}

\pubyear{2017}

\begin{document}
\label{firstpage}
\pagerange{\pageref{firstpage}--\pageref{lastpage}}
\maketitle

% Abstract of the paper (Aims, Methods, and Results)
\begin{abstract}

We present a study of observed Type II supernova (SN~II) colours using optical/near-infrared photometric data from the \textit{Carnegie Supernovae Project-I}. We analyse four colours ($B-V$, $u-g$, $g-r$, and $g-Y$) and find that SN~II colour curves can be described by two linear regimes during the photospheric phase. The first ($s_{\rm 1,colour}$) is steeper and has a median duration of $\sim 40$ days. The second, shallower slope ($s_{\rm 2,colour}$) lasts until the end of the ``plateau'' ($\sim 80$ days). The two slopes correlate in the sense that steeper initial colour curves also imply steeper colour curves at later phases. As suggested by recent studies, SNe~II form a continuous population of objects from the colour point of view as well. We investigate correlations between the observed colours and a range of photometric and spectroscopic parameters including the absolute magnitude, the $V$-band light-curve slopes, and metal-line strengths. We find that less luminous SNe~II appear redder, a trend that we argue is not driven by uncorrected host-galaxy reddening. While there is significant dispersion, we find evidence that redder SNe~II (mainly at early epochs) display stronger metal-line equivalent widths. Host-galaxy reddening does not appear to be a dominant parameter, neither driving observed trends nor dominating the dispersion in observed colours. Intrinsic SN~II colours are most probably dominated by photospheric temperature differences, with progenitor metallicity possibly playing a minor role. Such temperature differences could be related to differences in progenitor radius, together with the presence or absence of circumstellar material close to the progenitor stars.

\end{abstract}


%%%%%%%%%%%%%%%%%%%%%%%%%%%%%%%%%%%%%%%%%%%%%%%%%%

%%%%%%%%%%%%%%%%% BODY OF PAPER %%%%%%%%%%%%%%%%%%
\section{Introduction}

The community of supernova (SN) researchers is now experiencing a growing interest in the study of core-collapse supernovae (CC~SNe), after focusing more on the well-studied Type Ia supernovae (hereafter SNe~Ia; \citealt{min41,elias85}) for 20 years. CC~SNe are believed to be the explosions of massive stars at the end of their lives ($\geq 8~{\rm M}_{\odot}$; see \citealt{smartt09a} for a review). CC~SNe exhibit a large range of observed photometric and spectroscopic properties. They can be first subclassified according to the absence or the presence of \ion{H}{I} lines: SNe~Ib/c and SNe~II, respectively (see, e.g., \citealt{filippenko97}, and references therein). Secondly, hydrogen-rich CC~SNe can be grouped using spectroscopic properties: SNe~II with broad, long-lasting Balmer lines in their spectra; SNe~IIb, which evolve spectroscopically from SNe~II at early times to SNe~Ib several weeks past maximum brightness \citep{woosley87,filippenko88,filippenko93}, and SNe~IIn with relatively narrow emission lines in their spectra \citep{che81,fra82,sch90,filippenko91,chu94}.

SNe~II\footnote{Throughout the rest of this text we refer to SNe~II as the two historical groups, SNe~IIP (plateau) and SNe~IIL (linear), since recent studies indicate a relatively continuous distribution (\citealt{anderson14a,sanders15,valenti16,galbany16a}; see also \citealt{rubin16b}).}, the most frequently occurring CC~SNe in the Universe \citep{li2011,graur17}, are the explosions of stars that have retained a significant fraction of their hydrogen envelopes prior to exploding. Thanks to direct progenitor detections and earlier hydrodynamical modelling, most SN~II progenitors have been robustly established as the explosions of red supergiants (RSGs; $M \leq 25 {\rm M}_{\odot}$; \citealt{grassberg71}; \citealt{falk77}; \citealt{chevalier76}; \citealt{vandyk03}; \citealt{smartt09a}).

SNe~II are very useful tools for understanding the composition and evolution of our Universe because they have been established as metallicity \citep{dessart14,anderson16a} and distance indicators (e.g., \citealt{hamuy02}). Several methods exist to standardise SN~II peak brightness and make them useful for cosmology: the ``expanding photosphere method'' \citep{kirshner74,schmidt94,hamuy01,leonard03,dessart05,dessart06,jones09,enriquez11,gall16}, the ``standard candle method'' \citep{hamuy02,nugent06,andrea10,poznanski10,dejaeger17b,gall17}, the ``photospheric magnitude method'' \citep{rodriguez14}, and the ``photometric colour method'' \citep{dejaeger15b,dejaeger17a}.

Generally, SN~II standardisation is achieved by adding an observed colour correction to account for reddening of light caused by host-galaxy dust extinction or any intrinsic variation in the magnitude-colour relations. 
However, with this factor, it is assumed that all SNe~II follow the same magnitude-colour law, which has never been proved to be the case. 
Additionally, by assuming that the colour diversity comes exclusively from dust extinction (which may not be true), such a colour term correction has led the SN~Ia and SN~II communities to derive surprisingly low total-to-selective extinction ratios ($R_{V}$) as compared with the Milky Way Galaxy (1.5--2.5; \citealt{krisciunas07,eliasrosa08,goobar08,folatelli10,mandel11,phillips13,poznanski09,olivares10,burns14,rodriguez14,dejaeger15b}). These low $R_{V}$ values could reflect differences in the dust properties, or intrinsic magnitude-colour relations for SNe~Ia/SNe~II not taken into account in the colour corrections. 

Within this context, a more complete understanding of SN~II colours would be an asset for cosmological studies while also furthering our knowledge of SN~II progenitors and explosion properties. Additionally, as the $K$-correction \citep{oke68,hamuy93,kim96,nugent02} --- applied to take into account the expansion of the Universe --- strongly depends on colour, understanding SN~II colour diversity would help improve its accuracy. To date, there are very few statistical studies of SN~II observed colours and how these relate to intrinsic colours or the effects of reddening from the host galaxy. 

While the overall SN~II colour behaviour (through studies of colour curves) is relatively well known from investigations of individual events such as SN~1999em \citep{hamuy01,leonard02,elmhamdi03}, SN~1999gi \citep{leonard02b}, SN~2004et \citep{sahu06,maguire10b}, SN~2005cs \citep{pastorello09}, SN~2007od \citep{inserra11}, SN~2013by \citep{valenti15}, and SN~2013ej \citep{valenti14,bose15,huang15,mauerhan16,dhungana16}, these have yet to be put within the context of the full diversity of SNe~II and the progenitor parameters that influence intrinsic colours. 

General SN~II colour behaviour, as noted by \citet{patat94}, consists of two regimes with two different slopes and can be described as follows. At early times ($\sim 10$ days), a rapid increase of the $(B-V)$ colour is seen as the envelope expands and cools from an initial temperature of $> 10,000$~K to the recombination temperature of hydrogen ($\sim 5500$~K). After $\sim 35$ days from the explosion (i.e. during the recombination phase), the colour varies more slowly, since the photospheric temperature is assumed to be the hydrogen recombination temperature. For the redder bands (e.g., $V-I$), the colour increases more slowly than $(B-V)$ during the photospheric phase, mainly because the spectral energy distribution is less sensitive to temperature changes in the red than in the blue part of the spectrum (see \citealt{galbany16a}).

Understanding the origin of the observed diversity in SN~II colours is also important for deriving the host-galaxy extinction. One of the commonly used methods to estimate the reddening is to assume that all SN~II $(B-V)$ colour curves evolve similarly, and the objects which appear to be offset by a constant are reddened by host-galaxy extinction \citep{schmidt92}. Despite the fact that the colours evolve differently, if one assumes that all SN~II colours at a given epoch are intrinsically similar (which we argue is not actually the case), the offset is directly related to the amount of dust. To increase the colour uniformity, one should pick an epoch where the physical conditions (mainly the temperature) are similar. However, even if a such an epoch is chosen (e.g., the plateau phase; \citealt{faran14a}), intrinsic differences still exist. For example, \citet{faran14b} used an average colour of (a maximum of) 11 slow-declining SNe~II (historically SNe~IIP) and (a maximum of) 8 fast-declining SNe~II (historically SNe~IIL). They showed that the fast-declining SNe~II are, on average, redder than the slow-declining events. \citet{pastorello04} and \citet{spiro14a}, using a sample of low-luminosity SNe~II and comparing the colour with a sample of normal SNe~II, found that low-luminosity SNe~II have intrinsic colours that are unusually red.

Using theoretical models, \citet{kasen09} confirmed the results of \citet{pastorello04}; they found a slight trend for brighter models (fast-declining SNe~II) to be bluer than fainter models at 50 days after the explosion.
Also from a theoretical point of view, \citet{dessart13} have tried to explain the intrinsic colour variations with differences in progenitor properties. They showed that the colour evolution of SNe~II is primarily driven by different progenitor radii, while the explosion energies have negligible influence. SNe~II with more extended progenitors are subject to weaker cooling through expansion and thus tend to be bluer. This latter conclusion was
supported by \citet{lisakov17}, who studied the low-luminosity SN~II family and showed that their model with the highest mixing-length parameter employed during stellar evolution produced SNe~II that appear redder. The mixing-length parameter and the radius anticorrelate; therefore, this again showed that SNe~II with more compact progenitors display redder colours, as the ejecta cool more quickly owing to the effects of expansion. In addition, \citet{lisakov17} suggested that the mass-loss rate suffered by the progenitor does not influence the colour.

An additional intrinsic progenitor parameter that is thought to play an important role in SN~II colour diversity is metallicity. Progenitor metallicity can affect the observed colours of SNe~II in two ways, as shown and discussed by \citet{dessart13}. First, the metallicity plays 
a role in the path taken by stellar evolution to produce the pre-SN progenitor. A lower metallicity progenitor will produce a more compact star prior to explosion. Following the above discussion, this will lead to redder SNe~II at early times.
Secondly, the abundance of metals in the progenitor star's envelope directly affects 
the strength of metal lines within the spectrum and hence the SN's colour.
At lower metallicity the flux at blue wavelengths is significantly higher owing to the reduced effects of line blanketing, leading to bluer colours for lower metallicity progenitors
throughout the photospheric phase.

Finally, during the last few years a number of studies have claimed that some (and maybe many) SNe~II show signs of interaction with circumstellar material (CSM) during 
the first few days post explosion \citep[e.g.,][]{khazov16,morozova16,yaron17,moriya17,dessart17,morozova17}. Such interaction with previously unaccounted material (neither in stellar evolution models nor in SN~II explosion models)
would provide an additional source of energy and is likely to keep the photosphere at higher temperatures (and therefore bluer colours) for longer periods of time. 

In summary, there appear to be three key progenitor parameters that are expected to play important roles in producing intrinsic diversity in SN~II colours and their evolution: (1) progenitor radius, (2) progenitor metallicity, and (3) the presence or absence of CSM close to the progenitor stars. How these parameters relate between different progenitors is not clear. Neither is the degree to which the effects of host-galaxy extinction play an important or secondary role in driving observed SN~II colours. Such questions can only be answered through the analysis of large samples of events where one can hope to cover as large an area of Nature's parameter space as possible. This drives the motivation of the current study: using a large sample of well-observed SNe~II with both photometric and spectral coverage to understand the underlying causes of differences in observed SN~II colours, as well as to better understand the SN~II colour-term corrections used in cosmological analyses.

For this purpose, we use the Carnegie Supernova Project-I sample (CSP-I; \citealt{ham06}; PIs Phillips \& Hamuy), one of the most complete SN~II datasets, with optical and near-infrared (NIR) light curves in a well-understood photometric system. We focus our study on the photospheric phase, from the explosion until the SN falls onto the $^{56}$Ni decay slope. This is motivated mainly by the lack of data during the nebular phase and because for SN~II cosmology, the Hubble diagram is derived for an epoch $\sim 50$ days after the explosion (i.e. during the recombination phase).

The paper is organised as follows. Section 2 contains a brief description of the dataset. In Section 3, we show a detailed study of the $(B-V)$ colour-curve properties and search for possible correlations with the photometric/spectroscopic parameters presented by \citet{gutierrez17a}, which are a refined version of the parameters discussed by \citet{anderson14a,gutierrez14}. Three other colours [$(u-g)$, $(g-r)$, and $(g-Y)$] are analysed in Section 4, in the same vein as $(B-V)$. In Section 5, the host-galaxy extinction is examined. Finally, in Section 6 we more fully discuss our main results and try to understand them in terms of differences in progenitor properties. We provide our concluding remarks in Section 7. In the Appendix, the reader can find additional plots showing the correlations discussed in the main text.

\section{Data Sample}

\subsection{Data Reduction}\label{AKS}

In this work, the sample from \citet{dejaeger15b,dejaeger17a} is used. For this reason, only a brief description of the Carnegie Supernova Project-I (CSP-I; \citealt{ham06}) and the methods used to reduce the data are given in this section.

The CSP-I was a five-year (2004--2009) low-redshift SN survey providing optical and NIR light curves in a well-defined and well-understood photometric system. It was also designed to spectroscopically monitor the majority of the objects. Using the Las Campanas Observatory facilities (the Swope 1~m and the du Pont 2.5~m telescopes), CSP-I succeeded in building a large sample of 69 SNe~II, where 49 have both optical and NIR light curves with good temporal coverage (the other 20 SNe~II do not have NIR data). From this sample, as described by \citet{dejaeger15b,dejaeger17a}, we excluded four SNe~II (SN~2004ej, SN~2005hd, SN~2007X, and SN~2008K) from the analysis. The sample used in this work thus consists of the 65 SNe~II listed in Table \ref{table:table_BV}.

All optical images ($B$, $V$, $u$, $g$, and $r$) were reduced in a standard manner including bias subtraction, flat-field correction, a linearity correction, and an exposure-time correction for a shutter time delay. The final magnitudes were derived relative to a local sequence of stars and calibrated from observations of standard stars in the \citet{lan92} ($BV$) and \citet{smithja2002} ($ugri$) systems.

The NIR images were also reduced following standard steps consisting of dark subtraction, flat-field division, sky subtraction, and geometric alignment and combination of the dithered frames. NIR photometry ($Y$) was calibrated using standard stars in the \citet{persson04} system. For more information about the data processing technique and the different instruments/filters, the reader can refer to \citet{ham06}, \citet{contreras10}, \citet{stritzinger11}, and \citet{krisciunas17}.

All magnitudes are expressed in the natural CSP-I photometric system and are simultaneously corrected for Milky Way (MW) extinction and for the expansion of the Universe ($K$-correction; \citealt{oke68,hamuy93,kim96,nugent02}). All of the steps to make these corrections are fully outlined by \citet{dejaeger15b,dejaeger17a}. 

In this work, we use the recalibrated CSP-I photometry that will be published in a definitive CSP-I data paper by Contreras et al. (in prep.). The CSP-I SN~II spectroscopy was recently published by \citet{gutierrez17b}.

\begin{table*}
\caption{SN~II $(B-V)$ colour-curve parameters, host-galaxy information, and Milky Way extinction.}
\begin{threeparttable}
\scriptsize
\begin{tabular}{cccccc}
\hline
SN & host recession velocity & $E(B-V)_{\rm MW}$ & $s_{1,(B-V)}$ & $s_{2,(B-V)}$ & $T_{{\rm trans},(B-V)}$\\
&(km s$^{-1}$) & (mag) & (mag 100d$^{-1}$) & (mag 100d$^{-1}$) & (days)\\
\hline
\hline
2004er & 4411 & 0.0220 & 2.49 (0.10) & 0.32 (0.06) & 48.38 (1.36) \\
2004fb & 6100 &0.0546  & $\cdots$ ($\cdots$) & $\cdots$ ($\cdots$) &$\cdots$ ($\cdots$) \\
2004fc & 1831 & 0.0219 & 2.69 (0.11) & 0.87 (0.08) & 42.11 (2.00) \\
2004fx & 2673 & 0.0883 & 1.69 (0.24) & 0.50 (0.14) & 36.78 (3.36) \\
2005J  & 4183 & 0.0224 & 2.81 (0.08) & 0.71 (0.04) & 37.66 (0.84) \\
2005K  & 8204 & 0.0340 & 1.63 (0.20) & $\cdots$ ($\cdots$) &$\cdots$ ($\cdots$) \\
2005Z  & 5766 & 0.0236 & 3.02 (0.09) & 0.82 (0.11) & 35.13 (1.46) \\
2005af & 563 & 0.1588  & $\cdots$ ($\cdots$) & $\cdots$ ($\cdots$) &$\cdots$ ($\cdots$) \\
2005an & 3206 & 0.0819 & 3.41 (0.07) & 1.06 (0.10) & 37.76 (0.68) \\
2005dk & 4708 & 0.0409 & 3.34 (0.22) & 1.25 (0.05) & 37.31 (0.91) \\
2005dn & 2829 & 0.0437 & 1.16 (0.16) & 0.40 (0.03) & 34.23 (1.58) \\
2005dt & 7695 & 0.0248 & 2.28 (0.35) & 0.53 (0.15) & 43.16 (3.91) \\
2005dw & 5269 & 0.0194 & 2.38 (0.23) & 0.75 (0.12) & 38.25 (2.27) \\
2005dx & 8012 & 0.0205 & 3.64 (0.20) & 0.98 (0.64) & 30.88 (8.36) \\
2005dz & 5696 & 0.0691 & 2.22 (0.13) & 0.26 (0.08) & 43.91 (1.70) \\
2005es & 11287 & 0.0718 & 3.08 (0.44) & $\cdots$ ($\cdots$) &$\cdots$ ($\cdots$) \\
2005gk & 8773 & 0.0482 & 1.25 (0.08) & $\cdots$ ($\cdots$) &$\cdots$ ($\cdots$) \\
2005lw & 7710 & 0.0423 & 4.88 (0.52) & 0.62 (0.45) & 32.55 (2.43) \\
2005me & 6726 & 0.0203 & 3.18 (0.32) & 0.70 (0.20) & 33.60 (2.36) \\
2006Y & 10074 & 0.1114 & 1.67 (0.15) & $\cdots$ ($\cdots$) &$\cdots$ ($\cdots$) \\
2006ai & 4571 & 0.1080 & 2.03 (0.06) & $\cdots$ ($\cdots$) &$\cdots$ ($\cdots$) \\
2006bc & 1363 & 0.1762 & 2.34 (0.15) & 0.71 (0.32) & 41.19 (2.70) \\
2006be & 2145 & 0.0251 & 2.69 (0.05) & 1.01 (0.02) & 31.85 (0.39) \\
2006bl & 9708 & 0.0435 & 2.63 (0.09) & 1.25 (0.22) & 39.78 (3.64) \\
2006ee & 4620 & 0.0516 & 2.24 (0.18) & 0.81 (0.14) & 40.57 (2.81) \\
2006it & 4650 & 0.0850 & 3.30 (0.30) & $\cdots$ ($\cdots$) &$\cdots$ ($\cdots$) \\
2006iw & 9226 & 0.0426 & 3.07 (0.23) & 0.75 (0.20) & 37.87 (2.81) \\
2006ms & 4543 & 0.0300 & 3.75 (0.44) & $\cdots$ ($\cdots$) &$\cdots$ ($\cdots$) \\
2006qr & 4350 & 0.0384 & 3.06 (0.14) & 0.80 (0.11) & 36.78 (1.54) \\
2007P  & 12224 & 0.0349 & 2.17 (0.15) & $\cdots$ ($\cdots$) &$\cdots$ ($\cdots$) \\
2007U  & 7791 & 0.0449 & 3.30 (0.07) & 0.95 (0.24) & 34.75 (2.56) \\
2007W  & 2902 & 0.0443 & 2.73 (0.22) & 0.70 (0.11) & 33.39 (2.30) \\
2007aa & 1465 & 0.0227 & 1.77 (0.10) & 0.69 (0.06) & 35.89 (1.52) \\
2007ab & 7056 & 0.2292 & $\cdots$ ($\cdots$) & $\cdots$ ($\cdots$) &$\cdots$ ($\cdots$) \\
2007av & 1394 & 0.0311 & $\cdots$ ($\cdots$) & $\cdots$ ($\cdots$) &$\cdots$ ($\cdots$) \\
2007hm & 7540 & 0.0549 & 1.55 (0.20) & 0.47 (0.20) & 32.38 (4.10) \\
2007il & 6454 & 0.0403 & 2.12 (0.11) & 0.55 (0.09) & 49.14 (2.33) \\
2007it & 1193 & 0.0994 & 2.64 (0.33) & $\cdots$ ($\cdots$) &$\cdots$ ($\cdots$) \\
2007ld & 8994 & 0.0810 & 2.42 (0.34) & $\cdots$ ($\cdots$) &$\cdots$ ($\cdots$) \\
2007oc & 1450 & 0.0173 & 2.06 (0.11) & 0.72 (0.08) & 37.06 (1.52) \\
2007od & 1734 & 0.0312 & 2.97 (0.09) & 0.93 (0.03) & 27.07 (0.66) \\
2007sq & 4579 & 0.1769 & $\cdots$ ($\cdots$) & 1.41 (0.13) &$\cdots$ ($\cdots$) \\
2008F  & 5506 & 0.0422 & 1.89 (0.21) & $\cdots$ ($\cdots$) &$\cdots$ ($\cdots$) \\
2008M  & 2267 & 0.0389 & 2.30 (0.15) & 0.89 (0.17) & 45.43 (3.52) \\
2008W  & 5757 & 0.0837 & 2.64 (0.39) & 0.98 (0.11) & 32.54 (2.38) \\
2008ag & 4439 & 0.0713 & 1.38 (0.19) & 0.39 (0.04) & 58.89 (3.49) \\
2008aw & 3110 & 0.0351 & 2.54 (0.05) & 1.30 (0.06) & 48.14 (1.02) \\
2008bh & 4345 & 0.0187 & 3.11 (0.24) & 0.98 (0.20) & 37.56 (2.76) \\
2008bk & 230  & 0.0168 & 2.11 (0.32) & 0.46 (0.04) & 49.38 (1.93) \\
2008bp & 2723 & 0.0596 & $\cdots$ ($\cdots$) & $\cdots$ ($\cdots$) &$\cdots$ ($\cdots$) \\
2008br & 3019 & 0.0800 & 3.22 (0.22) & 1.07 (0.05) & 24.56 (1.19) \\
2008bu & 6630 & 0.3523 & 3.09 (0.45) & $\cdots$ ($\cdots$) &$\cdots$ ($\cdots$) \\
2008ga & 4639 & 0.5639 & $\cdots$ ($\cdots$) & $\cdots$ ($\cdots$) &$\cdots$ ($\cdots$) \\
2008gi & 7328 & 0.0549 & 2.90 (0.19) & 0.77 (0.31) & 37.84 (2.87) \\
2008gr & 6831 & 0.0113 & 2.60 (0.12) & 0.73 (0.22) & 38.06 (2.29) \\
2008hg & 5684 & 0.0158 & 3.31 (0.16) & $\cdots$ ($\cdots$) &$\cdots$ ($\cdots$) \\
2008ho & 3082 & 0.0163 & 2.20 (0.18) & $\cdots$ ($\cdots$) &$\cdots$ ($\cdots$) \\
2008if & 3440 & 0.0281 & 1.96 (0.08) & $\cdots$ ($\cdots$) &$\cdots$ ($\cdots$) \\
2008il & 6276 & 0.0142 & $\cdots$ ($\cdots$) & $\cdots$ ($\cdots$) &$\cdots$ ($\cdots$) \\
2008in & 1566 & 0.0193 & 2.75 (0.10) & 0.78 (0.15) & 39.86 (2.08) \\
2009N  & 1036 & 0.0182 & 3.06 (0.14) & 0.98 (0.12) & 37.92 (1.84) \\
2009ao & 3339 & 0.0332 & 1.91 (0.17) & $\cdots$ ($\cdots$) &$\cdots$ ($\cdots$) \\
2009au & 2819 & 0.0785 & 1.91 (0.05) & $\cdots$ ($\cdots$) &$\cdots$ ($\cdots$) \\
2009bu & 3494 & 0.0217 & 2.12 (0.24) & 0.88 (0.12) & 37.44 (3.04) \\
2009bz & 3231 & 0.0340 & 2.17 (0.08) & 0.95 (0.09) & 36.23 (1.48) \\
\hline
\hline
\end{tabular}
\scriptsize
Notes --- Column 1, SN name; Column 2, host-galaxy heliocentric recession velocity taken from the NASA Extragalactic Database (NED: \url{http://ned.ipac.caltech.edu/}); Column 3, reddening due to the Milky Way \citep{schlafly11}, from the NASA/IPAC infrared science archive website (\url{http://irsa.ipac.caltech.edu/cgi-bin/bgTools/nph-bgExec}); Columns 4, 5, and 6, the measured colour-curve parameters $s_{1,(B-V)}$, $s_{2,(B-V)}$, and $T_{{\rm trans},(B-V)}$ (respectively).
\label{table:table_BV}
\end{threeparttable}
\end{table*}

\subsection{Host-Galaxy Extinction}

As mentioned in Section \ref{AKS}, all magnitudes are corrected for Milky Way Galaxy extinction ($A_{VG}$) and for the Universe's expansion ($K$-correction), but not for host-galaxy extinction ($A_{Vh}$). In the literature, there are a handful of methods to estimate reddening. These include the \ion{Na}{i}~D equivalent width \citep{munari97,tur03,poznanski11} or the $(V-I)$ colour excess at the end of the plateau \citep{olivares10}. However, none of these methods seems to be satisfactory. The accuracy of using the \ion{Na}{i}~D line in low-resolution spectra (which is indeed the case for the current sample) has been questioned \citep{poznanski11,phillips13,galbany17}. For the colour method, we must assume that all SNe~II have the same intrinsic colour, which seems not to be true, as we will demonstrate in this work. The validity of these methods has also been questioned by \citet{faran14a}; using a variety of different host-galaxy extinction estimation methods, they found that when applied to SN~II correlations, the dispersion increased. This latter observation was also recently outlined by \citet{gutierrez17a}. Therefore, we decided to not correct for $A_{Vh}$ because (a) we believe that the extinction is probably not significant for most SNe~II in the sample, and (b) any attempts to correct for extinction are likely to simply add additional errors to our colours, given the very uncertain nature of the corrections. 

In Section \ref{txt:discussion_avh}, we discuss the effect of the dust extinction in more detail, and in Section \ref{txt:cc_intrin} the ``intrinsic'' colour dispersion of our sample is further explored. Note that as the $K$-corrections depend on the colours, we test whether the trends presented in this work are stronger with or without $K$-corrections. We find that all of the correlations are stronger when the $K$-corrections are applied, justifying and validating our inclusion of these corrections.

\section{$(B-V)$ Colour Curves}\label{txt:txt_BV}
 
In this section, we first analyse the morphology of the $(B-V)$ colour curves, and then compare observed colours at different epochs with a range of $V$-band photometric and optical spectroscopic parameters as presented by \citet{gutierrez17a}. The choice of the $(B-V)$ colour is motivated by the fact that historically, $B$ and $V$ photometry were the most widely obtained, hence allowing easier comparison with previous studies found in the literature.  Additionally, in the literature, the host-galaxy extinction is regularly given as the colour excess $E(B-V)$. Finally, note that \citet{anderson14a} used the $V$-band light curves in their study.

In this analysis, we use the Pearson test to determine the existence and strength of correlations between our defined colour parameters and other photometric and spectral properties. We perform 10,000 Monte Carlo bootstrapping simulations, and then the mean Pearson value ($r$) of these 10,000 simulations is calculated and presented in each figure, together with its standard deviation ($\sigma$). As in \citet{anderson14a}, we take a conservative approach: in each figure, only the upper limit (i.e. $p$-value calculated for $r$-$\sigma$) of the probability of finding such a correlation strength by chance is plotted\footnote{To obtain the lower limit we use the free $p$-value calculator provided by Dr. Daniel Soper: \url{http://www.danielsoper.com/statcalc/calculator.aspx?id=44 .}}. Finally, it is worth noting that in this paper, the colours themselves are linearly interpolated to different epochs.

\subsection{Measured Parameters}\label{txt:params}

\begin{figure}
\includegraphics[width=9.0cm]{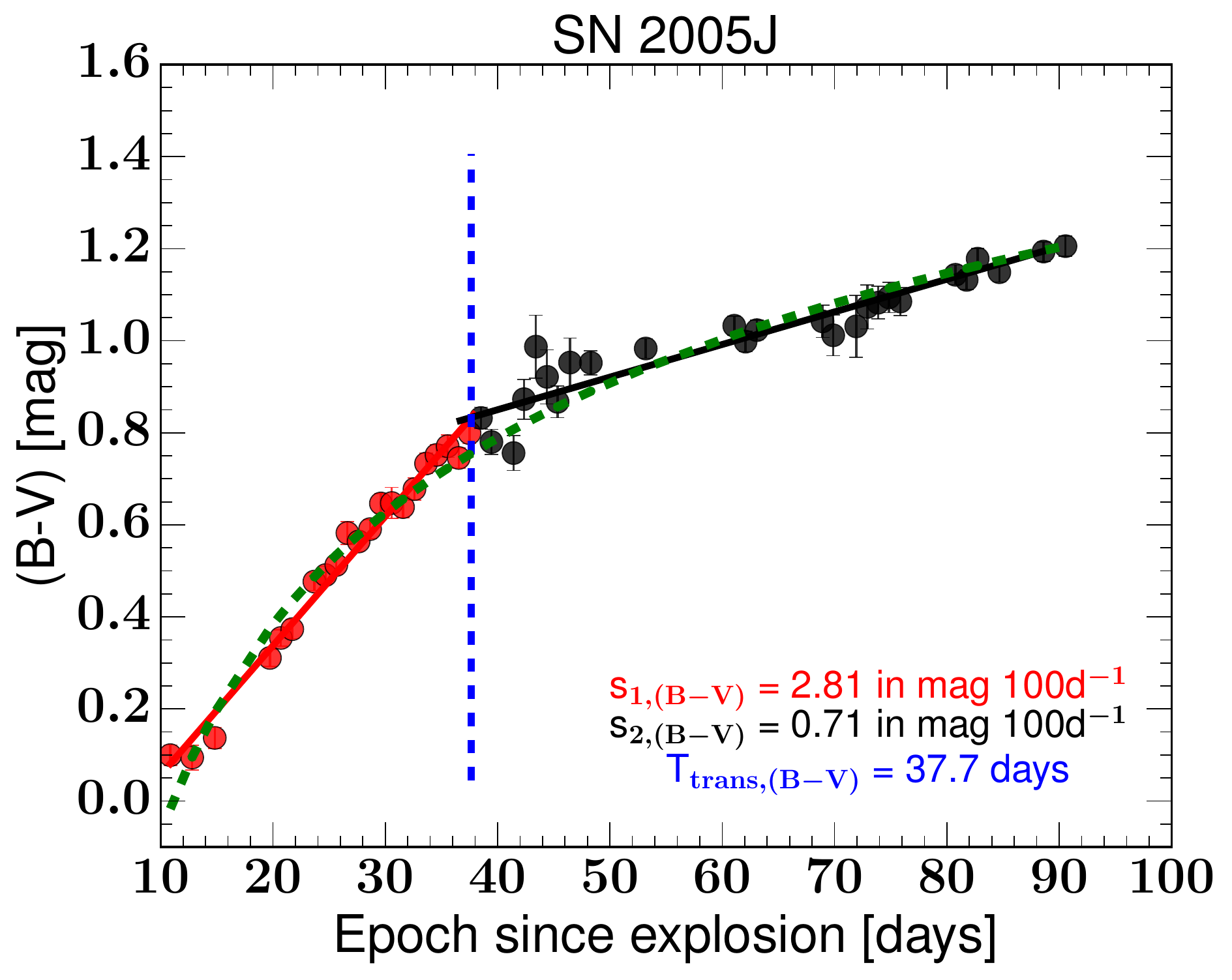}
\caption{An example of the $(B-V)$ colour-curve parameters measured for each SN as applied to SN~2005J. The two slopes, $s_{1,(B-V)}$ and $s_{2,(B-V)}$, are respectively represented in red and black. The epoch of the transition, $T_{{\rm trans},(B-V)}$, is indicated with a vertical blue dashed line. These parameters for our full sample are listed in Table \ref{table:table_BV}. The dashed green line represents the power-law fit, $(B-V) = 3.894 - 5.694\, t^{-0.177}$.}
\label{fig:CC_parameters}
\end{figure}

The SN~2005J $(B-V)$ light curve is shown in Figure \ref{fig:CC_parameters} together with a schematic of the $(B-V)$ colour-curve parameters chosen for measurement. Looking at the $(B-V)$ colour curves of the whole sample and as noted by \citet{patat94}, two regimes composed of two different slopes are apparent (see Figure \ref{fig:CC_parameters} and Appendix \ref{appendix:colour_fit}). During the first $\sim 30$--40 days, the object becomes quickly redder as the SN envelope expands and cools. After this first phase, the SN colour changes more slowly as the rate of cooling decreases. Following the notation from \citet{anderson14a}, we defined the following parameters.

\begin{enumerate}
\item {$s_{1,(B-V)}$: the first colour rate of change in magnitudes per 100 days since the explosion date (hereafter mag 100d$^{-1}$).} 
\item {$s_{2,(B-V)}$: the second colour rate of change in mag 100d$^{-1}$ after $T_{{\rm trans},(B-V)}$ and until the end of the plateau phase. $s_{2,(B-V)}$ is always observed to be flatter than $s_{1,(B-V)}$ (as first observed by \citealt{patat94}).}
\item {$T_{{\rm trans},(B-V)}$: the epoch in days from explosion of the transition between $s_{1,(B-V)}$ and $s_{2,(B-V)}$.} 
\end{enumerate}

\indent To measure the free parameters $s_{1,(B-V)}$, $s_{2,(B-V)}$, and $T_{{\rm trans},(B-V)}$, we use a Python program which consists of performing a weighted least-squares fit of the ``$AK$-corrected'' colour curves (Milky Way extinction and $K$-correction) with one and two slopes. To choose between one or two slopes, an $F$-test\footnote{Fisher test or Wald test.} is performed. The fit is only done from the explosion date \citep{gutierrez17a} to the end of the plateau (i.e. during the optically thick phase). When the epoch of the end of the plateau is unknown, we choose 80 days, which is the average value of our sample \citep{anderson14a}. If the $F$-test favoured one slope but the first point starts after 35 days past the explosion, we consider the slope as $s_{2,(B-V)}$ and not $s_{1,(B-V)}$, as the transition occurs on average at $\sim 38$ days. Note that we also tried to fit the colour curves using a power law; however, in the majority of cases ($> 55$\%), the one-line/two-lines fitting is statistically better than the power-law fitting. The measured parameters are listed in Table \ref{table:table_BV}. From the colour fit, we identify two very noisy SNe~II (SN~2007ab and SN~2008ga) for which only a cloud of points is seen. In the rest of the $(B-V)$ section, these two SNe~II are omitted from our analysis. Note also that 5 SNe (SN~2004fb, SN~2005af, SN~2007av, SN~2008bp, and SN~2008il) have no values for their colour slopes and transition epochs owing to a lack of data (fewer than four epochs before the end of the plateau).

\subsection{Distributions}\label{txt:BV_correlation}

\begin{figure}
\includegraphics[width=9.0cm]{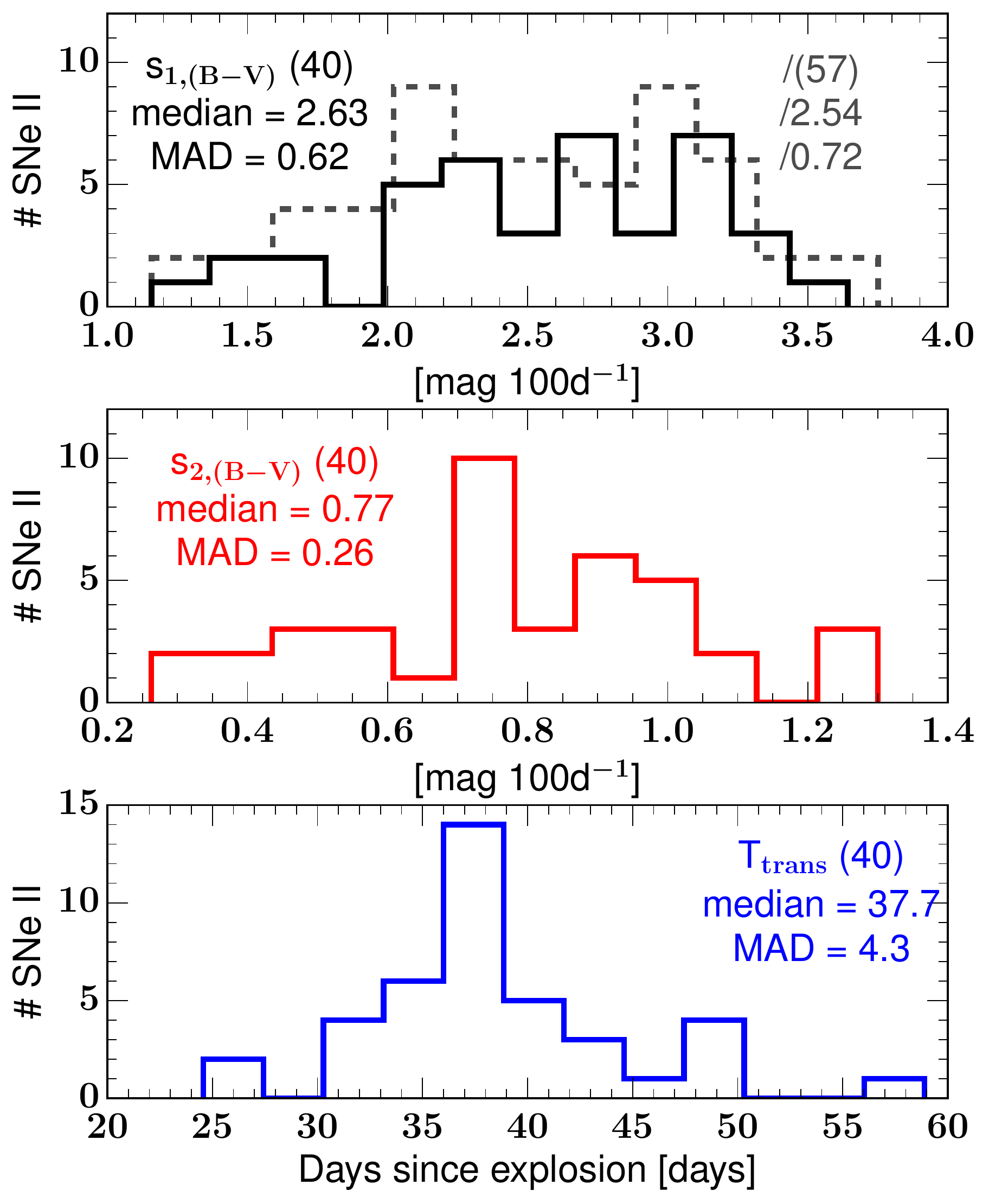}
\caption{Histograms of the three measured parameters of SN~II $(B-V)$ colour curves. \textit{Top:}
initial colour curves from explosion, $s_{1,(B-V)}$. The grey dashed histogram includes SNe which do not show two slopes but only one. \textit{Middle:} second colour curves after transition, $s_{2,(B-V)}$. 
\textit{Bottom:} epoch of the transition, $T_{{\rm trans},(B-V)}$. In each panel, the number of SNe is listed, together with
the median and the median absolute deviation (MAD).}
\label{fig:BV_params_distribution}
\end{figure}

\indent In Figure \ref{fig:BV_params_distribution}, histograms of the three $(B-V)$ colour-curve parameter distributions ($s_{1,(B-V)}$, $s_{2,(B-V)}$, and $T_{{\rm trans},(B-V)}$) are displayed. After explosion, the colour increases with a median slope of $2.63 \pm 0.62$ mag 100d$^{-1}$ followed by a less steep rise at a rate of $0.77 \pm 0.26$ mag 100d$^{-1}$. The average epoch of the transition between the two regimes is $37.7 \pm 4.3$ days after the explosion. If we also include SNe~II which do not show two slopes but only one, the $s_{1,(B-V)}$ slope decreases to $2.54 \pm 0.72$ mag 100d$^{-1}$, consistent within the uncertainties (grey dashed histogram in Figure \ref{fig:BV_params_distribution}, top panel). An important result from the slope distributions is that they are not bimodal, confirming the result found by \citet{anderson14a,sanders15,valenti16,galbany16a} --- that is, SNe~II form a continuous class. This was also observed by \citet{faran14b}, where the authors found similar evolution between the colours of slow- and fast-declining SNe~II.

\begin{figure}
\includegraphics[width=9.0cm]{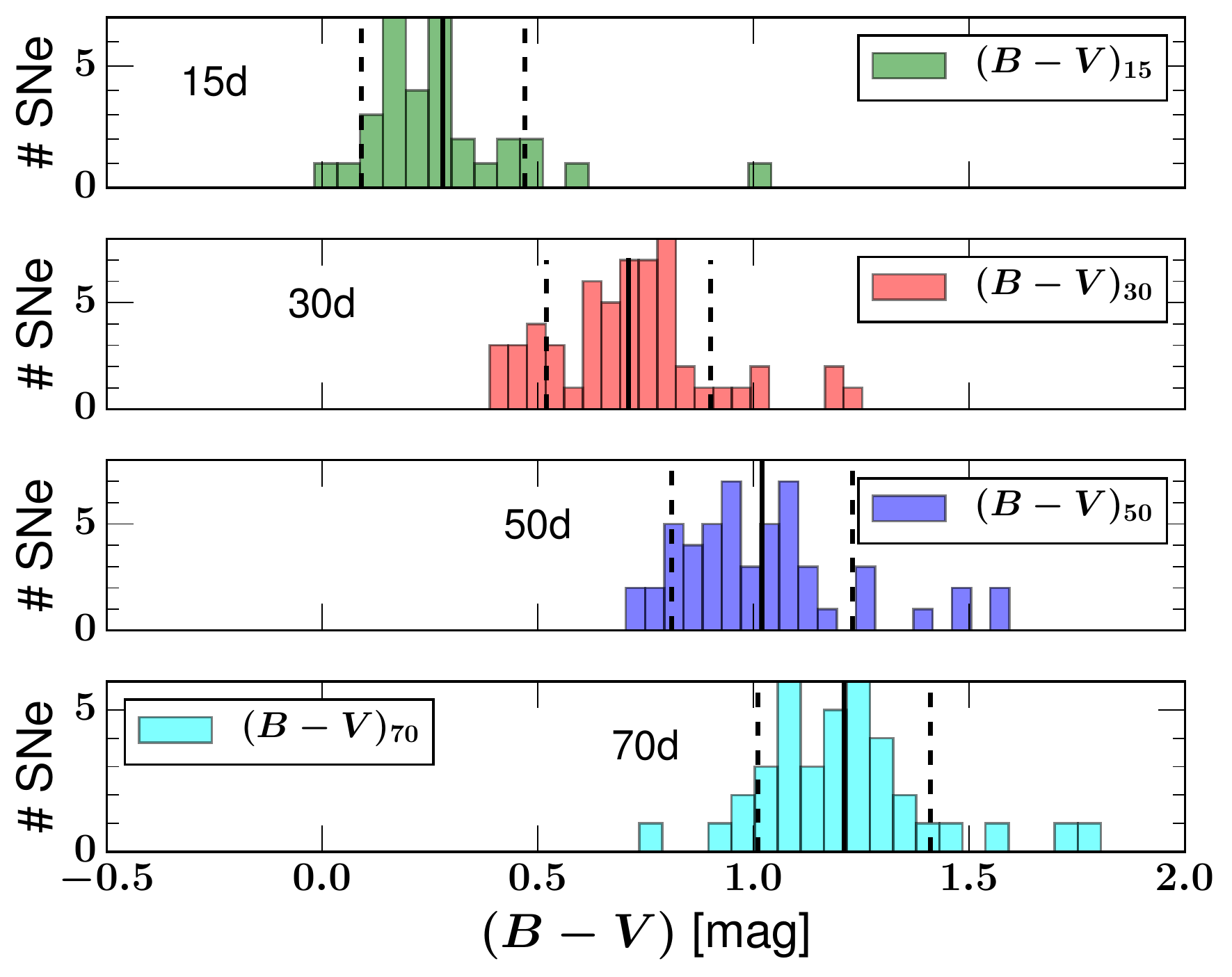}
\caption{$(B-V)$ colour distributions at four different epochs: 15d (first from top), 30d (second from top), 50d (third from top), and 70d (bottom) after the explosion. The solid line is the average value while the dashed lines are the $1\sigma$ uncertainty.}
\label{fig:BV_distribution}
\end{figure}

\indent Figure \ref{fig:BV_distribution} illustrates the $(B-V)$ colour distribution at four epochs: 15, 30, 50, and 70 days after the explosion (hereafter 15d, 30d, 50d, and 70d, respectively). As expected, the distribution at early times is bluer ($<(B-V)_{15}> = 0.28 \pm 0.19$ mag) and quickly becomes redder with time ($<(B-V)_{30}> = 0.71 \pm 0.19$ mag) until the plateau phase ($<(B-V)_{50}> = 1.02 \pm 0.21$ mag), and finally evolves slowly during the recombination phase ($<(B-V)_{70}> = 1.21 \pm 0.20$ mag). The fast evolution at early epochs is expected, as the temperature drops more quickly than at late epochs when slower evolution is seen \citep{faran17}.

\begin{figure}
\includegraphics[width=9.0cm]{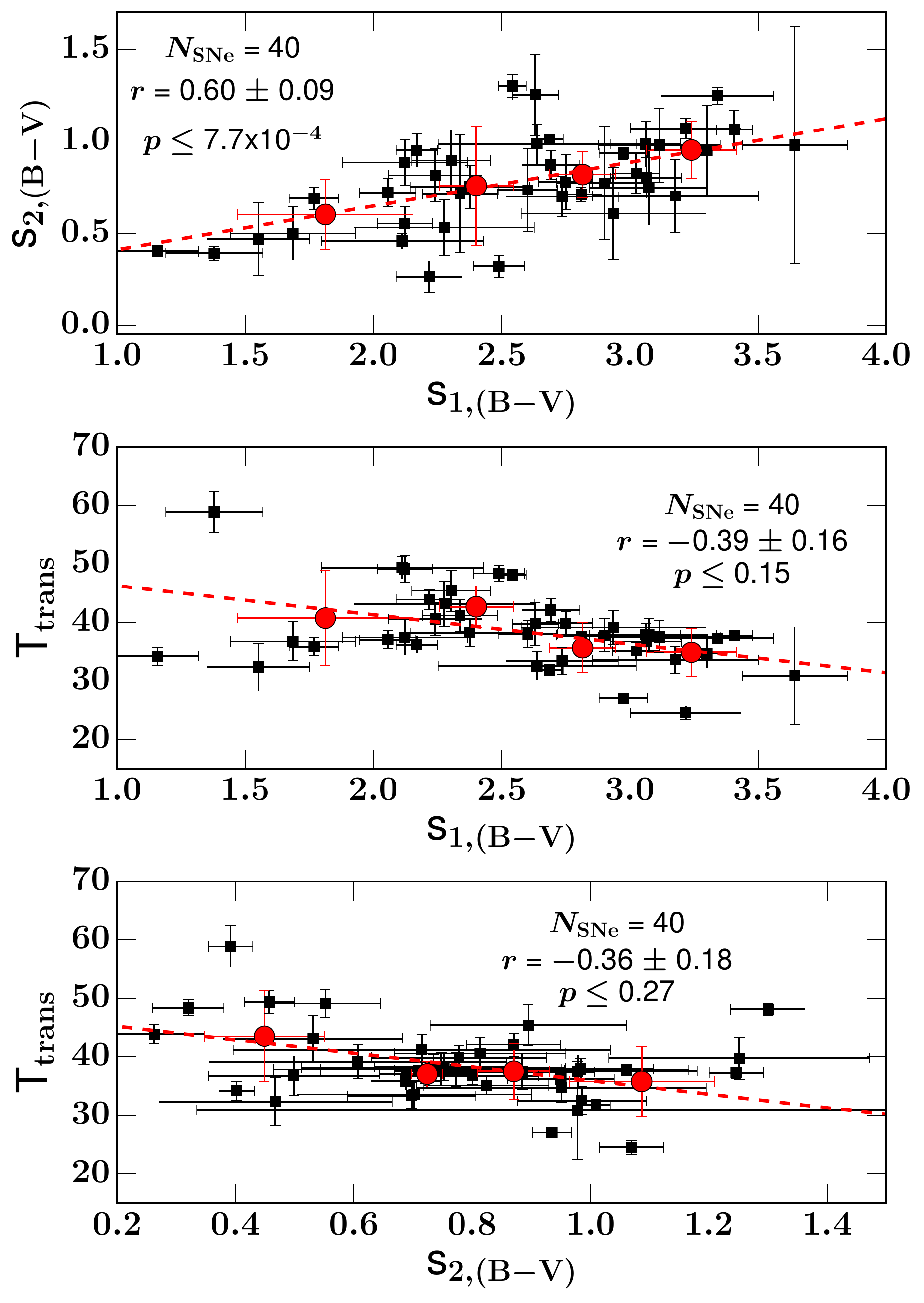}
\caption{Correlations between the three measured parameters of the SN~II $(B-V)$ colour curves. {\it Top:} $s_{1,(B-V)}$ with $s_{2,(B-V)}$, both expressed in mag 100d$^{-1}$. {\it Middle:} $s_{1,(B-V)}$ with $T_{{\rm trans},(B-V)}$ (in days). {\it Bottom:} $s_{2,(B-V)}$ with $T_{{\rm trans},(B-V)}$. In each panel the number of SNe is listed, together with the Pearson factor and the $p$-value ($p$), which indicates the probability of an uncorrelated system producing datasets that have a Pearson factor at least as extreme as the one computed from these datasets.}
\label{fig:BV_correlations}
\end{figure}

\indent In Figure \ref{fig:BV_correlations}, we show how the measured parameters correlate. A strong correlation between the two slopes $s_{1,(B-V)}$ and $s_{2,(B-V)}$ is found with a Pearson factor ($r$) of $0.60 \pm 0.09$ ($p \leq 7.7 \times 10^{-4}$). This implies that SN~II $(B-V)$ colour curves with a slower initial rise also have a slower cooling after transition. Two other trends are also observed between $s_{1,(B-V)}/T_{{\rm trans},(B-V)}$ and $s_{2,(B-V)}/T_{{\rm trans},(B-V)}$ with a Pearson factor of $-0.39 \pm 0.16$ ($p \leq 0.15$) and $-0.36 \pm 0.18$ ($p \leq 0.27$), respectively.
This suggests that SN~II $(B-V)$ colour curves with a faster evolution (steeper $s_{1,(B-V)}$ or $s_{2,(B-V)}$) also have a transition at earlier epochs, although at low statistical significance.

\indent We also search for correlation between these three $(B-V)$ colour parameters and the colour at four different epochs: 15d, 30d, 50d, and 70d after the explosion. A slight anticorrelation between $T_{{\rm trans},(B-V)}$ and the colour $(B-V)$ at 15d ($r = -0.53 \pm 0.14$, $p \leq 0.11$) is seen: bluer SNe~II at early epochs have an earlier transition. However, this is not statistically seen at later epochs (30d, 50d, and 70 d). Finally, as expected, the colours at different epochs correlate strongly: bluer SNe~II at 15d are also bluer at 70d.\\

\subsection{Photometric/Spectroscopic Correlations}\label{txt:BV_spec_phot_correlation}

\indent In this section, we investigate correlations between the $(B-V)$ colour-curve parameters defined in Section \ref{txt:params} with spectroscopic \citep{gutierrez17a} and photometric parameters \citep{gutierrez17a}.

\subsubsection{Photometric Parameters}\label{txt:BV_phot_corr}

\begin{figure}
\includegraphics[width=9.0cm]{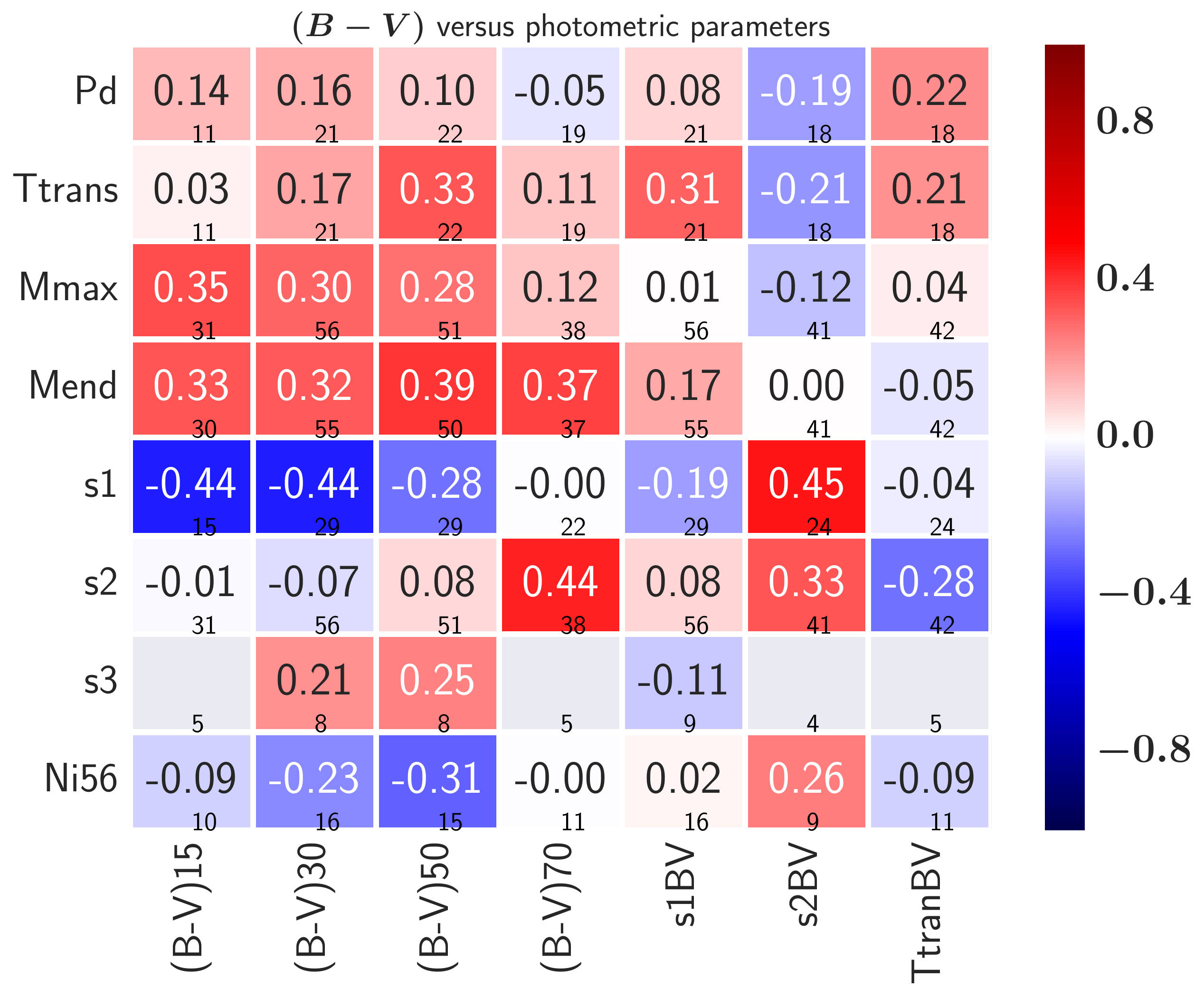}
\caption{The matrix correlation between the seven parameters measured from the SN~II $(B-V)$ colour curves ($B-V$ at 15d, 30d, 50d, 70d; $s_{1,(B-V)}$; $s_{2,(B-V)}$; $T_{\rm trans,(B-V)}$) versus the $V$-band light-curve parameters presented by \citet{gutierrez17a}. For each pair, the Pearson factor and the size of the sample are given. The colour bar represents the strength of the relationship. Note that a larger negative number indicates a stronger anticorrelation, while a larger positive number indicates a stronger correlation.}
\label{fig:matrix_BV_photo}
\end{figure}

We use a selection of 8 photometric parameters discussed by \citet{gutierrez17a}: plateau duration ($Pd$); transition between $s_{1}/s_{2}$ ($T_{\rm trans}$); absolute $V$-band magnitude at maximum brightness ($M_{\rm max}$); absolute $V$-band magnitude measured 30~d before $t_{PT}$ ($M_{\rm end}$), where $t_{PT}$ is the midpoint of the transition from plateau to radioactive tail; decline rate of the initial slope of the $V$-band light curve ($s_{1}$); slope of the plateau phase ($s_{2}$); slope of the radiative tail ($s_{3}$); and $^{56}$Ni mass including lower limits ($^{56}$Ni; see \citealt{anderson14a} for more details). In Figure \ref{fig:matrix_BV_photo}, the correlation matrix between these photometric parameters and the $(B-V)$ colour-curve parameters defined above is displayed. From this figure the following trends emerge.

\begin{enumerate}
\item{$(B-V)$ at both 15d and 30d shows some (low significance) correlations with the absolute magnitude, $M_{\rm max}$ (respectively $r = 0.35 \pm 0.16$, $p \leq 0.31$; $r = 0.30 \pm 0.16$, $p \leq 0.30$) and $M_{\rm end}$ (respectively $r = 0.33 \pm 0.18$, $p \leq 0.43$; $r = 0.32 \pm 0.17$, $p \leq 0.27$): more luminous SNe~II are generally bluer (Figure \ref{fig_appendix:redder_fainter}). The later epochs (50d--70d) also correlate with the absolute magnitude, but mainly with $M_{\rm end}$ (respectively $r = 0.39 \pm 0.08$, $p \leq 2.8 \times 10^{-2}$; $r = 0.37 \pm 0.17$, $p \leq 0.23$). The correlations between the colour at early epochs and the absolute magnitude at maximum brightness could be caused by dust effects (redder SNe~II are fainter); however, the strength of this relationship decreases at later epochs (50d and 70d), suggesting that it is driven by intrinsic SN colour differences rather than uncorrected extinction. This is also supported by the analysis presented in Section \ref{txt:discussion_avh}, where we show that the strength of these correlations increases when one removes SNe most likely suffering from significant host-galaxy reddening. Diagrams presenting these magnitude-colour trends are found in Figure \ref{fig_appendix:redder_fainter}.\\}

\item{$(B-V)$ at 30d and $s_{1}$ anticorrelate ($r = -0.45 \pm 0.17$, $p \leq 9.9 \times 10^{-2}$) in the sense that bluer SNe~II at early epochs show a steeper decline in the $V$ band after maximum brightness (Figure \ref{fig:appendix_redder_IIL}).\\}

\item{$(B-V)$ at 70d and $s_{2}$ correlate ($r = 0.44 \pm 0.07$, $p \leq 2.2 \times 10^{-2}$) in the sense that redder SNe~II at later epochs have a steeper plateau slope; they are fast-declining SNe~II (Figure \ref{fig:appendix_redder_IIL}).}

\end{enumerate}

From this analysis, we conclude that low-luminosity SNe~II are on average redder, and that the colours are dominated by intrinsic differences, since (in addition to the above arguments) the degree of host-galaxy extinction is unlikely to correlate with the decline rates ($s_{1}$ and $s_{2}$). 

\subsubsection{Spectroscopic Parameters}\label{txt:BV_spec_corr}

\begin{figure}
\includegraphics[width=9.0cm]{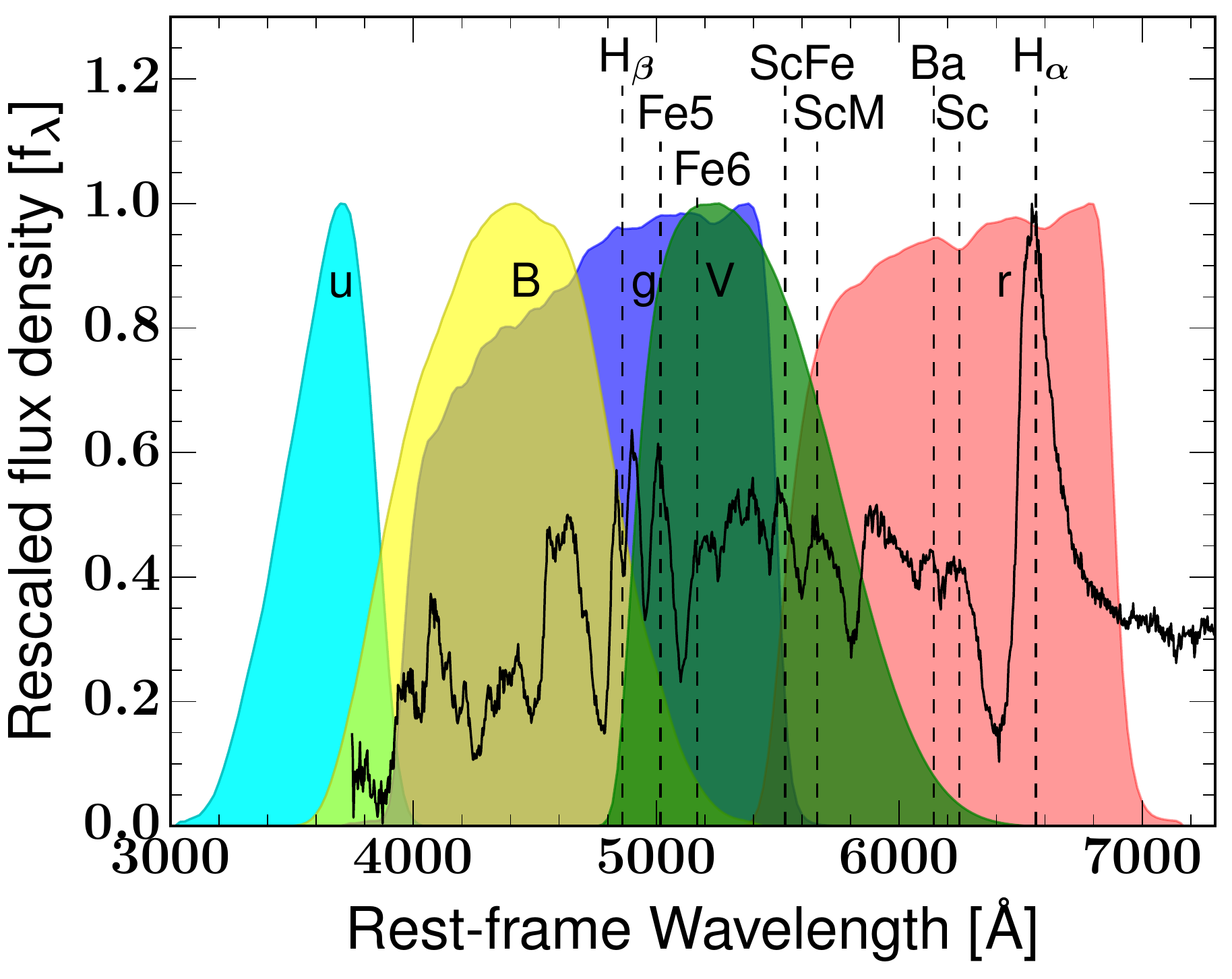}
\caption{SN~2005J spectrum at 65~d after the explosion. The rest-frame wavelengths (not the absorption minima) of H$\alpha$ $\lambda 6563$, H$\beta$ $\lambda 4861$, \ion{Fe}{II} $\lambda 5018$ (Fe5), \ion{Fe}{II} $\lambda 5169$ (Fe6), \ion{Sc}{II}/\ion{Fe}{II} $\lambda 5531$ (ScFe), \ion{Sc}{II} multiplet $\lambda 5663$ (ScM), \ion{Ba}{II} $\lambda 6142$ (Ba), and \ion{Sc}{II} $\lambda 6247$ (Sc) are marked with dashed lines. Optical filter transmission curves ($u$, $B$, $g$, $V$, and $r$) are respectively shown in cyan, yellow, blue, green, and red.}
\label{fig:spect_filter}
\end{figure}

\begin{figure}
\includegraphics[width=9.0cm]{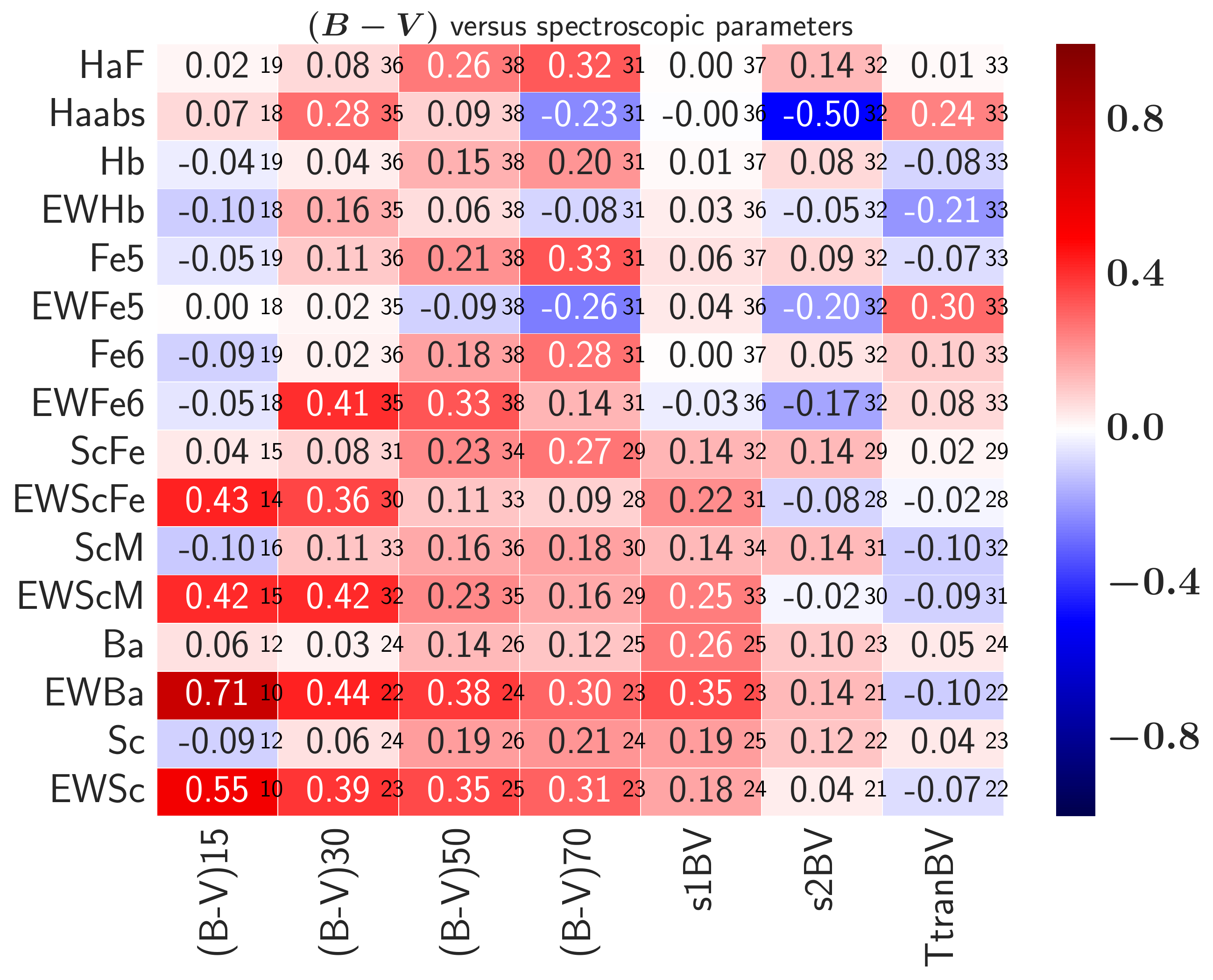}
\caption{The correlation matrix between the seven parameters measured from the SN~II $(B-V)$ colour curves ($B-V$ at 15d, 30d, 50d, 70d; $s_{1,(B-V)}$, $s_{2,(B-V)}$, $T_{{\rm trans},(B-V)}$) versus the spectroscopic parameters presented by \citet{gutierrez17a}. For each pair, the Pearson factor and the size of the sample are given. The colour bar represents the strength of the relationship.}
\label{fig:matrix_BV_spec}
\end{figure}

In Figure \ref{fig:matrix_BV_spec}, the correlation matrix between our 7 parameters derived from the $(B-V)$ colour curves and 16 spectroscopic parameters from \citet{gutierrez17a} is displayed. We used the following spectroscopic parameters: H$\alpha$ $\lambda 6563$ velocity obtained from the full width at half-maximum intensity of the emission line (HaF), equivalent width of the H$\alpha$ absorption (Haabs), H$\beta$ $\lambda 4861$ velocity (Hb), H$\beta$ equivalent width (EWHb), and the velocity (measured from the absorption minimum) and the equivalent width of various lines: \ion{Fe}{II} $\lambda 5018$ (Fe5), \ion{Fe}{II} $\lambda 5169$ (Fe6), \ion{Sc}{II}/\ion{Fe}{II} $\lambda 5531$ (ScFe), \ion{Sc}{II} multiplet $\lambda 5663$ (ScM), \ion{Ba}{II} $\lambda 6142$ (Ba), and \ion{Sc}{II} $\lambda 6247$ (Sc). All of the velocities and equivalent widths are measured at 50d. Figure \ref{fig:spect_filter} shows an example of a SN~II spectrum (SN~2005J) at $\sim 65$~d after the explosion with all the used lines and the different filters displayed. In this analysis, we only include equivalent-width measurements when each specific line is detected (i.e. we do not include nondetections). From this figure the following trends are of interest.

\begin{enumerate}
\item{$(B-V)$ colour at different epochs correlates with the equivalent width of different elements. These trends are expected given that metal-line strengths are strongly dependent on the temperature of the line-forming region. At early times (15d), the $(B-V)$ colour correlates with EWBa ($r = 0.71 \pm 0.07$, $p \leq 5.1 \times 10^{-2}$), and slightly with EWSc ($r = 0.54 \pm 0.19$, $p \leq 0.31$). Later, at 30d, the colours appear to correlate with EWBa ($r = 0.44 \pm 0.11$, $p \leq 0.13$), EWScM ($r = 0.42 \pm 0.14$, $p \leq 0.11$), and EWFe6 ($r = 0.41 \pm 0.15$, $p \leq 0.13$), while still later (50d) only a trend with EWBa ($r = 0.38 \pm 0.19$, $p \leq 0.37$) is seen. All of the correlations are plotted in Figure \ref{fig_appen:EW_redder} and indicate that redder SNe~II have stronger metal-line equivalent widths.\\}

\item{$s_{2,(B-V)}$ anticorrelates with the strength of the H$\alpha$ absorption line (Haabs) in the sense that SNe~II with a rapid cooling after transition exhibit a weaker H$\alpha$ absorption line: $r = -0.50 \pm 0.07$ and $p \leq 1.4 \times 10^{-2}$ (Figure \ref{fig:Haabs_s2colour}).\\}

\item{No statistically significant evidence for a linear relation between the colour and the different velocities is seen.}

\end{enumerate}

\section{$(\lowercase{u-g})$, $(\lowercase{g-r})$, and $(\lowercase{g}-Y)$ Colour Curves}\label{txt:other_colours}

In this section, we analyse three others colour curves [$(u-g)$, $(g-r)$, and $(g-Y)$] and investigate the same trends as previously presented using $(B-V)$. The colour selection was motivated by the aim of studying the possible relation between the metallicity and the colour, H$\alpha$ line effects, and one colour that takes advantage of the NIR. We use only Sloan filters (+$Y$), as this is currently the most widely adopted photometric system and future projects such as the Large Synoptic Survey Telescope will use similar filters. For the three colours, the procedure used in Section \ref{txt:txt_BV} is followed; nevertheless, we do not show all of the figures here, displaying some of them in the Appendix \ref{AppendixA}. Note also that for the three colours, various examples of colour fitting are displayed in Figure \ref{fig:append_colour_fit}.

\subsection{$(\lowercase{u-g})$}

\subsubsection{Distributions}

The parameters measured from $(u-g)$ colour curves have a median $s_{1,(u-g)}$ value of $6.66 \pm 2.14$ mag 100d$^{-1}$, followed by a second slope with a median $s_{2,(u-g)}$ value of $2.53 \pm 1.08$ mag $^{-1}$, and a transition $T_{{\rm trans},(u-g)}$ at $35.3 \pm 4.7$ d after the explosion. As expected, the slopes $s_{1,(u-g)}$ and $s_{2,(u-g)}$ are steeper than those found with $(B-V)$. This is clearly due to the combination of the temperature (cooling with time) and the strongest sensitivity of the bluest bands to temperature changes. The effect is also seen in observed SN~II light curves, where the $u$ band exhibits a faster decline than the other bands \citep{galbany16a}, as well in theoretical models \citep{kasen09,dessart13}. Note also, in our SN~II sample, only 17 SNe~II have $u$-band data at later epochs, and thus for only 17 SNe~II is $s_{2,(u-g)}$ measured.\\
\indent 
Regarding the correlations between the two colour-curve parameters (as for $(B-V)$), $s_{1,(u-g)}$ correlates with $s_{2,(u-g)}$ with a Pearson factor of $0.70 \pm 0.11$ ($p \leq 1.3 \times 10^{-2}$), and trends are found between $s_{1,(u-g)}/T_{{\rm trans},(u-g)}$ and $s_{2,(u-g)}/T_{{\rm trans},(u-g)}$ with respective coefficients of $-0.49 \pm 0.18$ ($p \leq 0.22$) and $-0.34 \pm 0.22$ ($p \leq 0.65$). Although these trends are weaker than those derived for $(B-V)$ (Section \ref{txt:BV_correlation}), they are consistent, confirming that SNe~II with a slower initial rise/cooling in their colour curves also show a slower cooling after the transition. These correlations are depicted in Figure \ref{appendix:colour_params_corre}.
 
\indent As for the $(B-V)$ colour, we find an anticorrelation between $T_{{\rm trans},(u-g)}$ and $(u-g)$ at 15d, but the number of SNe~II is small ($r = -0.67 \pm 0.17$, $p \leq 0.25$ for 7 SNe~II). Finally, the $(u-g)$ colour distributions at four epochs evolve faster than $(B-V)$ as they become quickly redder with time: $<(u-g)_{15}> = 0.49 \pm 0.27$ mag, $<(u-g)_{30}> = 1.53 \pm 0.35$ mag, $<(u-g)_{50}> = 2.36 \pm 0.29$ mag, and $<(u-g)_{70}> = 2.79 \pm 0.31$ mag at epochs 15d, 30d, 50d, and 70d, respectively.

\subsubsection{Photometric Correlations}\label{txt:photo_ug}

The relationships found earlier between the absolute magnitude at maximum brightness ($M_{\rm max}$) and the colours are also seen using the $(u-g)$ colour. However, this correlation is only apparent at epoch 30d (Figure \ref{fig:redder_fainter}): $r = 0.51 \pm 0.12$ ($p \leq 2.0 \times 10^{-2}$). A trend is also seen for this colour and $M_{\rm end}$ ($r = 0.36 \pm 0.25$, $p \leq 0.43$, Figure \ref{fig_appendix:redder_fainter}), while at the other epochs trends are weak or nonexistent.

We also recover the anticorrelation found previously between the colour at 30d with $s_{1}$: $r = -0.42 \pm 0.15$, $p \leq 0.21$ (Figure \ref{fig:appendix_redder_IIL}). Regarding the correlation found with $(B-V)$ between the colour at 70d and $s_{2}$, here our analysis is restricted by the low number of available events ($N=7$).

\subsubsection{Spectroscopic Correlations}\label{ug_spec_corr}

The comparison between $(u-g)$ colours and spectroscopic parameters may reveal interesting trends, given that $u$-band observations are known to be strongly affected by line-blanketing effects.

The $(u-g)$ colour also correlates with the equivalent width of various elements at different epochs. The correlations seen are on average stronger than for the $(B-V)$ colour. At 15d with EWScM ($r = 0.57 \pm 0.15$, $p\leq 0.13$); at 30d with EWFe6 ($r = 0.53 \pm 0.18$, $p \leq 7.9 \times 10^{-2}$), EWScFe ($r = 0.49 \pm 0.15$, $p \leq 0.12$), and EWScM ($r = 0.48 \pm 0.12$, $p \leq 8.3 \times 10^{-2}$); at 50d with the same elements but weaker, EWFe6 ($r = 0.42 \pm 0.17$, $p\leq 0.25$), EWScFe ($r = 0.40 \pm 0.14$, $p\leq 0.25$), and EWScM ($r = 0.40 \pm 0.19$, $p\leq 0.34$). At later epochs (70d), the number of SNe is too low to enable any clear conclusions. Specific correlations are shown in Figure \ref{fig_appen:EW_redder}.

We find that $s_{1,(u-g)}$ seems to correlate with different equivalent widths such as EwFe5, EWScM, EWBa, and EWScM. However, in each case, the uncertainty in the Pearson coefficient is high, and thus the $p$-value derived from the upper limit is also high. For example, for EWScFe we have $r =0.43 \pm 0.21$ ($p\leq 0.26$), and for EWBa we have $r = 0.35 \pm 0.34$ ($p\leq 0.97$). For all these correlations, the low-luminosity SN~2008bk seems to be an outlier. If we remove this SN, for example in the EWBa relationship, the Pearson factor increases to $r = 0.73 \pm 0.12$ with a very low $p$-value of $p \leq 9.3 \times 10^{-3}$. This suggests that SNe~II with faster initial cooling have larger absorption-line equivalent widths. 

\subsection{$(\lowercase{g-r})$}

\subsubsection{Distributions}\label{distri_gr}

The $(g-r)$ colour curves are characterised by an initial slope ($s_{1,(g-r)}$) with a median value of $1.90 \pm 0.53$ mag 100d$^{-1}$ lasting $37.9 \pm 4.3$ d ($T_{{\rm trans},(g-r)}$), followed by a second slope with a median $s_{2,(g-r)}$ of $0.68 \pm 0.25$ mag 100d$^{-1}$. As for the previous colours, a strong correlation between $s_{1,(g-r)}$ and $s_{2,(g-r)}$ is found ($r = 0.43 \pm 0.11$, $p\leq 4.7 \times 10^{-2}$), and a weaker anticorrelation between $s_{1,(g-r)}$ and $T_{{\rm trans},(g-r)}$ ($r = -0.37 \pm 0.17$, $p\leq 0.22$) is also visible. Again, for $(g-r)$ colour curves the SNe~II with a slow initial rise/cooling also show a slower cooling after the transition which appears at later epochs. Between $s_{2,(g-r)}$ and $T_{{\rm trans},(g-r)}$ we do not find a correlation ($r = -0.07 \pm 0.20$, $p \leq 0.99$).

With this colour, we do not recover previous correlations between $T_{{\rm trans},(g-r)}$ and the $(g-r)$ colour at 15d but, one between $(g-r)$ at 70d and $s_{2,(g-r)}$ is seen ($r = 0.45 \pm 0.10$, $p \leq 3.3 \times 10^{-2}$). Finally, at the same epoch, a correlation with $s_{1,(g-r)}$ ($r = 0.52 \pm 0.07$, $p \leq 5.2 \times 10^{-3}$) is also seen. Redder SNe~II at late epochs have a faster cooling before and after the transition in their colour curves. However, it is important to note that these two correlations are seen only using $(g-r)$.

We also look at the $(g-r)$ colour distributions at four epochs. As expected, $(g-r)$ becomes redder with time: $<(g-r)_{15}> = 0.24 \pm 0.28$ mag, $<(g-r)_{30}> = 0.55 \pm 0.25$ mag, $<(g-r)_{50}> = 0.79 \pm 0.21$ mag, and $<(g-r)_{70}> = 0.94 \pm 0.24$ mag at epochs 15d, 30d, 50d, and 70d, respectively. The colour evolution is much slower than for $(u-g)$ or $(B-V)$, as the shape of the red part of the spectrum is less affected by the temperature change during the photospheric phase.

\subsubsection{Photometric Correlations}\label{txt:photo_gr}

We recover the correlations between the colour and the absolute magnitude found with $(B-V)$ and $(u-g)$: $(g-r)$ at 15d and 30d with the absolute magnitude $M_{\rm max}$ (respectively $r = 0.53 \pm 0.11$, $p \leq 1.5 \times 10^{-2}$; $r = 0.38 \pm 0.17$, $p \leq 0.11$) or $M_{\rm end}$ (respectively $r = 0.58 \pm 0.11$, $p \leq 6.6 \times 10^{-3}$; $r = 0.48 \pm 0.12$, $p \leq 5.0 \times 10^{-3}$); see Figure \ref{fig_appendix:redder_fainter}. In these diagrams, the outlier is SN~2008bp, which was also identified as an outlier by \citet{anderson14a}. Finally, the absence of correlation at later epochs (50d and 70d) suggests an intrinsic origin rather than dust extinction.

For this colour, two relations found previously are also seen. First, the colour at later epochs correlates with $s_{2}$. A weak correlation is seen at 50d ($r = 0.33 \pm 0.19$, $p \leq 0.32$), which becomes stronger 20 days later ($r = 0.64 \pm 0.004$, $p \leq 7.0 \times 10^{-6}$). This correlation means that redder SNe~II at later epochs have a steeper plateau decline rate; see Figure \ref{fig:appendix_redder_IIL}. Second, the anticorrelation between the colour at early epochs and $s_{1}$ is also found ($r = -0.60 \pm 0.05$, $p \leq 2.7 \times 10^{-2}$): bluer SNe~II have a steeper initial decline --- they are fast-declining SNe~II (Figure \ref{fig:bluer_s1}).

\subsubsection{Spectroscopic Correlations}\label{gr_spec_corr}

The $(g-r)$ colour at 30d also correlates slightly or strongly with different equivalent widths such as Haabs ($r = 0.48 \pm 0.19$, $p \leq 0.49$) or EWFe6 ($r = 0.48 \pm 0.05$, $p \leq 9.9 \times 10^{-3}$), in the sense that bluer SNe~II have smaller equivalent widths. Note also that $(g-r)$ at 50d correlates with EWFe6 ($r = 0.45 \pm 0.08$, $p \leq 1.9 \times 10^{-2}$). These correlations are shown in Figure \ref{fig_appen:EW_redder}. 

We also see a weak anticorrelation between the strength of the H$\alpha$ absorption line (Haabs) and $s_{2,(g-r)}$ in the sense that SNe~II with a rapid cooling after transition show a weaker H$\alpha$ absorption line: $r = -0.41 \pm 0.16$ and $p \leq 0.15$ (Figure \ref{fig:Haabs_s2colour}).

Even if these correlations are not seen for the previous colours, we think it is important to mention that $(g-r)$ at 50d and 70d correlate with different velocities in the sense that redder SNe~II at later epochs have faster velocities: HaF (respectively $r = 0.51 \pm 0.03$, $p \leq 1.7 \times 10^{-3}$; $r = 0.58 \pm 0.02$, $p \leq 5.7 \times 10^{-4}$), Hb (respectively $r = 0.42 \pm 0.09$, $p \leq 3.7 \times 10^{-2}$; $r = 0.50 \pm 0.05$, $p \leq 7.5 \times 10^{-3}$), Fe5 (respectively $r = 0.48 \pm 0.06$, $p \leq 7.0 \times 10^{-3}$; $r = 0.59 \pm 0.03$, $p \leq 5.7 \times 10^{-4}$), ScFe (respectively $r = 0.49 \pm 0.04$, $p \leq 5.9 \times 10^{-3}$; $r = 0.60 \pm 0.02$, $p \leq 5.0 \times 10^{-4}$), ScM (respectively $r = 0.44 \pm 0.08$, $p \leq 2.6 \times 10^{-2}$; $r = 0.47 \pm 0.09$, $p \leq 2.9 \times 10^{-2}$), and Sc (respectively $r =0.50 \pm 0.09$, $p \leq 3.3 \times 10^{-2}$; $r = 0.60 \pm 0.07$, $p \leq 6.4 \times 10^{-3}$). The correlations are displayed in Figure \ref{fig_appen:velo_colour}.

Finally, it is also important to note that $(g-r)$ correlates better with the spectroscopic parameters derived from the H$\alpha$ line than $(B-V)$ or $(u-g)$. This is not surprising, as in the rest frame the H$\alpha$ line falls directly inside the $r$ filter (effective wavelength $\lambda 6223.3$; see \url{http://csp.obs.carnegiescience.edu/data/filters}).

\subsection{$(\lowercase{g}-Y)$}

\subsubsection{Distributions}

The $(g-Y)$ colour curves are characterised by an initial cooling ($s_{1,(g-Y)}$) with a median value of $2.95 \pm 0.49$ mag 100d$^{-1}$ followed by a second slope with a median $s_{2,(g-Y)}$ of $0.83 \pm 0.31$ mag 100d$^{-1}$, and a transition $T_{{\rm trans},(g-Y)}$ at $39.6 \pm 4.3$ d after the explosion.

While for the other colours, strong correlations between the two colour slopes are seen (see Figure \ref{appendix:colour_params_corre}), for $(g-Y)$ no statistically significant trend is found ($r = 0.19 \pm 0.18$, $p \leq 0.97$). Additionally, the anticorrelation between $s_{1,(g-Y)}/T_{{\rm trans},(g-Y)}$ is weak, with $r = -0.49 \pm 0.21$ ($p \leq 0.23$), while the one between $s_{2,(g-Y)}/T_{{\rm trans},(g-Y)}$ is strong, with $r = -0.55 \pm 0.12$ ($p \leq 5.8 \times 10^{-2}$). We also try to correlate these three parameters with the colours at four different epochs as achieved in the previous sections. No correlations were found as for the previous colour sets.

\indent Repeating our previous analysis, we look at the $(g-Y)$ colour distributions at four epochs. We find that $(g-Y)$ becomes redder with time: $<(g-Y)_{15}> = 0.49 \pm 0.37$ mag, $<(g-Y)_{30}> = 1.01 \pm 0.31$ mag, $<(g-Y)_{50}> = 1.43 \pm 0.43$ mag, and $<(g-Y)_{70}> = 1.62 \pm 0.44$  mag at epoch 15d, 30d, 50d, and 70d, respectively. At early times (15d) the statistics are very low, with only 14 SNe~II (versus 37, 37, and 25 SNe~II for 30d, 50d, and 70d, respectively).

\subsubsection{Photometric Correlations}\label{photo_gY}

As for other colour combinations, we find that more luminous SNe~II are bluer. Although only a weak trend is seen with $M_{\rm max}$, with $M_{\rm end}$ the correlation is stronger at different epochs. The correlation factors are $r = 0.66 \pm 0.14$ ($p \leq 6.8 \times 10^{-2}$), $r = 0.41 \pm 0.11$ ($p \leq 4.1 \times 10^{-2}$), and $r = 0.48 \pm 0.13$ ($p \leq 3.1 \times 10^{-2}$) for 15d, 30d, and 50d, respectively (Figure \ref{fig_appendix:redder_fainter}).

Regarding the other photometric correlations seen in the previous sections, here we only recover the correlation between $(g-Y)$ at 70d and $s_{2}$ ($r = 0.50 \pm 0.06$, $p \leq 6.1 \times 10^{-2}$) --- that is, redder SNe~II have steeper $s_{2}$ (Figure \ref{fig:appendix_redder_IIL}).

\subsubsection{Spectroscopic Correlations}\label{gY_spec_corr}

For the $(g-Y)$ colour, only one slight correlation between the colour at 30d and the equivalent width EWFe6 is seen (Figure \ref{fig_appen:EW_redder}), in the sense that redder SNe~II show larger equivalent widths ($r = 0.41 \pm 0.11$, $p \leq 0.12$). However, we do not recover the correlation between $s_{2,(g-Y)}$ and Haabs. Finally, note that as with $(g-r)$, a correlation between the velocities and the colours at late epochs (50d and 70d) is seen: redder SNe~II at late epochs have faster ejecta velocities.

\section{Host-Galaxy Extinction Effects}\label{txt:discussion_avh}

\subsection{Colour-Brightness Correlations}

As mentioned previously, in this work we do not correct for host-galaxy extinction, as no currently available method seems to improve the uniformity of the SN~II sample \citep{faran14a,gutierrez17a}. However, here we discuss the possible effects of host-galaxy extinction on the colour-brightness correlations. For this purpose, we use three different subsamples as well as the full sample.

\indent First, we assume that in those SNe showing the reddest colours, the majority of this colour is dominated by uncorrected host-galaxy extinction. While throughout this article we argue that most of the diversity in SN~II colours is intrinsic, it is expected that some SNe~II do suffer from considerable host-galaxy extinction, given that galactic environments with a high dust content are known to exist. In this context, for each colour-brightness correlation presented, we remove SNe~II that have colours falling within the $\sim 10$\% reddest of the sample. For each correlation, the exact colour and the epoch are different, and therefore the exact colour cut value changes. However, using the $(B-V)$ colour (for example), the cut occurs at $\sim 0.45$, 0.9, 1.3, and 1.4 mag for 15d, 30d, 50d, and 70d, respectively. In Table \ref{tab:col_bright_avh}, a comparison of the correlation strengths without (first column) and with (second column) the colour cut is shown. The strength of correlations generally increased after removing these reddest SNe~II, suggesting an intrinsic explanation of the colour-brightness relation (as discussed below). While the choice of 10\% was initially arbitrary, we also tested cuts of $\sim 20$\% and $\sim 30$\%. However, in both of these latter cases the strengths of correlations generally decreased, and thus our choice of 10\% appears to be justified.

\indent Second, even if the validity of the use of \ion{Na}{i}~D in low-resolution spectra for deriving precise host-galaxy extinction has been challenged \citep{poznanski11,phillips13,galbany17}, the presence of significant \ion{Na}{i}~D absorption does usually imply some degree of host-galaxy extinction. The absence of this \ion{Na}{i}~D absorption line also probably indicates a low level of host-galaxy extinction. So, here we remove all SNe~II that have an \ion{Na}{i}~D equivalent width $\geq 1$~\AA\ \citep{anderson14a}. In the third column of Table \ref{tab:col_bright_avh}, the new correlations for this sample are listed; again, we generally find an increase in correlation strength when removing those events most likely to be significantly affected by dust extinction.

\indent Finally, following the procedure of \citet{phillips99}, \citet{folatelli10}, and \citet{stritzinger17a}, we choose SNe~II which are located at significant distances away from host-galaxy nuclei, together with those that do not have \ion{Na}{i}~D absorption detections ($A_{V}$(host) $= 0.00$ from Table 6 in \citealt{anderson14a}). This final sample consists of 19 SNe~II (13 SNe~II with NIR photometry). The new correlations are presented in the fourth column of Table \ref{tab:col_bright_avh}. 

\indent Comparing the strength of correlations in the first two subsamples to those of our full sample, we can assess the likelihood of host-galaxy extinction producing our observed trends. If any of the trends are produced by the effects of uncorrected reddening, one should expect the strength of correlations to \textit{decrease} in the subsamples where we remove SNe~II affected by significant host-galaxy extinction. We see that the majority of the correlations (mostly at earlier epochs) are stronger for the culled samples (colour or \ion{Na}{i}~D). This confirms our hypothesis of an intrinsic origin for the dominant colour diversity rather that dust effects, in the majority of our sample. Note also that the differences between the full sample and subsamples are generally stronger with $M_{\rm max}$ than $M_{\rm end}$. Regarding the subsample with both \ion{Na}{i}~D absorption and position cuts, the statistics are too low to enable meaningful conclusions (although there is no evidence that correlation strengths \textit{decrease} in this sample); however, this subsample will be useful for investigating the intrinsic colour of our SNe~II (Section \ref{txt:cc_intrin}).

\begin{table*}
\caption{Comparison of correlation strengths between colours and absolute magnitudes.}
\begin{threeparttable}
\scriptsize
\begin{tabular}{ccccccccccccc}
\hline
\hline
Correlations & \multicolumn{3}{c}{All sample} & \multicolumn{3}{c}{Sample cut} & \multicolumn{3}{c}{Sample cut} & \multicolumn{3}{c}{Sample cut} \\
 & \multicolumn{3}{c}{} & \multicolumn{3}{c}{based on colours} & \multicolumn{3}{c}{based on \ion{Na}{i}~D} & \multicolumn{3}{c}{based on \ion{Na}{i}~D $+$ position} \\
\hline
 & $N_{\mathrm{SNe}}$ & $r$ & $p\leq$ & $N_{\mathrm{SNe}}$ & $r$ & $p\leq$ & $N_{\mathrm{SNe}}$ & $r$ & $p\leq$ & $N_{\mathrm{SNe}}$ & $r$ & $p\leq$ \\
\hline
$(B-V)15$ vs. $M_{\rm max}$ & 31 & 0.35 $\pm$ 0.16 & 0.31 & 28 & 0.50 $\pm$ 0.10 & 3.5 $\times$ 10$^{-2}$ & 22 & 0.49 $\pm$ 0.10 & 7.3 $\times$ 10$^{-2}$ & 12 & 0.49 $\pm$ 0.22 & 0.40 \\
$(B-V)15$ vs. $M_{\rm end}$ & 30 & 0.33 $\pm$ 0.18 & 0.43 & 27 & 0.44 $\pm$ 0.14 & 0.13 & 21 & 0.49 $\pm$ 0.09 & 7.2 $\times$ 10$^{-2}$ & 12 & 0.59 $\pm$ 0.18 & 0.19 \\
$(B-V)30$ vs. $M_{\rm max}$ & 56 & 0.30 $\pm$ 0.16 & 0.30 & 50 & 0.47 $\pm$ 0.03 & 1.4 $\times$ 10$^{-3}$ & 36 & 0.39 $\pm$ 0.14 & 0.14 & 18 & 0.30 $\pm$ 0.27 & 0.92\\
$(B-V)30$ vs. $M_{\rm end}$ & 55 & 0.32 $\pm$ 0.17 & 0.27 & 49 & 0.40 $\pm$ 0.13 & 6.1 $\times$ 10$^{-2}$ & 35 & 0.35 $\pm$ 0.19 & 0.36 & 18 & 0.24 $\pm$ 0.29 & 0.83\\
$(B-V)50$ vs. $M_{\rm end}$ & 50 & 0.39 $\pm$ 0.08 & 2.8 $\times$ 10$^{-2}$ & 45 & 0.33 $\pm$ 0.16 &0.26 & 30 & 0.34 $\pm$ 0.21 & 0.49 & 15 & 0.42 $\pm$ 0.26 & 0.57 \\
$(B-V)70$ vs. $M_{\rm end}$ & 37 & 0.37 $\pm$ 0.17 & 0.23 &33 & 0.23 $\pm$ 0.29 & 0.74 & 24 & 0.10 $\pm$ 0.30 & 0.35 & 11 & 0.34 $\pm$ 0.31 & 0.92\\
$(u-g)30$ vs. $M_{\rm max}$ & 35 & 0.51 $\pm$ 0.12 & 2.0 $\times$ 10$^{-3}$ & 31 & 0.56 $\pm$ 0.03 & 2.1 $\times$ 10$^{-3}$ & 26 & 0.54 $\pm$ 0.06 & 1.3 $\times$ 10$^{-2}$ & 12 & 0.39 $\pm$ 0.22 & 0.59 \\
$(u-g)30$ vs. $M_{\rm end}$ & 34 & 0.36 $\pm$ 0.22 & 0.43 & 31 & 0.41 $\pm$ 0.24 & 0.36 & 25 & 0.49 $\pm$ 0.12 & 6.9 $\times$ 10$^{-2}$ & 12 & 0.25 $\pm$ 0.31 & 0.86\\
$(g-r)15$ vs. $M_{\rm max}$ & 33 & 0.53 $\pm$ 0.11 & 1.5 $\times$ 10$^{-3}$ & 30 & 0.57 $\pm$ 0.03 & 2.1 $\times$ 10$^{-3}$ & 23 & 0.69 $\pm$ 0.01 & 3.6 $\times$ 10$^{-4}$ & 12 & 0.52 $\pm$ 0.19 & 0.29\\
$(g-r)15$ vs. $M_{\rm end}$ & 32 & 0.58 $\pm$ 0.11 & 6.6 $\times$ 10$^{-3}$ &29 & 0.46 $\pm$ 0.14 & 9.0 $\times$ 10$^{-2}$ & 22 & 0.77 $\pm$ 0.02 & 5.8 $\times$ 10$^{-5}$ & 12 & 0.43 $\pm$ 0.27 & 0.62\\
$(g-r)30$ vs. $M_{\rm max}$ & 57 & 0.38 $\pm$ 0.17 & 0.11 & 51 & 0.34 $\pm$ 0.11 & 0.10 & 37 & 0.47 $\pm$ 0.16 & 6.2 $\times$ 10$^{-2}$ & 18 & 0.18 $\pm$ 0.28 & 0.69\\
$(g-r)30$ vs. $M_{\rm end}$ & 56 & 0.49 $\pm$ 0.12 & 5.0 $\times$ 10$^{-3}$ & 50 & 0.36 $\pm$ 0.16 & 0.16 & 36 & 0.55 $\pm$ 0.18 & 2.6 $\times$ 10$^{-2}$ & 18 & 0.08 $\pm$ 0.27 & 0.44\\
$(g-Y)15$ vs. $M_{\rm end}$ & 13 & 0.66 $\pm$ 0.14 & 6.8 $\times$ 10$^{-3}$ & 12 & 0.73 $\pm$ 0.09 & 3.1 $\times$ 10$^{-2}$ & 8 & 0.59 $\pm$ 0.21 & 0.35 &5 & $\cdots$ $\pm$ $\cdots$ &$\cdots$\\
$(g-Y)30$ vs. $M_{\rm end}$ & 39 & 0.41 $\pm$ 0.11 & 6.3 $\times$ 10$^{-3}$ & 35 & 0.54 $\pm$ 0.02 & 1.4 $\times$ 10$^{-3}$ & 25 & 0.27 $\pm$ 0.26 & 0.96  &14 & 0.25 $\pm$ 0.30 &0.88\\
$(g-Y)50$ vs. $M_{\rm end}$ & 38 & 0.48 $\pm$ 0.13 & 3.1 $\times$ 10$^{-2}$ & 34 & 0.30 $\pm$ 0.23 & 0.69 & 23 & 0.21 $\pm$ 0.29 & 0.72 &11 & 0.06 $\pm$ 0.30 &0.47\\
\hline
\end{tabular}
Notes --- Four different samples are used: total (all), removing 10\% of the reddest objects (based on colours), only the SNe~II with an \ion{Na}{i}~D equivalent width smaller than 1~\AA\ (based on \ion{Na}{i}~D), and only the SNe~II away from the host-galaxy nuclei with no \ion{Na}{i}~D equivalent width (based on \ion{Na}{i}~D + position).
\label{tab:col_bright_avh}
\end{threeparttable}
\end{table*}

\subsection{Other Correlations}

Following the procedure from the previous section, we investigate the effect of host-galaxy extinction in all of the correlations found in this work. A summary of the Pearson factor between the full sample and three subsamples (colour, \ion{Na}{i}~D equivalent width, and \ion{Na}{i}~D equivalent width $+$ position) are respectively displayed in the first, second, third, and fourth columns of Table \ref{tab:correlations_avh}. 

\indent This analysis provides additional support for our conclusion that the correlations are driven mostly by intrinsic origin, as in the majority of the cases the Pearson factor is stronger for the subsamples than for the full sample. Note that the strength increase is more likely seen at earlier epochs (15d--30d) than at later epochs (50d--70d). This could explain why the strengths of the colour/velocity correlations are not greater for the culled sample, as these correlations are only seen at later epochs.

\begin{table*}
\caption{Comparison of correlation strengths between colours and spectroscopic/photometric parameters.}
\begin{threeparttable}
\centering
\small
\resizebox{\textwidth}{!}{%	       
\begin{tabular}{lllllllllllll}
\hline
\hline
Correlations & \multicolumn{3}{c}{All sample} & \multicolumn{3}{c}{Sample cut} & \multicolumn{3}{c}{sample cut}  & \multicolumn{3}{c}{sample cut} \\
 & \multicolumn{3}{c}{} & \multicolumn{3}{c}{based on colours} & \multicolumn{3}{c}{based on \ion{Na}{i}~D} & \multicolumn{3}{c}{based on \ion{Na}{i}~D $+$ position} \\
\hline
 & $N_{\mathrm{SNe}}$ & $r$ & $p\leq$ & $N_{\mathrm{SNe}}$ & $r$ & $p\leq$ & $N_{\mathrm{SNe}}$ & $r$ & $p\leq$ & $N_{\mathrm{SNe}}$ & $r$ & $p\leq$ \\
\hline
$(B-V)70$ vs. $s_{2}$ & 38 & 0.44 $\pm$ 0.07 & 2.2 $\times$ 10$^{-2}$ & 34 &0.33 $\pm$ 0.20 & 0.46 &24 & 0.22 $\pm$ 0.28 &0.78 &11 & 0.14 $\pm$ 0.26 &0.71\\
$(g-r)50$ vs. $s_{2}$ & 53 & 0.33 $\pm$ 0.19 & 0.32 & 48 & 0.38 $\pm$ 0.12 &7.4 $\times$ 10$^{-2}$ & 31 & 0.39 $\pm$ 0.15 &0.19 & 15 & 0.46 $\pm$ 0.24 & 0.42\\
$(g-r)70$ vs. $s_{2}$ & 41 & 0.64 $\pm$ 0.005 & 7.0 $\times$ 10$^{-6}$ & 37 & 0.60 $\pm$ 0.01 & 1.2 $\times$ 10$^{-4}$ &25 & 0.61 $\pm$ 0.06 & 4.3 $\times$ 10$^{-3}$ & 11 & 0.34 $\pm$ 0.27 & 0.85\\
$(g-Y)70$ vs. $s_{2}$ & 25 & 0.50 $\pm$ 0.12 & 6.1 $\times$ 10$^{-2}$ & 23 & 0.54 $\pm$ 0.09 & 3.1 $\times$ 10$^{-2}$ & 14 & 0.51 $\pm$ 0.19 & 0.26 &6 & 0.40 $\pm$ 0.53 &0.81\\
$(B-V)15$ vs. $s_{1}$ & 15 & $-$0.44 $\pm$ 0.23 & 0.45 &13 & $-$0.31 $\pm$ 0.28 & 0.93 &12 &$-$0.41 $\pm$ 0.26 &0.64 & 5 & $\cdots$ $\pm$ $\cdots$ &$\cdots$\\
$(B-V)30$ vs. $s_{1}$ & 29 & $-$0.44 $\pm$ 0.13 & 9.9 $\times$ 10$^{-2}$ &26 & $-$0.57 $\pm$ 0.08 & 1.1 $\times$ 10$^{-2}$ &20 &$-$0.53 $\pm$ 0.13 & 8.1 $\times$ 10$^{-2}$ & 7 & $-$0.57 $\pm$ 0.31 & 0.65\\
$(u-g)30$ vs. $s_{1}$ & 23 & $-$0.42 $\pm$ 0.15 & 0.21 &21 & $-$0.52 $\pm$ 0.10 & 6.5 $\times$ 10$^{-2}$ &17 &$-$0.37 $\pm$ 0.22 & 0.56 & 5 & $\cdots$ $\pm$ $\cdots$ & $\cdots$\\
$(g-r)15$ vs. $s_{1}$ & 16 & $-$0.60 $\pm$ 0.05 & 2.7 $\times$ 10$^{-2}$ &14 & $-$0.58 $\pm$ 0.10 & 8.2 $\times$ 10$^{-2}$ & 13 & $-$0.57 $\pm$ 0.11 & 0.11 5 & $\cdots$ &$\cdots$ $\pm$ $\cdots$ & $\cdots$\\
$(B-V)15$ vs. EWSc & 10 & 0.55 $\pm$ 0.19 & 0.31 &9 & 0.63 $\pm$ 0.19 & 0.24 &7 &0.67 $\pm$ 0.20 & 0.29 & 7 & 0.73 $\pm$ 0.28 & 0.31\\
$(B-V)15$ vs. EWBa & 10 & 0.71 $\pm$ 0.08 & 5.1 $\times$ 10$^{-2}$ &9 & 0.62 $\pm$ 0.13 & 0.18 &7 &0.67 $\pm$ 0.16 & 0.24 & 7 & 0.73 $\pm$ 0.28 & 0.31\\
$(u-g)15$ vs. EWScM & 14 & 0.57 $\pm$ 0.15 & 0.13 &13 & 0.62 $\pm$ 0.09 & 6.2 $\times$ 10$^{-2}$ & 5 & $\cdots$ $\pm$ $\cdots$ & $\cdots$ &5 &$\cdots$ $\pm$ $\cdots$  &$\cdots$\\
$(B-V)30$ vs. EWBa & 22 & 0.44 $\pm$ 0.11 & 0.13 &20 & 0.53 $\pm$ 0.10 & 5.8 $\times$ 10$^{-2}$ &15 &0.51 $\pm$ 0.17 & 0.21 & 11 & 0.28 $\pm$ 0.39 & 0.75\\
$(B-V)30$ vs. EWScM & 32 & 0.42 $\pm$ 0.13 & 0.11 &29 & 0.54 $\pm$ 0.04 & 5.7 $\times$ 10$^{-3}$ &22 &0.43 $\pm$ 0.17 & 0.24 & 11 & 0.58 $\pm$ 0.17 & 0.11\\
$(B-V)30$ vs. EWFe6 & 35 & 0.41 $\pm$ 0.15 & 0.13 &31 & 0.30 $\pm$ 0.28 & 0.91 &25 &0.31 $\pm$ 0.28 &0.89 & 11 & $-$0.28 $\pm$ 0.33 & 0.89\\
$(u-g)30$ vs. EWFe6 & 26 & 0.53 $\pm$ 0.18 & 7.9 $\times$ 10$^{-2}$ & 23 & 0.39 $\pm$ 0.26 & 0.55 & 18 & 0.33 $\pm$ 0.28 & 0.84 & 9 & 0.06 $\pm$ 0.26 & 0.62\\
$(u-g)30$ vs. EWScM & 24 & 0.48 $\pm$ 0.12 & 8.3 $\times$ 10$^{-2}$ &22 & 0.58 $\pm$ 0.03 & 8.0 $\times$ 10$^{-3}$ & 18 & 0.50$\pm$ 0.13 & 0.13 & 9 & 0.54 $\pm$ 0.26 & 0.46\\
$(u-g)30$ vs. Haabs & 26 & 0.46 $\pm$ 0.16 & 0.14 & 23 & 0.51 $\pm$ 0.15 &9.1 $\times$ 10$^{-2}$ & 18 &0.26 $\pm$ 0.29 &0.91 & 9 & 0.28 $\pm$ 0.29 & 0.97\\
$(u-g)30$ vs. EWScFe & 22 & 0.49 $\pm$ 0.15 & 0.12 & 20 & 0.61 $\pm$ 0.03 &7.3 $\times$ 10$^{-3}$ & 16 & 0.49 $\pm$ 0.19 &0.26 & 9 & 0.79 $\pm$ 0.12 & 5.0 $\times$ 10$^{-2}$\\
$(g-r)30$ vs. Haabs & 35 & 0.31 $\pm$ 0.19 & 0.49 & 31 & 0.41 $\pm$ 0.16 & 5.3 $\times$ 10$^{-2}$ & 25 & 0.30 $\pm$ 0.23 &0.74 & 11 & 0.06 $\pm$ 0.35 &0.40\\
$(g-r)30$ vs. EWFe6 & 35 & 0.48 $\pm$ 0.05 & 9.9 $\times$ 10$^{-3}$ & 31 & 0.46 $\pm$ 0.14 & 7.9 $\times$ 10$^{-2}$ & 25 & 0.42 $\pm$ 0.21 & 0.31 & 11 & 0.01 $\pm$ 0.4 &0.23\\
$(g-Y)30$ vs. EWFe6 & 28 & 0.41 $\pm$ 0.11 & 0.12 & 25 & 0.32 $\pm$ 0.19 & 0.53 & 19 & 0.37 $\pm$ 0.28 & 0.71 &8 & 0.65 $\pm$ 0.24 &0.31\\
$(B-V)50$ vs. EWBa & 24 & 0.38 $\pm$ 0.19 & 0.37 &22 & 0.65 $\pm$ 0.02 & 1.7 $\times$ 10$^{-3}$ &17 &0.46 $\pm$ 0.25 & 0.42 & 13 & 0.39 $\pm$ 0.30 & 0.78\\
$(u-g)50$ vs. EWFe6 & 23 & 0.42 $\pm$ 0.17 & 0.25 & 21 & 0.38 $\pm$ 0.23 & 0.52 & 18 & 0.33 $\pm$ 0.28 & 0.84 & 9 & 0.31 $\pm$ 0.27 & 0.91\\
$(u-g)50$ vs. EWScFe & 21 & 0.40 $\pm$ 0.14 & 0.25 & 19 & 0.33 $\pm$ 0.23 & 0.68 & 16 & 0.35 $\pm$ 0.25 & 0.71 & 9 & 0.67 $\pm$ 0.22 & 0.22\\
$(u-g)50$ vs. EWScM & 23 & 0.40 $\pm$ 0.19 & 0.34 & 21 & 0.41 $\pm$ 0.23 & 0.43 & 18 &0.32 $\pm$ 0.29 & 0.91 & 9 & 0.30 $\pm$ 0.34 & 0.91\\
$(g-r)50$ vs. EWFe6 & 40 & 0.45 $\pm$ 0.08 & 1.9 $\times$ 10$^{-2}$ & 36 & 0.44 $\pm$ 0.07 & 4.2 $\times$ 10$^{-2}$ & 27 & 0.39 $\pm$ 0.20 &0.34 & 13 & 0.12 $\pm$ 0.30 &0.56\\
s$_{2,(B-V)}$ vs. Haabs & 32 & $-$0.50 $\pm$ 0.07 & 1.4 $\times$ 10$^{-2}$ &29 & $-$0.39 $\pm$ 0.18 & 0.27 &24 &$-$0.50 $\pm$ 0.15 &9.4 $\times$ 10$^{-2}$ & 13 & $-$ 0.59 $\pm$ 0.22 & 0.36\\
s$_{2,(g-r)}$ vs. Haabs & 35 & $-$0.41 $\pm$ 0.16 & 0.15 &31 & $-$0.27 $\pm$ 0.25 & 0.91 & 24 & $-$0.59 $\pm$ 0.08 & 1.1 $\times$ 10$^{-2}$ & 12 & $-$ 0.56 $\pm$ 0.17 &0.16\\
$(g-r)50$ vs. HaF & 40 & 0.51 $\pm$ 0.03 & 1.5 $\times$ 10$^{-3}$ & 36 & 0.46 $\pm$ 0.09 &3.6 $\times$ 10$^{-2}$ & 27 & 0.40 $\pm$ 0.19 & 0.29 & 13 & 0.40 $\pm$ 0.22 & 0.57\\
$(g-r)50$ vs. Hb & 40 & 0.42 $\pm$ 0.09 & 3.7 $\times$ 10$^{-2}$ & 36 & 0.36 $\pm$ 0.17 &0.27 & 27 & 0.32 $\pm$ 0.24 &0.37 & 13 & 0.37 $\pm$ 0.24 &0.66\\
$(g-r)50$ vs. Fe5 & 40 & 0.48 $\pm$ 0.06 & 7.0 $\times$ 10$^{-3}$ & 36 & 0.44 $\pm $0.12 & 5.7 $\times$ 10$^{-2}$ &27 &0.41 $\pm $0.19 &0.27 & 13 & 0.47 $\pm$ 0.25 &0.49\\
$(g-r)50$ vs. ScFe & 36 & 0.49 $\pm$ 0.04 & 5.9 $\times$ 10$^{-3}$ & 32 & 0.46 $\pm$ 0.10 & 4.3 $\times$ 10$^{-2}$ & 23 & 0.34 $\pm$ 0.24 & 0.65 & 11 & 0.47 $\pm$ 0.29 &0.61\\
$(g-r)50$ vs. ScM & 38 & 0.44 $\pm$ 0.08 & 2.6 $\times$ 10$^{-2}$ & 34 &0.39 $\pm$ 0.17 & 0.21 & 25 & 0.26 $\pm$ 0.27 & 0.97 & 12 & 0.44 $\pm$ 0.25 &0.56\\
$(g-r)50$ vs. Sc & 27 & 0.50 $\pm$ 0.09 & 3.3 $\times$ 10$^{-2}$ & 24 & 0.46 $\pm$ 0.14 & 0.12 & 17 & 0.41 $\pm$ 0.23 & 0.49 & 8 & 0.13 $\pm$ 0.50 &0.36\\
$(g-Y)50$ vs. HaF & 33 & 0.40 $\pm$ 0.13 & 0.13 & 30 & 0.36 $\pm$ 0.23 & 0.49 & 21 & 0.36 $\pm$ 0.28 &0.73 &9 & 0.42 $\pm$ 0.24 &0.65\\
$(g-r)70$ vs. Hb & 34 & 0.50 $\pm$ 0.05 & 7.5 $\times$ 10$^{-3}$ & 31 & 0.33 $\pm$ 0.19 & 0.45 & 22 & 0.38 $\pm$ 0.21 & 0.45 & 10 & 0.01 $\pm$ 0.43 &0.86\\
$(g-r)70$ vs. HaF & 34 & 0.58 $\pm$ 0.02 & 5.7 $\times$ 10$^{-4}$ & 31 & 0.45 $\pm$ 0.10 &5.3 $\times$ 10$^{-2}$ & 22& 0.44 $\pm$ 0.16 &0.21 & 10 & 0.07 $\pm$ 0.44 &0.30\\
$(g-r)70$ vs. Fe5 & 34 & 0.59 $\pm$ 0.03 & 5.7 $\times$ 10$^{-4}$ & 31 & 0.39 $\pm$ 0.15 & 0.19 & 22& 0.44 $\pm$
0.18 &0.24 & 10 & 0.08 $\pm$ 0.45 &0.28\\
$(g-r)70$ vs. ScFe & 32 & 0.60 $\pm$ 0.02 & 5.0 $\times$ 10$^{-4}$ & 29 & 0.45 $\pm$ 0.08 & 4.8 $\times$ 10$^{-2}$ & 20 & 0.42 $\pm$ 0.17 & 0.29 & 9 & 0.15 $\pm$ 0.48 &0.38\\
$(g-r)70$ vs. ScM & 33 & 0.47 $\pm$ 0.09 & 2.9 $\times$ 10$^{-2}$ & 30 & 0.29 $\pm$ 0.24 & 0.80 & 21 & 0.26 $\pm$ 0.28 & 0.93 & 9 & 0.01 $\pm$ 0.53 &0.25\\
$(g-r)70$ vs. Sc & 25 & 0.60 $\pm$ 0.07 & 6.4 $\times$ 10$^{-3}$ & 23 & 0.41 $\pm$ 0.16 & 0.25 & 16 & 0.43 $\pm$ 0.21 &0.41 & 7 & $-$0.29 $\pm$ 0.43 &0.82\\
$(g-Y)70$ vs. HaF & 21 & 0.52 $\pm$ 0.18 & 0.13 & 19 & 0.35 $\pm$ 0.28 & 0.77 & 13 & 0.48 $\pm$ 0.26 & 0.47 &5 &$\cdots$ $\pm$ $\cdots$  &$\cdots$\\
$(g-Y)70$ vs. Hb & 21 & 0.44 $\pm$ 0.17 & 0.24 & 19 & 0.28 $\pm$ 0.30 & 0.93 & 13 & 0.40 $\pm$ 0.28 & 0.70 &5 &$\cdots$ $\pm$ $\cdots$  &$\cdots$\\
$(g-Y)70$ vs. Fe5 & 21 & 0.55 $\pm$ 0.14 & 6.5 $\times$ 10$^{-2}$ & 19 & 0.43 $\pm$ 0.24 & 0.44 &13 & 0.58 $\pm$ 0.20 &0.20 &5 &$\cdots$ $\pm$ $\cdots$  &$\cdots$\\
\hline
\end{tabular}
}
Notes --- Four different samples are used: total (all), removing 10\% of the reddest objects (based on colours), only the SNe~II with an \ion{Na}{i}~D equivalent width smaller than 1~\AA\ (based on \ion{Na}{i}~D), and only the SNe~II away from the host-galaxy nuclei with no \ion{Na}{i}~D equivalent width (based on \ion{Na}{i}~D + position).
\label{tab:correlations_avh}
\end{threeparttable}
\end{table*}

\section{Discussion}\label{txt:discussions}

Using 65 SNe~II from the CSP-I sample, four different colours ($B-V$, $u-g$, $g-r$, and $g-Y$) and their evolution have been characterised. We define morphological colour-curve parameters, measure observed colours at different phases, and explore possible correlations between colour parameters and different spectroscopic and photometric properties. We now discuss these results in more detail, and we compare them to other observational and theoretical SN~II studies.

\subsection{Colour Evolution}\label{txt:Colour_evolution}

First, using a larger SN~II sample than that of \citet{patat94}, we confirm their finding that SN~II colour curves consist of two regimes with two different slopes (see Figure \ref{fig:CC_parameters} and Appendix \ref{fig:append_colour_fit}).While a linear fit (one or two slopes) describes the majority of SN~II colour curves better than a power-law fit ($55$\% for $B-V$ or $63$\% for $g-r$), it is important to note that for the $(g-Y)$ colour, half of the SN~II light curves favour a power-law fit ($51$\%). For this colour, only a very small fraction is described by two slopes, even if the SN colour curve is well sampled. Finally, for the $(u-g)$ colour, the majority of the SN~II colour curves are fitted by only one slope ($54$\%); however, this can be explained by the lack of $u$-band data at late epochs.

Second, the study of colour-curve parameters with four different colours leads to the overall conclusion that SNe~II which cool more quickly at early epochs also generally cool more quickly during the plateau phase. This correlation between $s_{1,{\rm colour}}$ and $s_{2,{\rm colour}}$ is clearly seen in Figure \ref{fig:BV_correlations} with a Pearson factor of $0.60 \pm 0.09$ ($p \leq 7.7 \times 10^{-4}$). This correlation is also found for $(u-g)$ and $(g-r)$; however, for $(g-Y)$, the correlation is not statistically significant (see Figure \ref{fig_appendix:s1_s2}).

Some trends are also observed between the epoch of the transition in the colour curve and the decline rate of the two slopes. While the trends are not statistically significant for all colours, in general SNe~II which cool faster (at both early and later times) have a transition at earlier epochs (see Figure \ref{fig:BV_correlations} and Figure \ref{fig_appendix:s1_s2}). It is possible that this can be understood in terms of the effects of additional energy (beyond the shock-deposited energy from the core-collapse explosion) from
interaction with CSM close to the progenitor star \citep[e.g.,][]{khazov16,morozova16,yaron17,moriya17,dessart17,morozova17}. Such interaction may slow the colour evolution, keeping the photosphere bluer for longer and in effect stretching the colour evolution in time. 

Finally, the colours at early and late epochs correlate in the sense that bluer SNe~II at 15d are generally also bluer at 70d. For example, $(B-V)$ at 15d correlates with $(B-V)$ at 30d, 50d, and 70d with a Pearson factor of $0.85 \pm 0.01$ ($p \leq 1.0 \times 10^{-6}$), $0.77 \pm 0.02$ ($p \leq 1.0 \times 10^{-5}$), and $0.79 \pm 0.02$ ($p \leq 7.8 \times 10^{-4}$), respectively. The same trend is also seen for $(u-g)$, $(g-r)$, and $(g-Y)$. If early-time colour differences are expected (in part) to be caused by differences in progenitor radii \citep[e.g.,][]{dessart13}, then one would not expect to see this result. Differences in progenitor radii are expected to affect the time taken for photospheric temperatures to reach those of hydrogen recombination (and thus the early-time colours and the transition between the different phases, although see \citealt{faran17}); however, one would then expect temperatures --- and therefore colours --- to converge to those required for hydrogen recombination. The correlation between colours at early and late photospheric epochs could then be understood either through the continued effect of CSM interaction, or the constant colour excess expected from any uncorrected host-galaxy reddenning\footnote{While here we generally argue that our results are \textit{not} dominated by uncorrected host-galaxy extinction, there is still the possibility that some specific findings are driven by the effects of dust extinction.}. 

\subsection{Are Redder SNe~II Fainter?}\label{txt:redder_fainter}

\begin{figure}
\includegraphics[width=9.0cm]{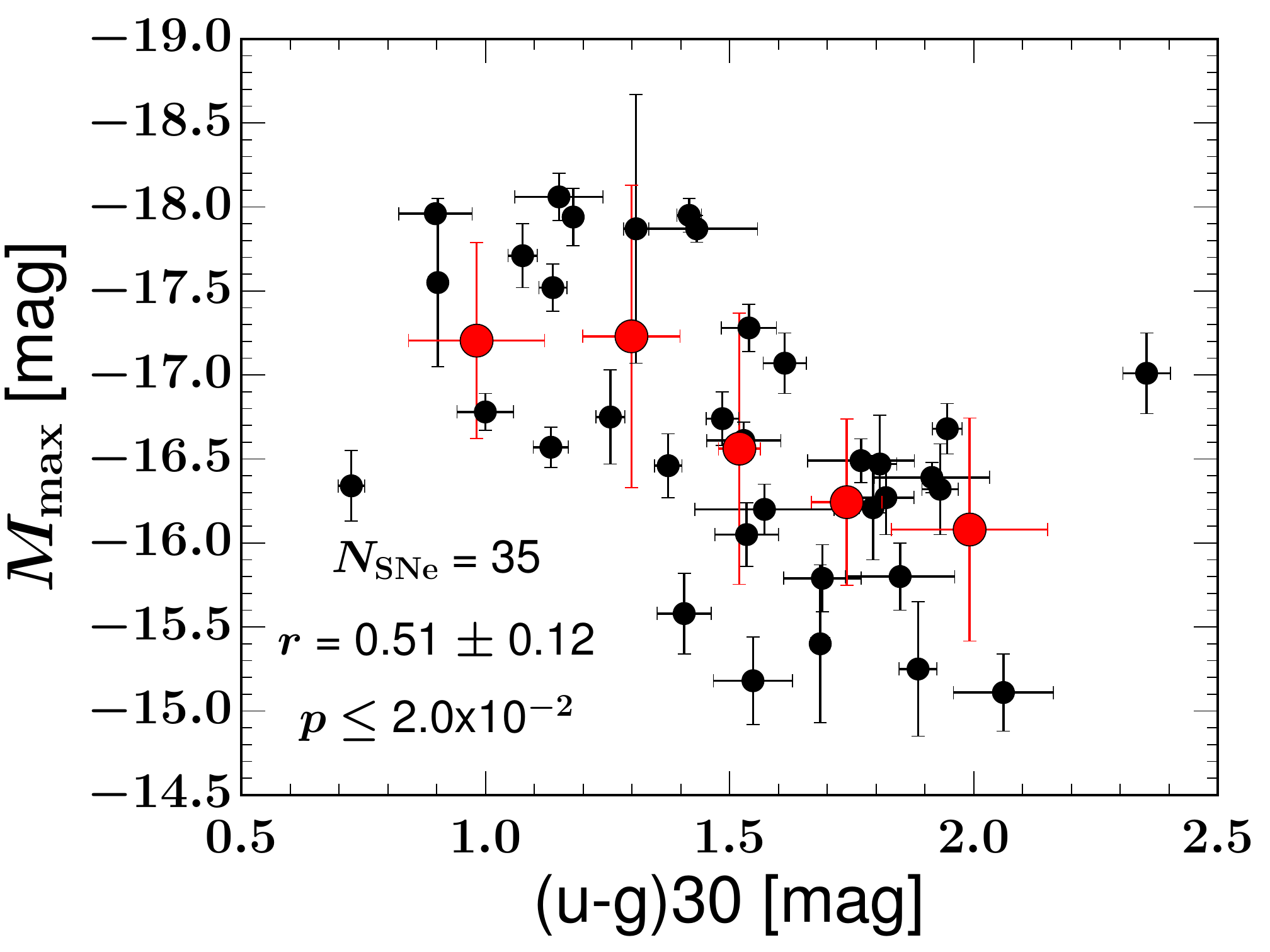}
\caption{Correlation between the absolute magnitude at maximum ($M_{\rm max}$) and the $(u-g)$ colour at epoch 30d. The results of Monte Carlo simulations on the statistics of these two variables are noted at the top of the figure: $N_{\rm SNe}$ (number of events), $r$ (Pearson's correlation coefficient), and $p$ (probability of detecting a correlation by chance). Binned data are shown in red circles, both here and throughout the paper.}
\label{fig:redder_fainter}
\end{figure}

For each colour combination at early epochs (15d--30d), redder SNe~II appear to be less luminous than bluer SNe~II. As an example, in Figure \ref{fig:redder_fainter}, the $(u-g)$ colour at epoch 30d and the absolute magnitude at maximum brightness are shown. This statement is not a new result for the SN community, as it has already been shown for SNe~Ia in a number of papers starting more than two decades ago \citep[e.g.,]{riess96,hamuy96,tripp98} and fully used for SN standardisation. However, to our knowledge, this is the first time that a study of SNe~II presents such correlations using different colours, epochs, and with a large sample (Appendix \ref{fig_appendix:redder_fainter}). Also, note that the $(B-V)$ colour comparison between our large SN~II luminosity distribution and literature low-luminosity SNe~II leads to the same conclusion. At an epoch of 60~d after explosion, \citet{spiro14a} derived a mean value of $(B-V) \approx 1.4$ mag for a sample of low-luminosity SNe~II [$(B-V) \approx 1.5$ mag found by \citealt{pastorello04}]. Using our sample of normal SNe~II, at this epoch we derive a smaller value [$(B-V) = 1.10 \pm 0.18$ mag], thus confirming that redder SNe~II are intrinsically fainter. (It is important to note here that the literature low-luminosity sample \textit{was} corrected for host-galaxy extinction.)

This correlation could be a direct effect of dust extinction that would make SNe both redder and fainter. Studies that have assumed such a hypothesis have concluded that the dust properties in SN host galaxies are incompatible with those in the Milky Way \citep{krisciunas07,eliasrosa08,goobar08,folatelli10,phillips13,poznanski09,olivares10,rodriguez14,dejaeger15b}. One possible explanation is the existence of an additional intrinsic relationship between SN colour and luminosity. 

Our results can be used to argue for an intrinsic origin of this luminosity vs. colour relation. At early epochs (15d--30d) the correlation between colour and $M_{\rm max}$ is stronger than at late epochs (50d--70d). For example, the $(g-r)$ colour at 15d correlates with $M_{\rm max}$ ($r \approx 0.53$), but not at 50d ($r \approx 0.04$). For $(B-V)$ the same is seen, with a Pearson factor at 15d of $\sim 0.35$ but only $r \approx 0.28$ and $r \approx 0.12$ at 50d and 70d, respectively. If the luminosity vs. colour relations were driven by reddenning from dust, one would not expect to see such differences, thus arguing for an intrinsic origin. This is also supported by our tests regarding the specific effects of host-galaxy extinction on our results (see Section~\ref{txt:discussion_avh}): if we remove those SNe~II that we expect to suffer from significant host-galaxy reddening, the correlations between the colour and $M_{\rm max}$ generally become stronger (see Table \ref{tab:col_bright_avh}) instead of weaker. An additional argument in favour of an intrinsic origin is that if all the colour diversity were simply due to extinction, then one would not expect to see any correlation with other parameters not affected by extinction, such as $s_{2}$. However, the colour at late times clearly correlates with $s_{2}$ (Figure \ref{fig:redder_IIL}).

In conclusion, there is strong evidence that brighter SNe~II are also bluer during the first few weeks after explosion. Such a result can be naturally explained through progenitor radii differences: SNe~II with more compact progenitors cool more quickly through rapid expansion. At the same time, in such compact progenitors significant internal energy is used to expand the star, and this leads to lower luminosities than the larger radius cases where the progenitor is already significantly expanded \citep[e.g.,][]{litvinova83,popov93,young04,utrobin07,bersten11,dessart13}. However, while such an explanation would have traditionally been accepted as the most plausible, a luminosity vs. colour relation could also be produced by CSM interaction. CSM interaction could cause both a boost in the initial SN~II luminosity and bluer colours because of the additional energy source. It is therefore difficult to disentangle effects due to progneitor radius and the presence/absence of CSM.

\subsection{Are Fast-Declining SNe~II Redder?}\label{txt:fig:redder_IIL}

Using a small sample of 11 slow-declining SNe~II (SNe~IIP) and 8 fast-declining SNe~II (SNe~IIL), \citet{faran14b} claimed that fast-declining SNe~II are redder, while \citet{patat94} had previously found ``Linears on average bluer than Plateaus,'' both at early and late epochs.

\begin{figure}
\includegraphics[width=9.0cm]{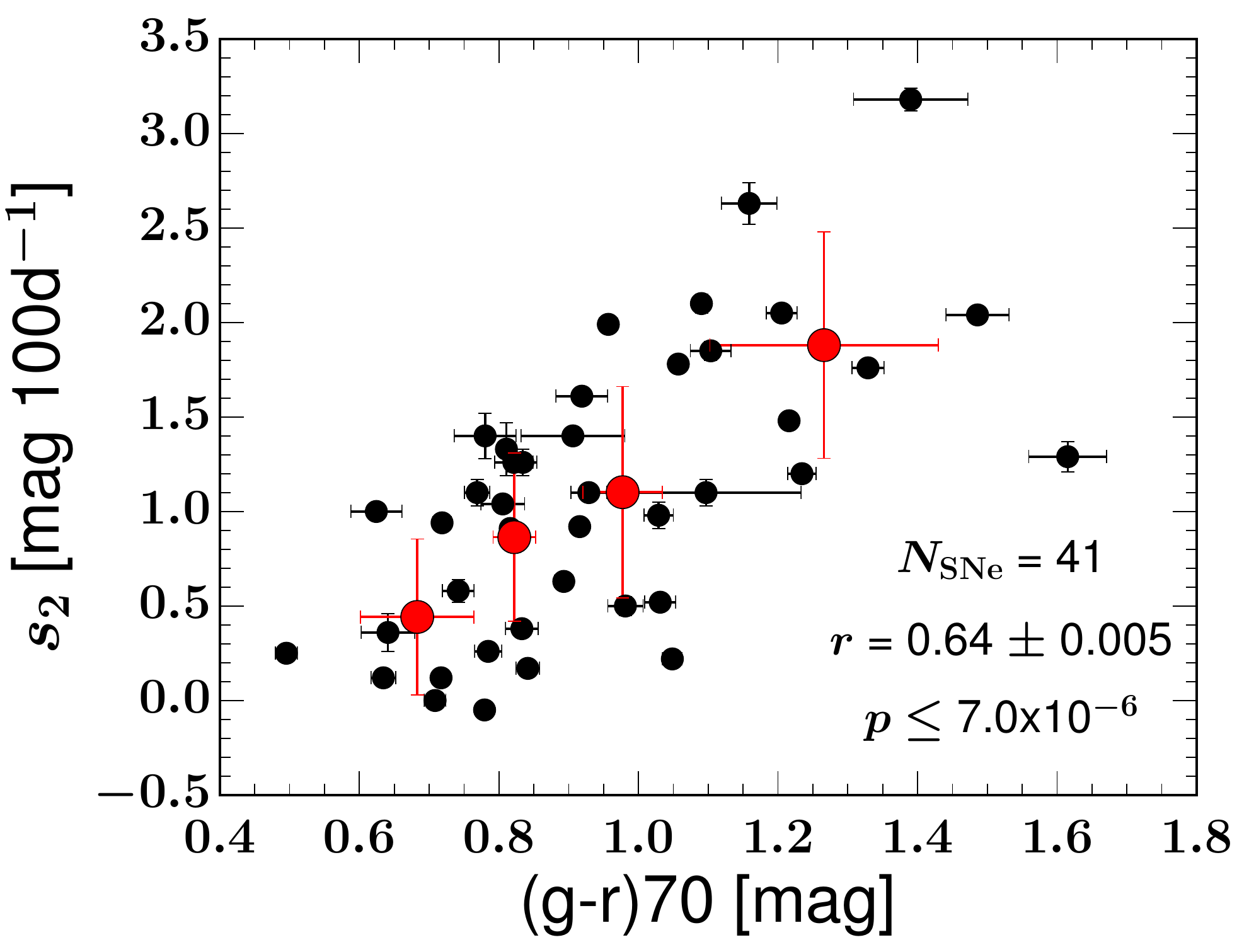}
\caption{Correlation between the slope of the plateau ($s_{2}$) in the $V$ band and the $(g-r)$ colour at epoch 70d. The results of Monte Carlo simulations on the statistics of these two variables are noted at the bottom of the figure, in a similar way to Figure \ref{fig:BV_correlations}.}
\label{fig:redder_IIL}
\end{figure}

We find that the redder SNe~II have steeper $s_{2}$ (i.e. fast-declining SNe~II), but only at later epochs (50d--70d). Figure \ref{fig:redder_IIL} illustrates our best example of this correlation (other examples, which show a similar trend, are displayed in Figure \ref{fig:appendix_redder_IIL}). 
This lack of correlation between SN~II decline rate and early-time colours suggests that the separation of SNe~II into fast (IIL) and slow (IIP) decliners is not related to progenitor radius 
(see also \citealt{valenti16}).

\begin{figure}
\includegraphics[width=9.0cm]{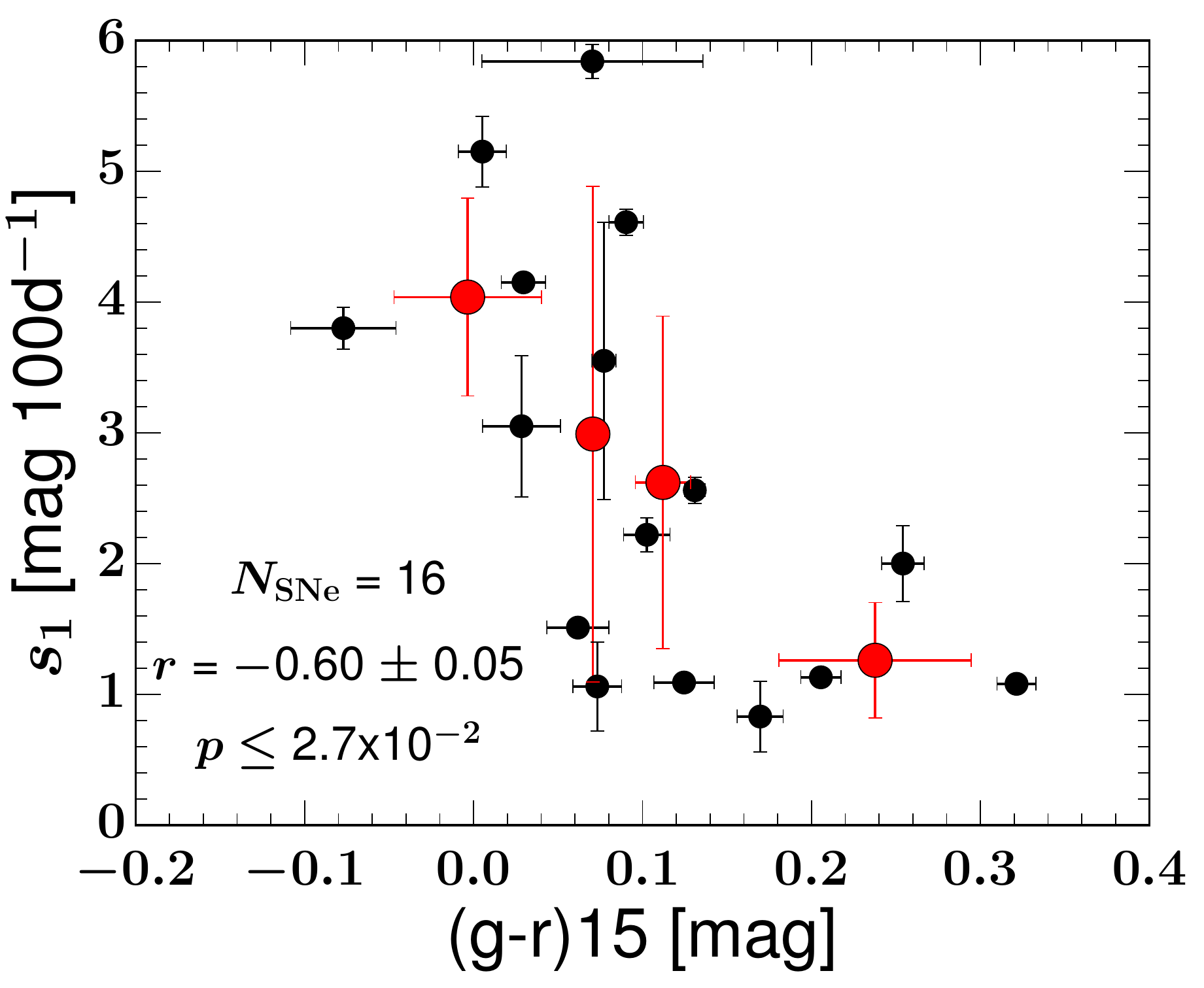}
\caption{Anticorrelation between the slope of the plateau ($s_{1}$) in the $V$ band and the $(g-r)$ colour at epoch 15d. The results of Monte Carlo simulations on the statistics of these two variables are noted at the bottom of the figure, in a similar way to Figure \ref{fig:BV_correlations}.}
\label{fig:bluer_s1}
\end{figure}

While a correlation is seen between $s_{2}$ and the colour at later epochs, at early times the initial slope after maximum brightness in the $V$-band light curve ($s_{1}$) seems to anticorrelate with the colour at early times: fast-declining SNe~II (steeper $s_{1}$) are bluer. Figure \ref{fig:bluer_s1} shows an example of this correlation using the $(g-r)$ colour at 15d (see also Figure \ref{fig:appendix_redder_IIL}). The physical cause of this correlation is not clear.

\subsection{Do Redder SNe~II have Stronger Metal-Line Equivalent Widths?}\label{txt:EW_colour}

\begin{figure}
\includegraphics[width=9.0cm]{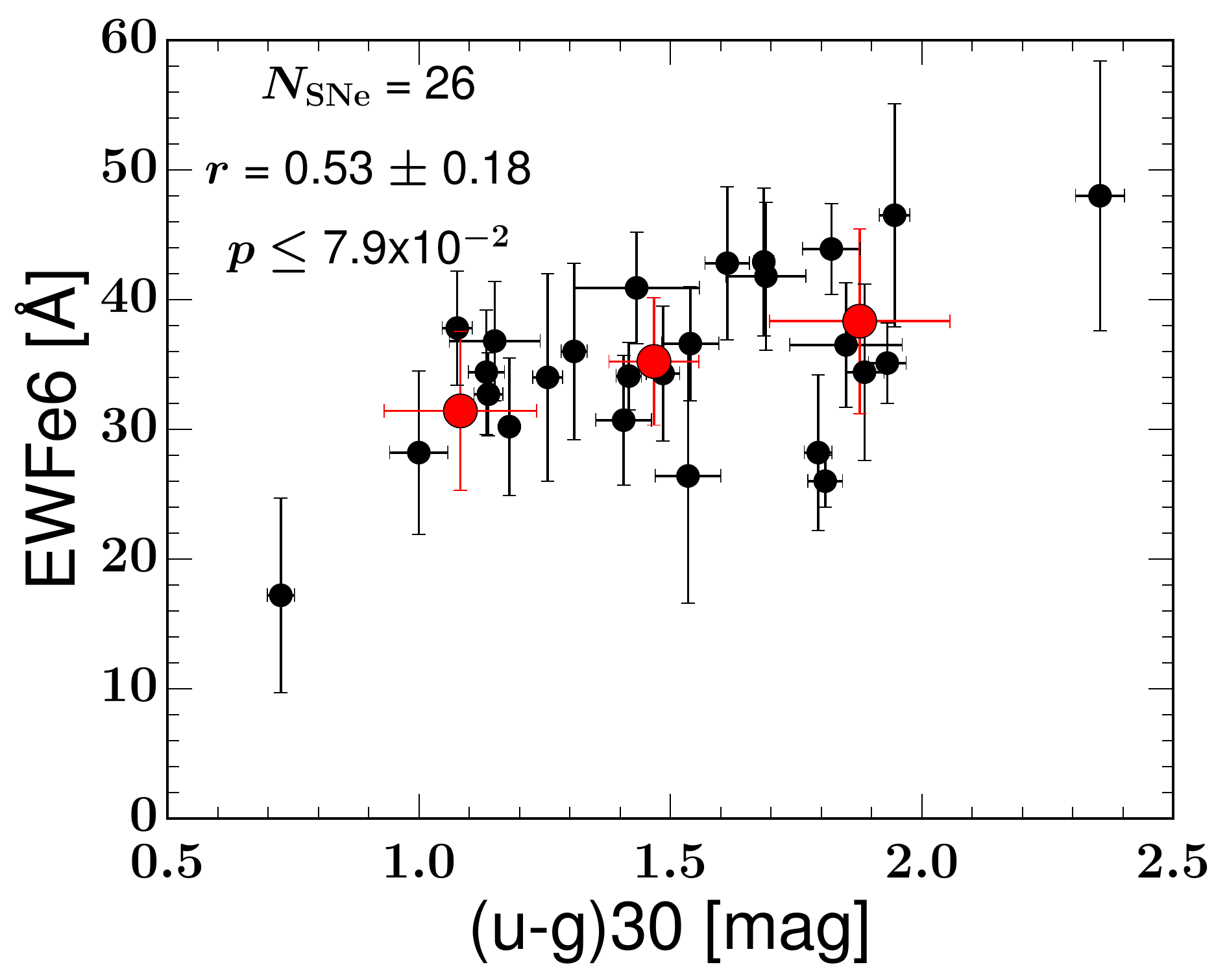}
\caption{Correlation between the \ion{Fe}{II} $\lambda 5169$ equivalent width (EWFe6) and the $(u-g)$ colour at epoch 30d. The results of Monte Carlo simulations on the statistics of these two variables are noted at the top of the figure, in a similar way to Figure \ref{fig:BV_correlations}.}
\label{fig:EW_redder}
\end{figure}

The comparisons between colours at different epochs and the equivalent widths of different absorption lines show that in general, bluer SNe~II tend to have smaller equivalent widths. In Figure \ref{fig:EW_redder}, one example is displayed using the $(u-g)$ colour at 15d and the \ion{Fe}{II} $\lambda 5169$ equivalent width (EWFe6). Correlations at different epochs, using other colours and with different spectral lines, can be found in Figure \ref{fig_appen:EW_redder}. 

The simplest explanation for these trends is that of an expected strong temperature depdendence of both the observed colours and the strength of spectral lines. Again, these temperature differences could be related to differences in progenitor radii and/or the presence/absence of CSM interaction. However, when interpreting these results, one needs to be aware of the effect of line strengths on SN~II magnitudes at specific bandpasses because of their presence within those bandpasses; see Figure~\ref{fig:spect_filter}. If, for example, we only see these correlations using the $(u-g)$ colour, it could be due a filter selection effect. If the Pearson factors are stronger with one colour than the others ($u-g$ in this work), one could argue that is simply because a deeper absorption line (larger equivalent width) reduces the brightness in the filter containing the line --- i.e. that the colour is caused by line blocking and not temperature effects. However, it is difficult to estimate this effect because a ``true'' temperature is not easily extracted from our data. We note that the $u$ band is strongly affected by line blanketing from millions of blended iron-group lines (mostly \ion{Ti}{iii} and \ion{Fe}{ii}; \citealt{kasen09}) which cannot be resolved.

Additionally, \citet{dessart14} show that a secondary effect affecting the colour is that caused by progenitor metallicity. Lower metallicity leads to smaller line strengths at a given temperature (colour). Since $(u-g)$ shows stronger correlations with the equivalent widths of different elements than do other colours, we test to see if a relation exists between $(u-g)$ and metallicity. As a metallicity proxy, we use the values from \citet{anderson16a}, which are derived using the N2 diagnostic (\citealt{marino13}, hereafter M13). We do not find statistical evidence of a such relationship. Nor do we detect a correlation between the metallicity and the slopes of the colour curves as theoretically suggested by \citet{dessart13}, where the authors showed that SNe~II with lower metallicity have faster colour evolution (due to the dependence of pre-SN radius on progeneitor
metallicity).

\subsection{Do Redder SNe~II have Faster Ejecta Velocities?}\label{velo_colour}

\begin{figure}
\includegraphics[width=9.0cm]{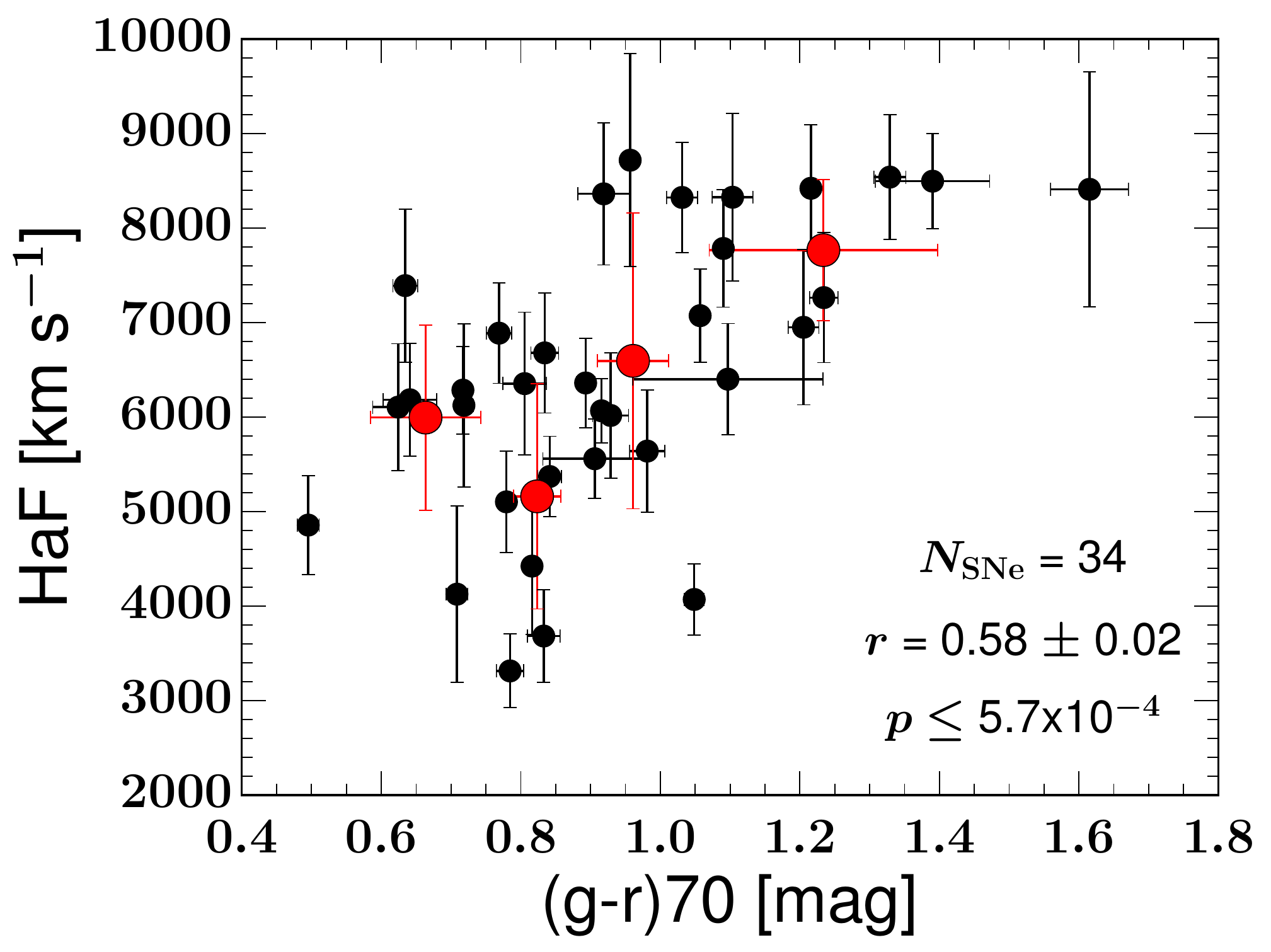}
\caption{Correlation between the H$\alpha$ velocity (HaF) and the $(g-r)$ colour at epoch 70d. The results of Monte Carlo simulations on the statistics of these two variables are noted at the bottom of the figure, in a similar way to Figure \ref{fig:BV_correlations}.}
\label{fig:velo_red}
\end{figure}

From our analysis, a correlation between the velocity at 50d of different lines (H$\alpha$, H$\beta$, FeII, ScFe, and ScM) and the $(g-r)$ colour (to a lesser extent $g-Y$) at later epochs (50d--70d) is found. Redder SNe~II have faster ejecta velocities. As an example, the correlation between the H$\alpha$ velocity and the colour $(g-r)$ at 70d is displayed in Figure \ref{fig:velo_red}. Such a correlation has not been predicted (to our knowledge) by any explosion models, and its physical origin is unclear. However, as above, we note the possibility of a direct influence of the spectral-line properties on the measured colour: Figure~\ref{fig:velo_red} correlates a colour using the $r$ band (which contains H$\alpha$) with a measured property of H$\alpha$.

\subsection{Intrinsic Colours}\label{txt:cc_intrin}

To support our hypothesis that the SN~II colour diversity is driven by intrinsic differences, in this section we propose a ``reductio ad absurdum''. If SN~II colour diversity were dominated by host-galaxy extinction, then a sample of SNe~II devoid of reddening should display roughly identical colours. To proceed, we follow the same steps as outlined in Section \ref{txt:discussion_avh}. First, for our low-reddening SN~II sample, we select SNe~II which are located away from host-galaxy nuclei and with no \ion{Na}{i}~D absorption, leading to a set of 19 SNe~II. 

\begin{figure}
\includegraphics[width=9.0cm]{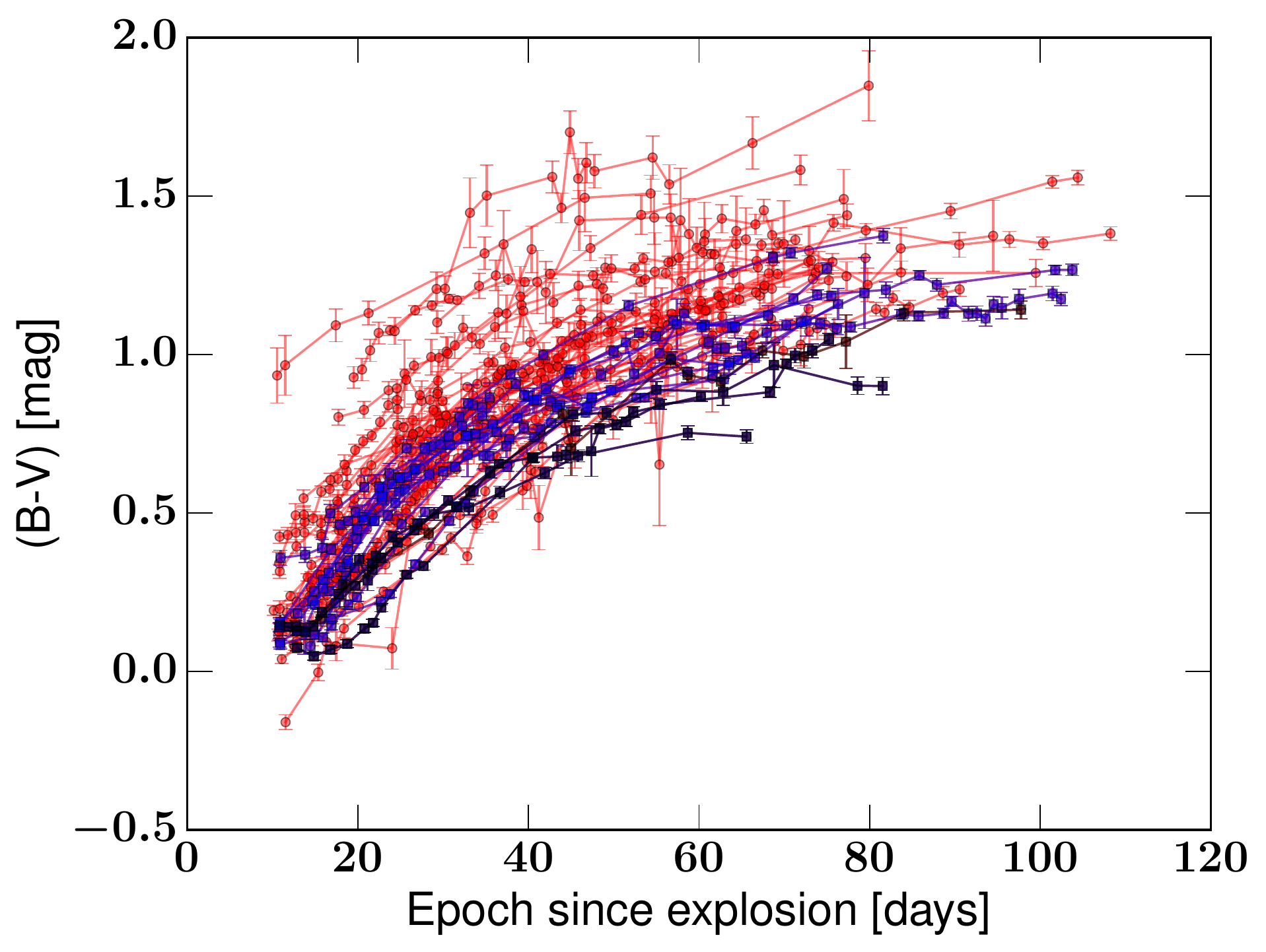}
\caption{$(B-V)$ curves for the entire sample (57 SNe~II) are shown in red. Blue light curves represent our unreddened subsample (19 SNe~II), while in black the bluest SNe from the unreddened subsample are shown (SN~2004fx, SN~2005dz, SN~2008M, and SN~2009bz).}
\label{fig:BV_colour_all}
\end{figure}

\indent Figure \ref{fig:BV_colour_all} shows $(B-V)$ colour curves for our full sample, along with the unreddened subsample composed of 19 SNe~II displayed in distinct colours (blue). None of the unreddened colour curves is located on the top of the plot; they do not have the reddest colours. This is to be expected given that extinction makes objects redder, and that we grant that \textit{some} SNe~II within the sample probably suffer from significant extinction. However, the important observation is that this ``unreddenned'' sample spans a wide range in SN~II colours, extending from $(B-V)$ (at 50d) of $\sim 0.70$ mag (SN~2004fx) to $(B-V)$ (at 50d) of $\sim 1.13$ mag (SN~2009N). Indeed, $\sim 75$\% of the full sample falls within this range of colours, suggesting that (at least) this same percentage of SNe~II suffers from negligible host-galaxy extinction. One can therefore conclude that significant dispersion exists in intrinsic SN~II colours and that this dispersion dominates observed SN~II colours.

With respect to colour evolution, for this subsample we derive a median value of $2.26 \pm 0.51$ mag 100d$^{-1}$, $0.74 \pm 0.22$ mag 100d$^{-1}$, and $37.9 \pm 7.4$ days for $s_{1,(B-V)}$, $s_{2,(B-V)}$, and $T_{{\rm trans},(B-V)}$, respectively. These values are consistent with the values derived for the whole sample ($2.65 \pm 0.62$, $0.77 \pm 0.26$, and $37.7 \pm 4.3$ days, respectively) but also with the ``reddened'' subsample ($2.69 \pm 0.66$, $0.81 \pm 0.26$, and $37.6 \pm 5.6$ days, respectively). This suggests that SNe~II display distinct colours and colour evolution independent of any host-galaxy extinction suffered by each event.

\indent Finally, to conclude our tests, as suggested by \citet{burns14}, \citet{stritzinger17a}, and \citet{galbany16a,galbany17} with respect to SNe~II, we can restrict our unreddened subsample to only the bluest objects (SN~2004fx, SN~2005dz, SN~2008M, and SN~2009bz). Again, if the colour were caused only by dust extinction, these four SNe~II with the bluest colours should have roughly identical colours. The colour-curve parameters for all the colours are displayed in Table \ref{table:table_intrin}; we note that the values obtained for this lowest-reddening subsample do not follow the same trend and also exhibit significant diversity. For example, SNe~II with higher $s_{1,(B-V)}$ values display both smaller or larger $s_{2,(B-V)}$ values (SN~2005dz and SN~2009bz, respectively). The fact that even for this restricted subsample, the colour curves also show diversity, is consistent with our previous conclusions and suggests that intrinsic colour diversity is the origin of most of the observed differences.

\begin{table*}
\begin{center}
\caption{Unreddened subsample colour-curve parameters.}
\begin{tabular}{ccccccc}
\hline
Parameters & SN~2004fx & SN~2005dz  & SN~2008M & SN~2009bz & mean \\
\hline
\hline
$s_{1,(B-V)}$ & 1.69 (0.24) & 2.22 (0.13)  & 2.30 (0.15) & 2.17 (0.08) & 2.10 (0.24)\\
$s_{2,(B-V)}$ & 0.50 (0.14) & 0.26 (0.08)  & 0.89 (0.17) & 0.95 (0.09) & 0.65 (0.28)\\
$T_{{\rm trans},(B-V)}$ & 36.8 (3.3) & 43.9 (1.7)  & 45.4 (3.5) & 36.2 (1.5) & 40.6 (4.2)\\
\hline
$s_{1,(u-g)}$ & 6.51 (0.48) & 5.50 (0.36)  & 6.38 (0.38) & 6.66 (0.26) & 6.26 (0.45)\\
$s_{2,(u-g)}$ & 1.63 (0.47) & $\cdots$ ($\cdots$)  & $\cdots$ ($\cdots$) & 3.33 (0.93) & 2.48 (0.85)\\
$T_{{\rm trans},(u-g)}$ & 29.9 (2.1) & $\cdots$ ($\cdots$) & $\cdots$ ($\cdots$) & 32.5 (4.1) & 31.2 (1.3)\\
\hline
$s_{1,(g-r)}$ & 1.34 (0.10) & 1.72 (0.14)  & 1.91 (0.08) & 1.72 (0.08) & 1.67 (0.21)\\
$s_{2,(g-r)}$ & 0.23 (0.07) & 0.06 (0.07)  & 0.77 (0.05) & 0.54 (0.06) & 0.40 (0.26)\\
$T_{{\rm trans},(g-r)}$ & 33.9 (1.7) & 41.1 (2.0) & 39.3 (1.7) & 33.4 (1.3) & 36.9 (3.5)\\
\hline
$s_{1,(g-Y)}$ & $\cdots$ ($\cdots$) & 2.35 (0.18)  & 2.92 (0.17) & 2.92 (0.16) & 2.73 (0.27)\\
$s_{2,(g-Y)}$ & $\cdots$ ($\cdots$) & 0.18 (0.13)  & 0.84 (0.08) & 0.82 (0.19) & 0.61 (0.30)\\
$T_{{\rm trans},(g-Y)}$ & $\cdots$ ($\cdots$) & 43.1 (1.0)  & 39.9 (1.6) & 35.1 (2.0) & 39.4 (3.3)\\
\hline
\hline
\end{tabular}
\label{table:table_intrin}
\end{center}
\end{table*}

\section{Conclusions}

An extensive analysis of SN~II colour curves using four different colours, together with their correlations with spectroscopic and photometric parameters, has been presented in this work. Here we list our main conclusions.

\begin{enumerate}

\item{SN~II colour curves consist of two regimes which correlate: SNe~II with steeper initial colour curves also have steeper colour curves during the plateau phase.\\}

\item{The colour evolution depends on the initial conditions. Bluer SNe~II at 15 d after explosion are also bluer at 70d after explosion.\\}

\item{SNe~II form a continuous population in observed colours, consistent with other recent light-curve analyses finding an absence of a clear separation of these events into distinct classes.\\}

\item{Redder SNe~II have fainter absolute magnitude at maximum. This correlation seems to originate from intrinsic colours, not from dust effects. Progenitor radius differences, together with the presence or absence of CSM close to progenitor stars, are the probabale causes of this relation.\\}

\item{An intrinsic range in SN~II colours of $\sim 0.70$--1.13 mag in $(B-V)$ is derived at 50 days post explosion, with $\sim 75$\% of the current sample falling within these limits. Such large intrinsic diversity is observed at all epochs.\\}

\item{Fast-declining SNe~II (SNe~IIL) are redder at later epochs (50d--70d); however, they appear bluer at early epochs (15d--30d).\\}

\item{At all epochs, redder SNe~II have larger metal-line equivalent widths.\\}

\item{The strength of the H$\alpha$ absorption line (Haabs) anticorrelates with the slope of the colour curve during the plateau phase ($s_{2,{\rm colour}}$).\\}

\item{Redder SNe~II at later epochs (50d--70d) seem to have faster ejecta velocities. However, the correlation is not seen in bluer filters.\\}

\end{enumerate}

The exact (intrinsic) physical origin of the observed colour differences and their correlation with other SN~II transient properties in still not well understood. Contemporary work --- especially at early epochs --- suggests that SN~II progenitors may not explode into a clean environment, but rather interact with CSM of various densities and extensions. When such interaction occurs, it is likely to complicate our understanding of SN~II colours and observations in general, especially with respect to constraints on progenitor radius.

Additional early-time multicolour observations can aid our understanding and attempt to isolate the dominant progenitor properties producing observed diversity at different epochs. In addition, higher-resolution spectra of SNe~II (than those generally obtained by the SN community) would help to definitively tie down the degree of host-galaxy reddening affecting observed colours. Such observations and advances in our understanding will bring added confidence to our use of SNe~II as distance indicators, allowing a better understanding of colour-term corrections. In this way, we will continue to better understand the progenitor systems of the most populous terminal stellar explosions, SNe~II, while also refining their use as astrophysical probes.

\section*{Acknowledgements}
We thank the referee for their thorough reading of the manuscript, which helped clarify and improve it. The work of the CSP-I has been supported by the NSF under grants AST--0306969, AST--0607438, and AST--1008343. Support for A.V.F.'s supernova research group at U.C. Berkeley has been provided by NSF grant AST--1211916, the TABASGO Foundation, Gary and Cynthia Bengier (T.d.J. is a Bengier Postdoctoral Fellow), the Christopher R. Redlich Fund, and the Miller Institute for Basic Research in Science (U.C. Berkeley). L.G. was supported in part by the NSF under grant AST--1311862. M.H. acknowledges support from the Ministry of Economy, Development, and Tourism's Millennium Science Initiative through grant IC120009, awarded to The Millennium Institute of Astrophysics (MAS). M.D.S. acknowledges funding by a research grant (13261) from the VILLUM FONDEN. E.Y.H. acknowledges support provided by NSF grant AST--1613472 and by the Florida Space Grant Consortium. 

This research has made use of the NASA/IPAC Extragalactic Database (NED), which is operated by the Jet Propulsion Laboratory, California Institute of Technology, under contract with the National Aeronautics and Space Administration (NASA), and of data provided by the Central Bureau for Astronomical Telegrams. The work is also based on observations obtained at the Gemini Observatory, which is operated by the Association of Universities for Research in Astronomy, Inc., under a cooperative agreement with the NSF on behalf of the Gemini partnership: the NSF, the STFC (United Kingdom), the National Research Council (Canada), CONICYT (Chile), the Australian Research Council (Australia), CNPq (Brazil), and CONICET (Argentina). This research used observations from Gemini programs GN-2005A-Q-11, GN-2005B-Q-7, GN-2006A-Q-7, GS-2005A-Q-11, GS-2005B-Q-6, and GS-2008B-Q-56.



%%%%%%%%%%%%%%%%%%%%%%%%%%%%%%%%%%%%%%%%%%%%%%%%%%

%%%%%%%%%%%%%%%%% APPENDICES %%%%%%%%%%%%%%%%%%%%%
\appendix

\section{}\label{AppendixA}
In the main body of this paper, we have shown an overview of our analysis and conclusions, without discussing all the correlations found in this SN~II colour-curve study. For completeness, in this Appendix several more figures are presented.

\subsection{Colour-Curve Fitting}\label{appendix:colour_fit}

In this section, additional examples of $(B-V)$ colour fitting are shown, together with examples of the $(u-g)$, $(g-r)$, and $(g-Y)$ colour fitting. As explained is the main body of this paper, when the data are well sampled, we clearly see two regimes with two different slopes as noted by \citet{patat94}, but in our case with a much larger set of objects.

% Put spaces around the ``='' signs in the three equations.
% Check the equations for the green-line fits: I have not tried some specific 
% times (t), but based on what happened in the figure in the main text, I
% think it would be wise to check these, to make sure they are correct.
%   Also, in the ordinate labels, change ``Mmax'' to ``M_max''
%  and ``Mend'' to ``M_end''
%  Also, in the ordinate labels, delete the space between ``100'' and ``d''
%  Don't italicize the Angstrom symbol.

\begin{figure*}
\includegraphics[width=5.6cm]{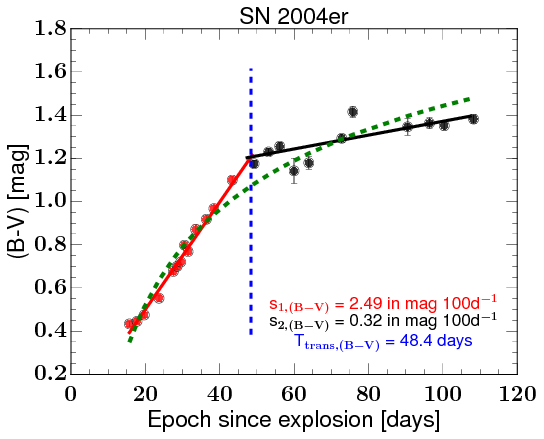}\includegraphics[width=5.6cm]{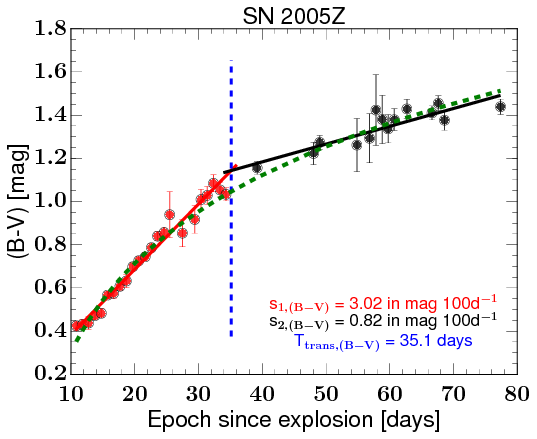}\includegraphics[width=5.6cm]{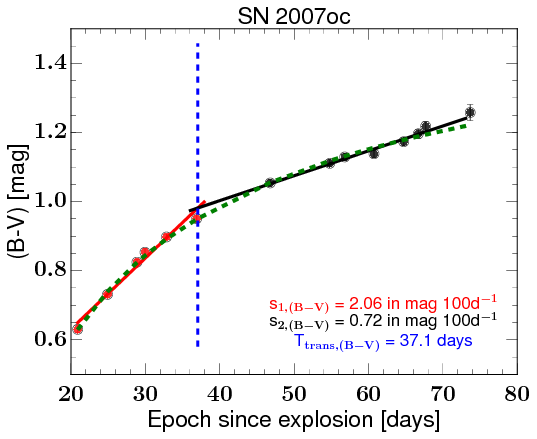}\\
\includegraphics[width=5.6cm]{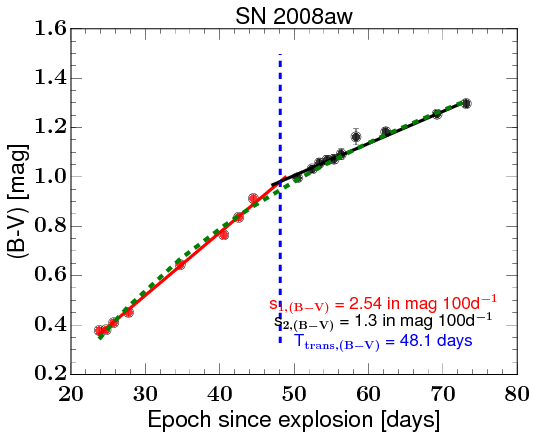}\includegraphics[width=5.6cm]{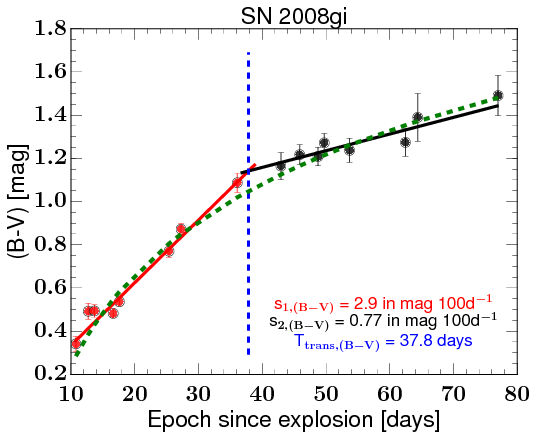}\includegraphics[width=5.6cm]{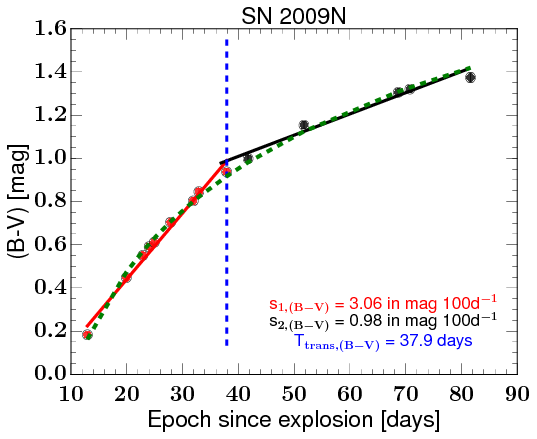}\\
\includegraphics[width=5.6cm]{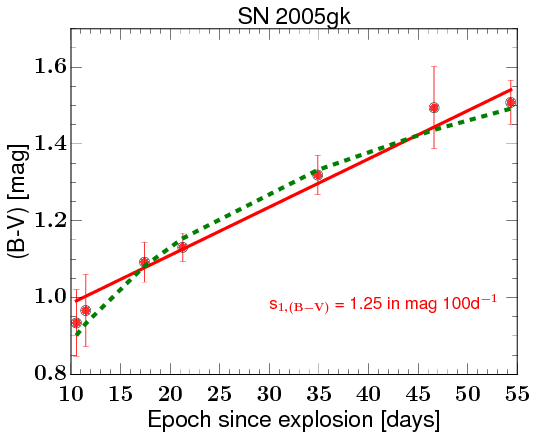}\includegraphics[width=5.6cm]{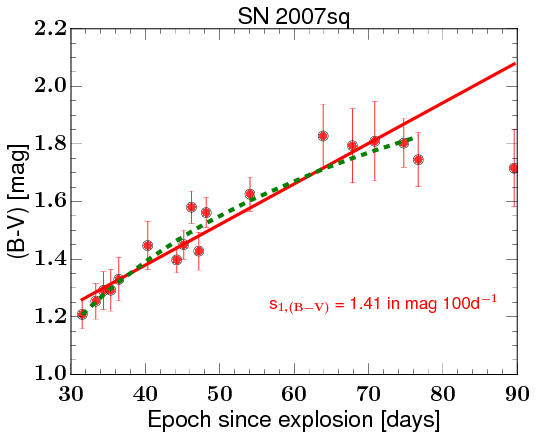}\includegraphics[width=5.6cm]{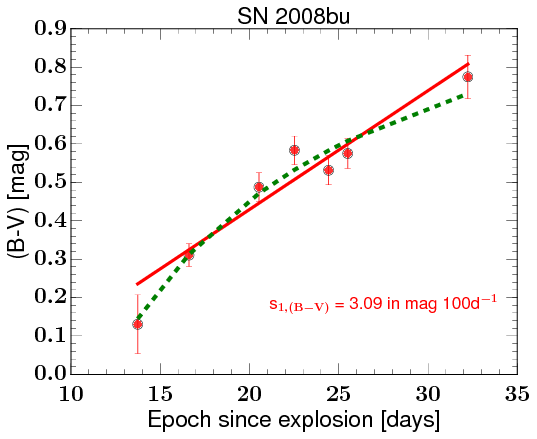}\\
\includegraphics[width=5.6cm]{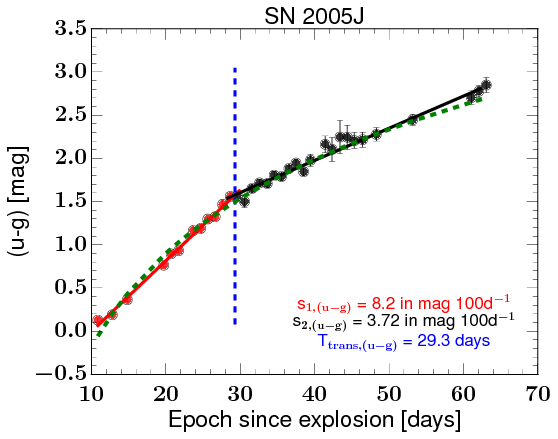}\includegraphics[width=5.5cm]{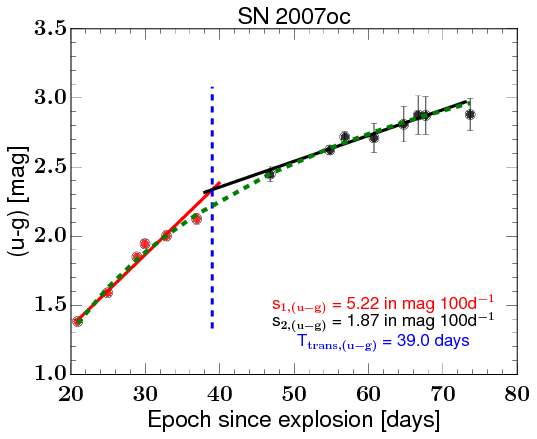}\includegraphics[width=5.7cm]{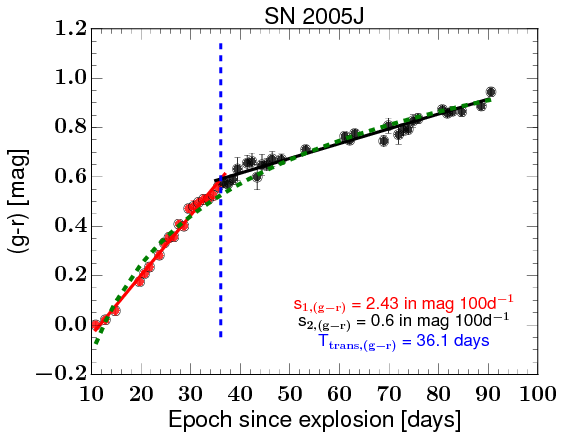}\\
\includegraphics[width=5.6cm]{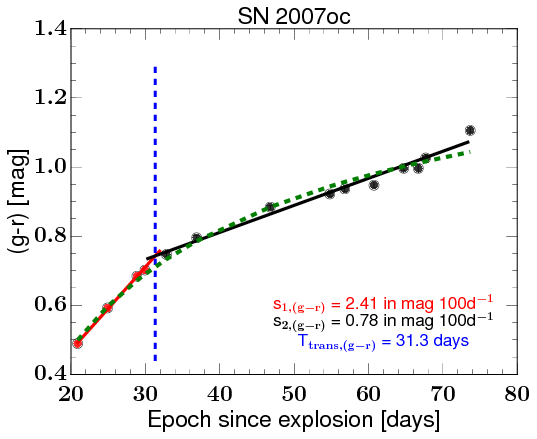}\includegraphics[width=5.6cm]{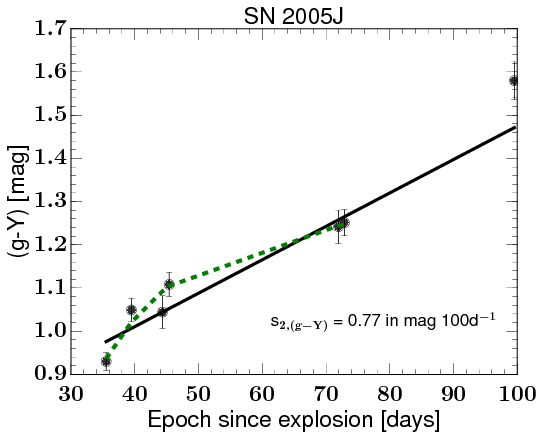}\includegraphics[width=5.6cm]{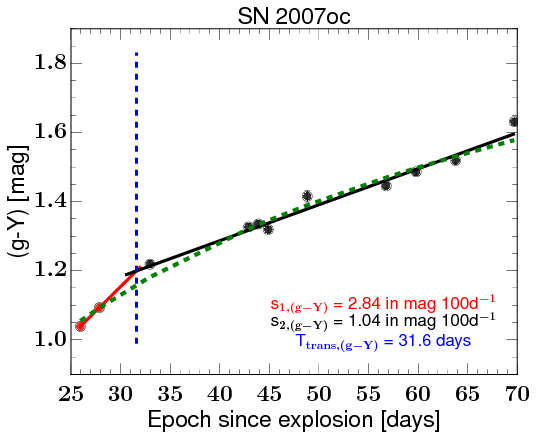}\\
\caption{Example of the colour-curve parameters measured for each SN. The two slopes, $s_{1,(X-Y)}$ and $s_{2,(X-Y)}$, are represented in red and black, respectively (when two slopes are seen). The epoch of the transition if it exists, $T_{{\rm trans},(X-Y)}$ is shown in blue. The colours $(B-V)$, $(u-g)$, $(g-r)$, and $(g-Y)$ are represented. Power-law fitting is shown with the green dashed line.}
\label{fig:append_colour_fit}
\end{figure*}

\subsection{Colour-Curve Parameter Correlation}\label{appendix:colour_params_corre}

In this section, more examples of the correlations between the two slopes seen in the colour curves are shown.

\begin{figure*}
\includegraphics[width=6.0cm]{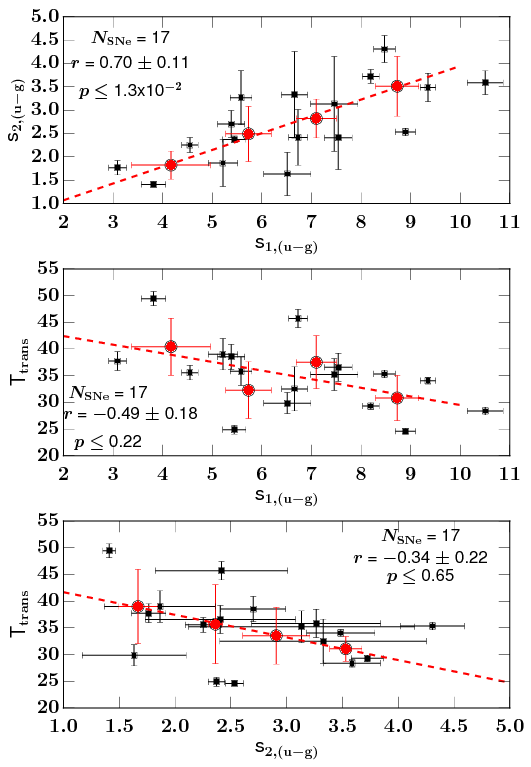}\includegraphics[width=6.2cm]{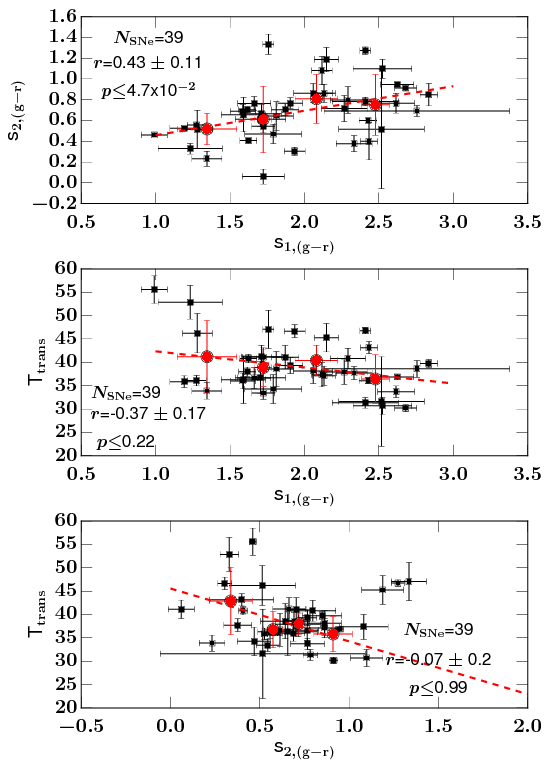}\includegraphics[width=6.2cm]{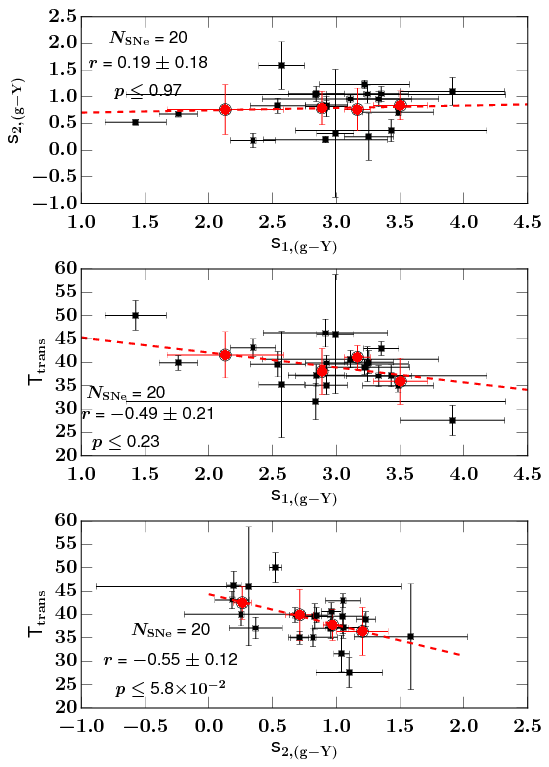}
\caption{Correlations between measured parameters of the SN~II colour curves: $(u-g)$ (\textit{left}), $(g-r)$(\textit{middle}), and $(g-Y)$ (\textit{right}). In all of the panels, the same structure is displayed: \textit{top}, $s_{\rm 1,colour}$ with $s_{\rm 2,colour}$ both expressed in mag (100 d)$^{-1}$; \textit{middle}, $s_{\rm 1,colour}$ with $T_{\rm trans,colour}$ (in days); \textit{bottom}: $s_{\rm 2,colour}$ with $T_{\rm trans,colour}$. In each panel the number of SNe is listed, together with the Pearson factor and the $p$-value ($p$).}
\label{fig_appendix:s1_s2}
\end{figure*}

\subsection{Are Redder SNe~II Less Luminous?}\label{appendix:redder_fainter}

\begin{figure*}
\includegraphics[width=6.1cm]{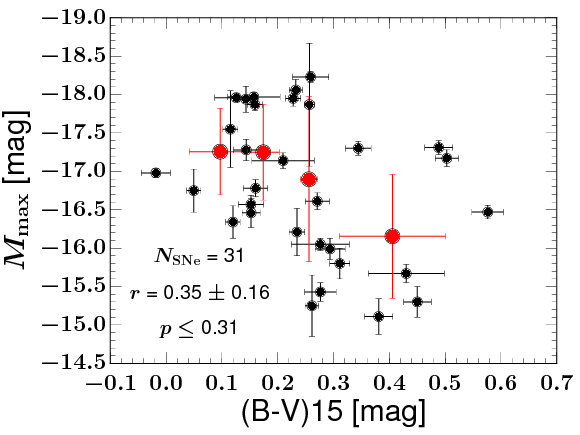}\includegraphics[width=6.1cm]{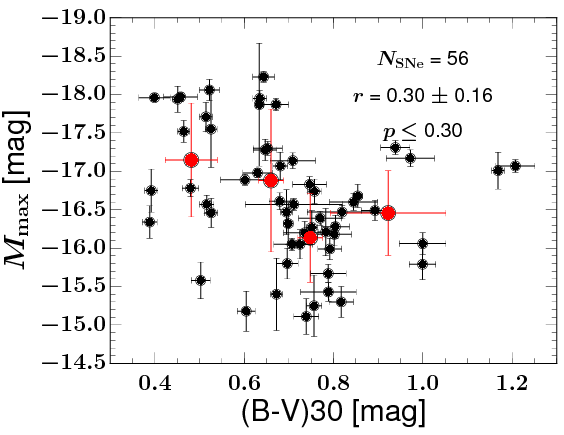}\includegraphics[width=6.1cm]{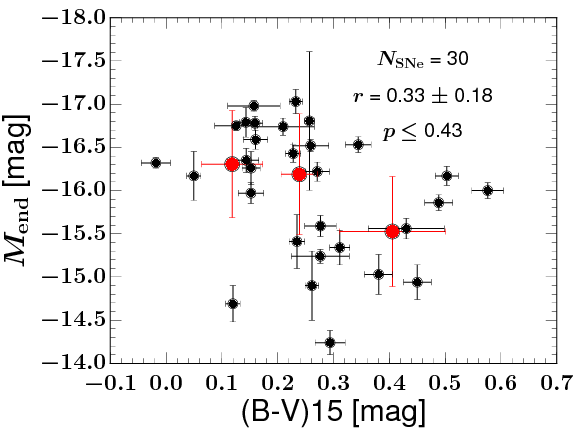}\\
\includegraphics[width=6.1cm]{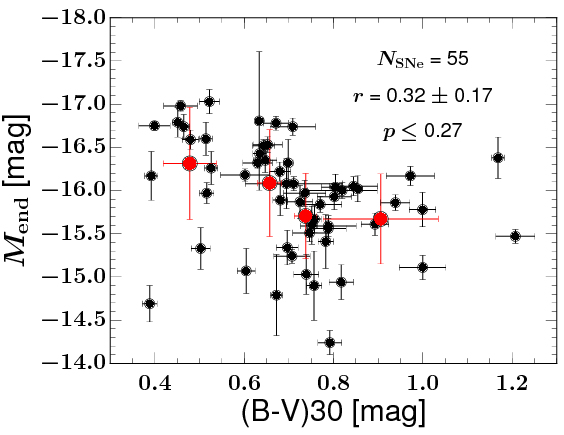}\includegraphics[width=6.1cm]{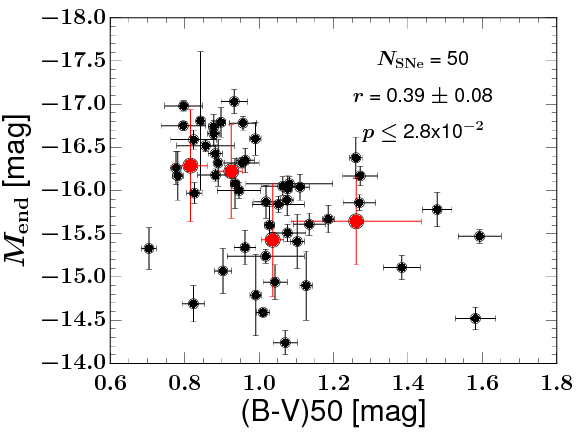}\includegraphics[width=6.1cm]{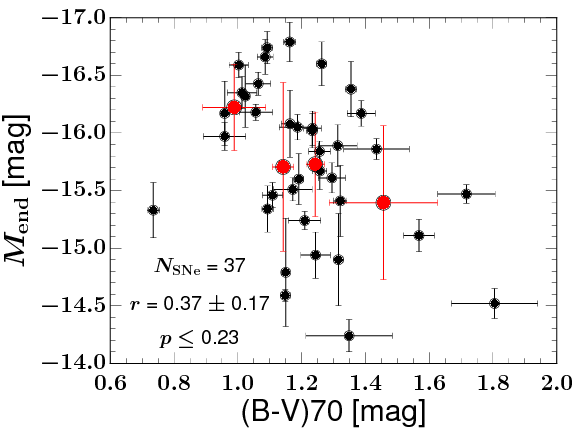}\\
\includegraphics[width=6.1cm]{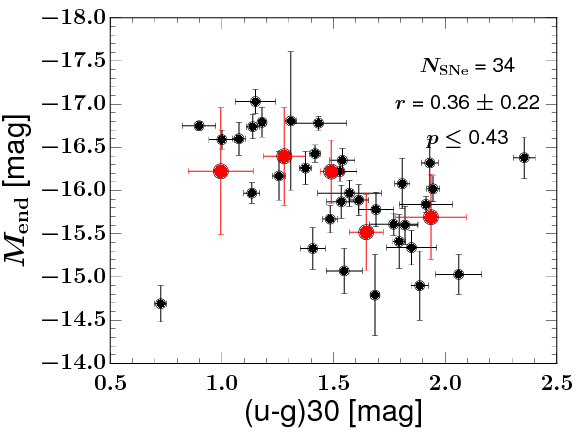}\includegraphics[width=6.1cm]{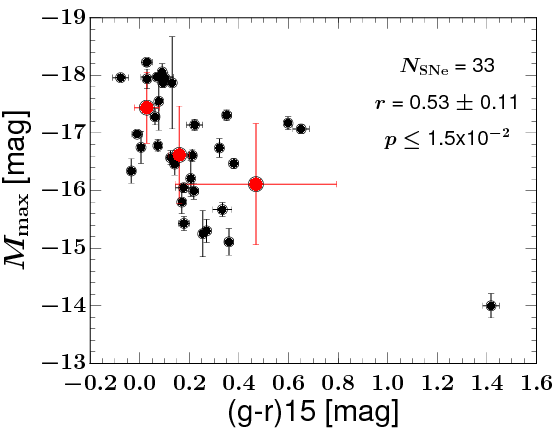}\includegraphics[width=6.1cm]{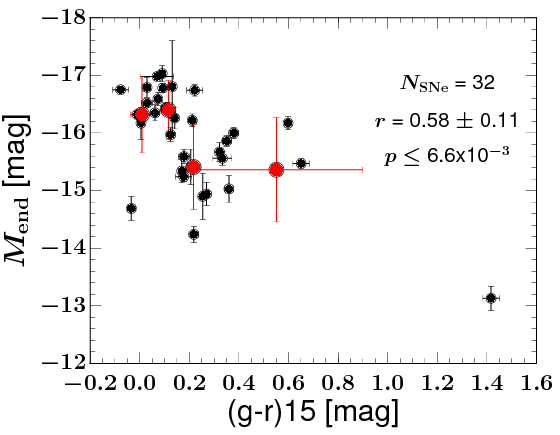}\\
\includegraphics[width=6.1cm]{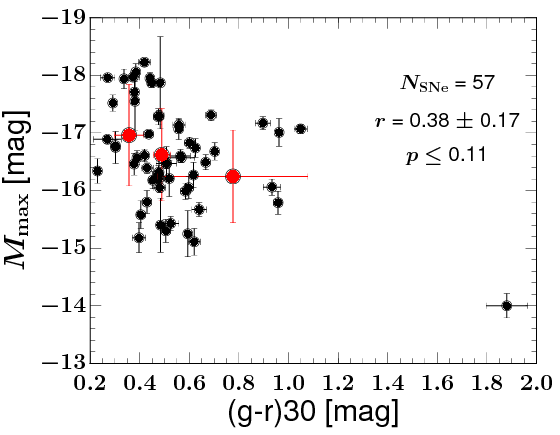}\includegraphics[width=6.1cm]{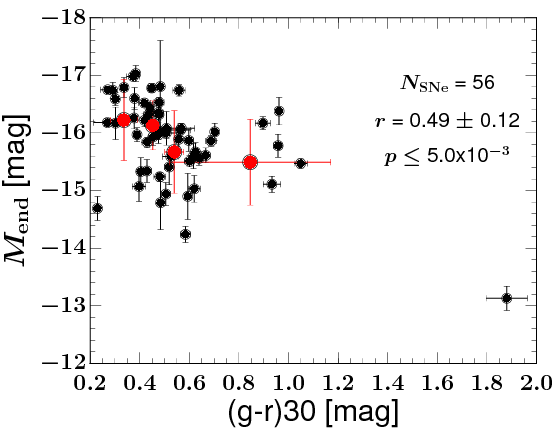}\includegraphics[width=6.1cm]{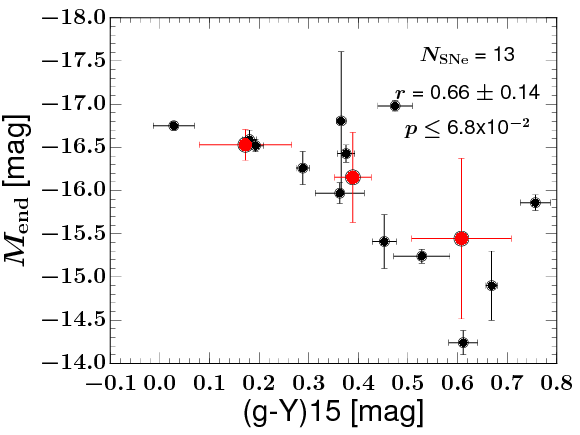}\\
\includegraphics[width=6.0cm]{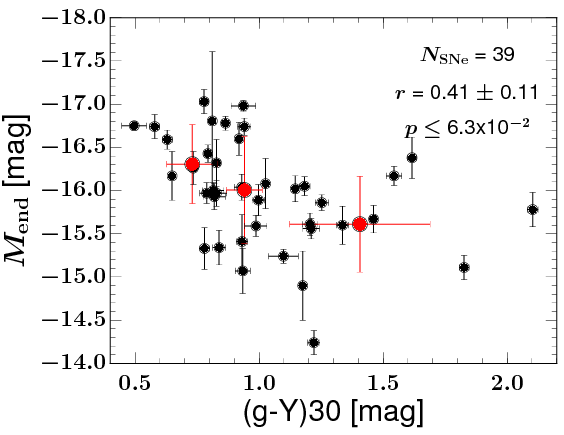}\includegraphics[width=6.0cm]{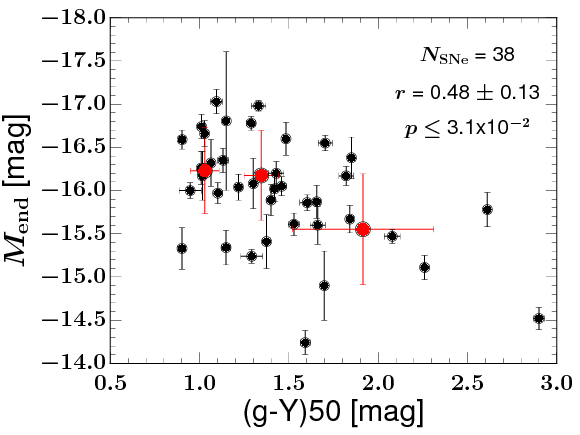}
\caption{Correlation between the colour and the absolute magnitude at different epochs. The results of Monte Carlo simulations on the statistics of the variables are noted, similarly to Figure \ref{fig:BV_correlations}.}
\label{fig_appendix:redder_fainter}
\end{figure*}

In this section, all correlations between the colour and the absolute magnitude are shown. The relation between the absolute magnitude at maximum brightness and the colour is also presented for different combinations of colour, as shown in Figure \ref{fig_appendix:redder_fainter}.

\subsection{Are Fast-Declining SNe~II Redder?}\label{Appendix_fig:redder_IIL}

\begin{figure*}
\includegraphics[width=6.3cm]{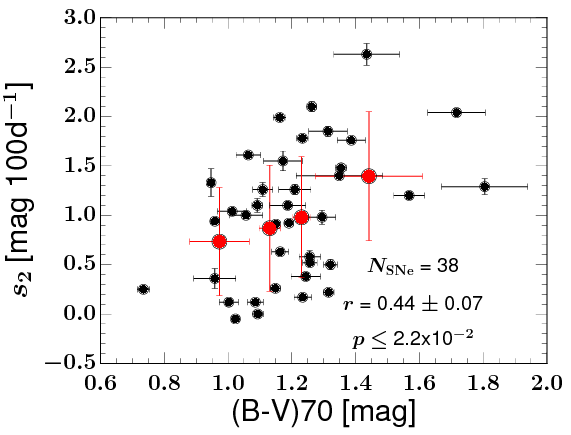}\includegraphics[width=6.3cm]{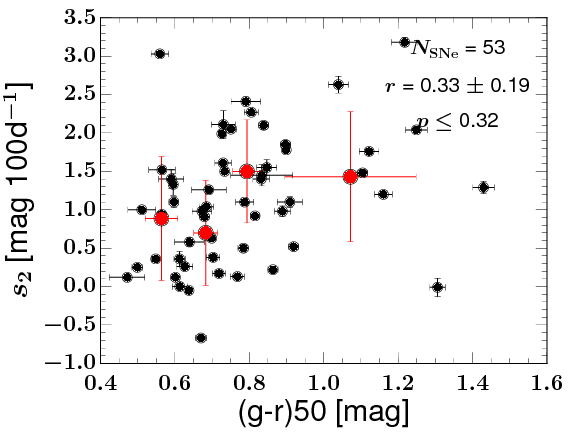}\includegraphics[width=6.3cm]{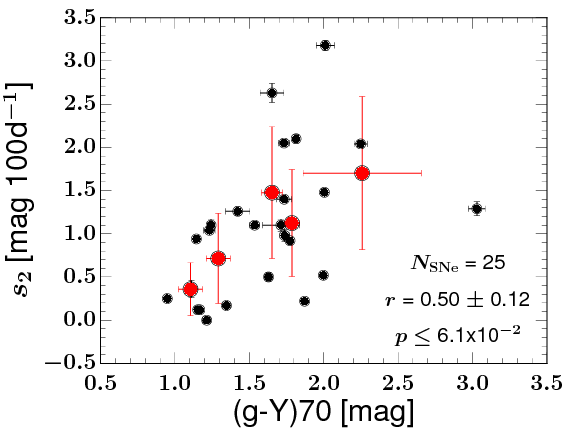}
\includegraphics[width=6.0cm]{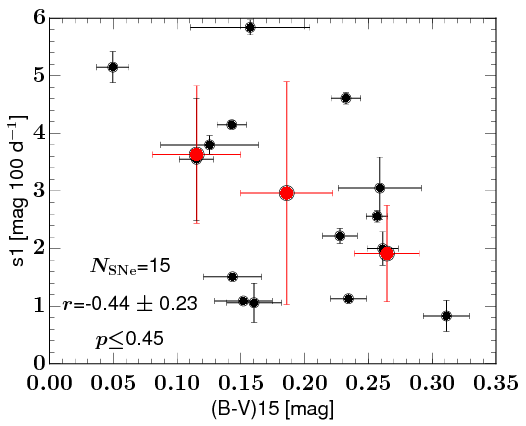}\includegraphics[width=6.0cm]{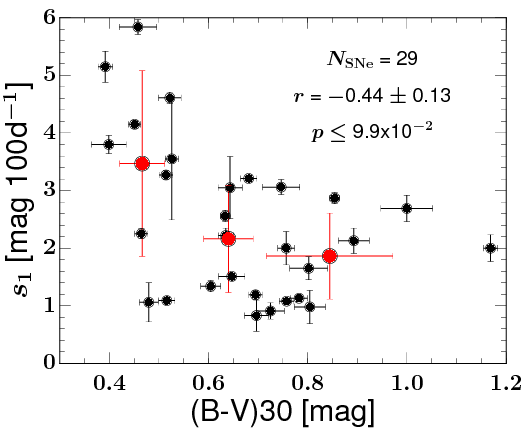}\includegraphics[width=6.0cm]{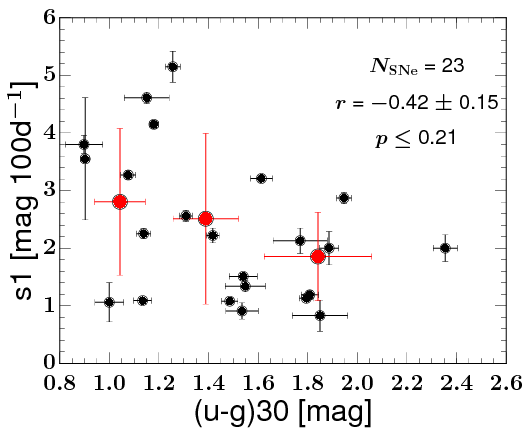}\\
\caption{Correlation between the slopes of the $V$-band light curve ($s_{1}$ and $s_{2}$) and the colours. The results of Monte Carlo simulations on the statistics of these two variables are noted, similarly to Figure \ref{fig:BV_correlations}.}
\label{fig:appendix_redder_IIL}
\end{figure*}

In this section, all correlations between the colour and the slopes of the $V$-band light curve are presented. First, at the top of Figure \ref{fig:appendix_redder_IIL}, the relation between $s_{2}$ and the colour at later epochs is displayed. In the bottom of the figure, correlations between the colour at early times and $s_{1}$ are displayed.

\subsection{Are SNe~II with Smaller Metal-Line Equivalent Widths Redder?}\label{Appendix_EW_colour}

\begin{figure*}
\includegraphics[width=4.9cm]{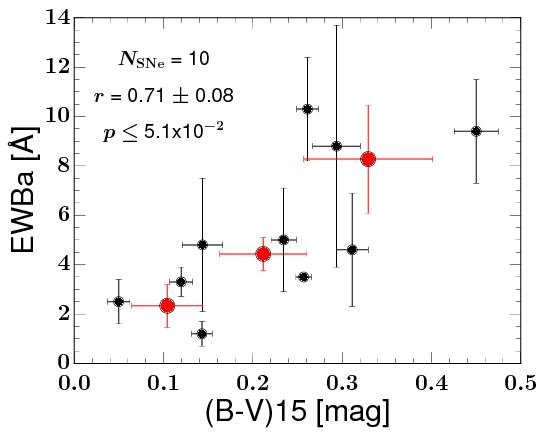}\includegraphics[width=4.8cm]{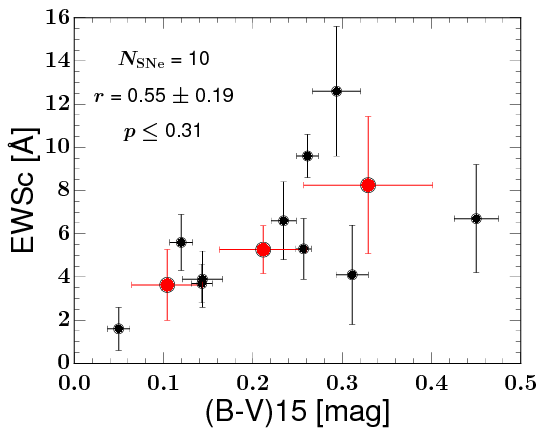}\includegraphics[width=4.8cm]{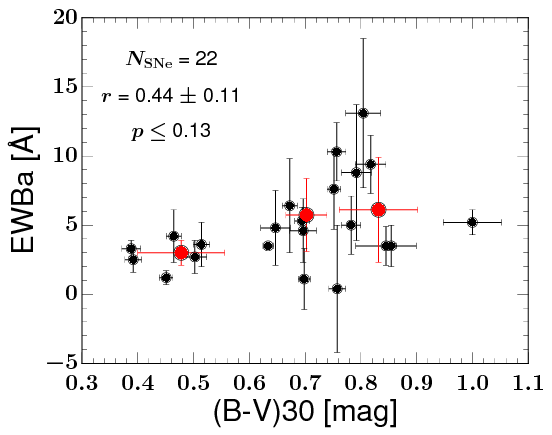}\\
\includegraphics[width=4.8cm]{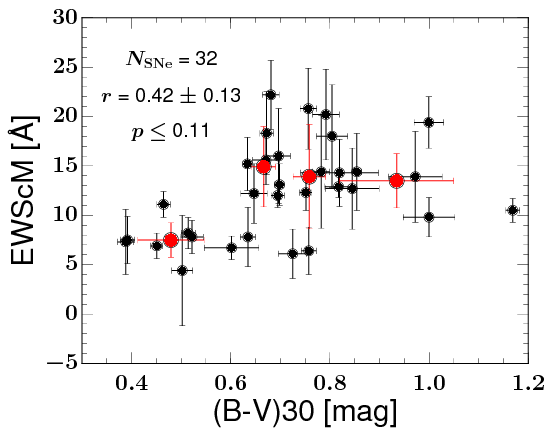}\includegraphics[width=4.8cm]{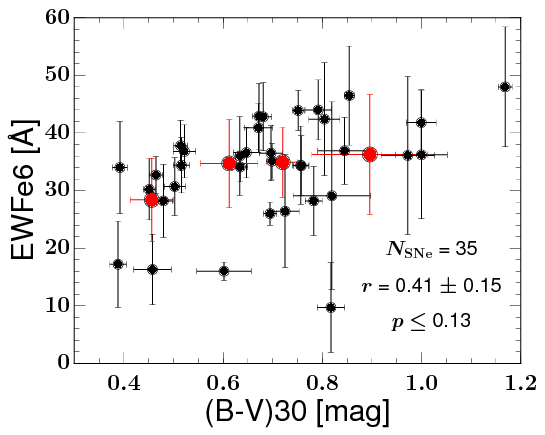}\includegraphics[width=4.8cm]{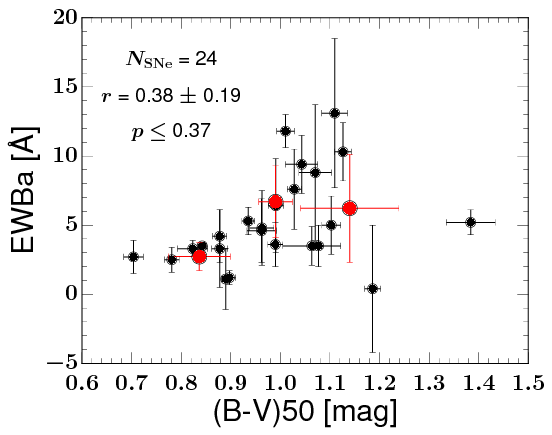}\\
\includegraphics[width=4.8cm]{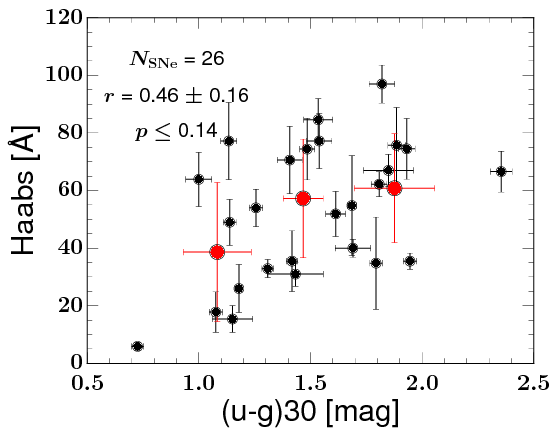}\includegraphics[width=4.8cm]{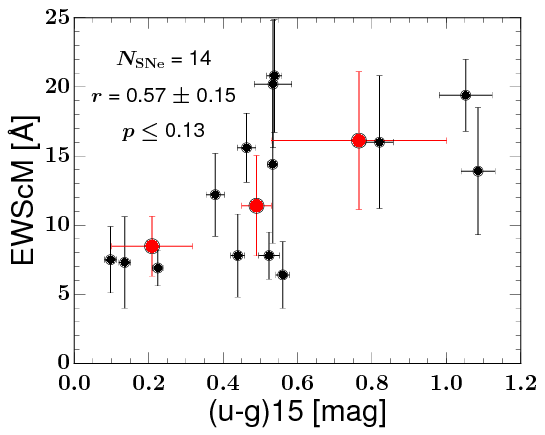}\includegraphics[width=4.8cm]{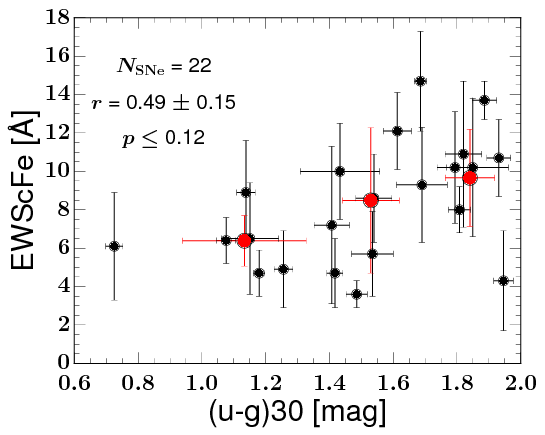}\\
\includegraphics[width=4.8cm]{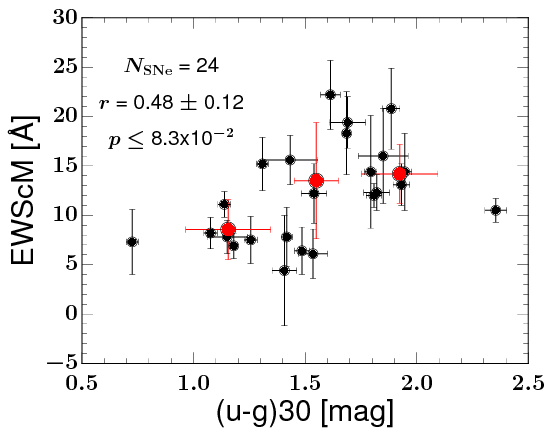}\includegraphics[width=4.8cm]{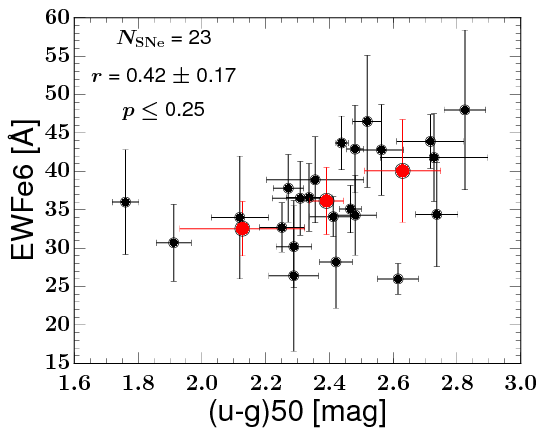}\includegraphics[width=4.8cm]{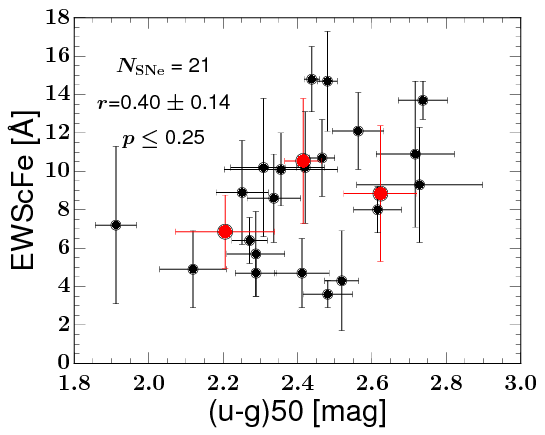}\\
\includegraphics[width=4.8cm]{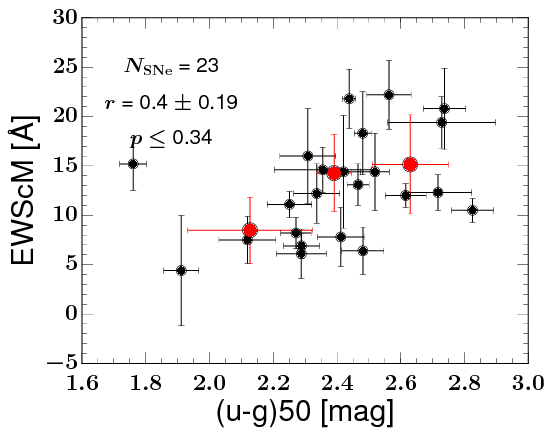}\includegraphics[width=4.8cm]{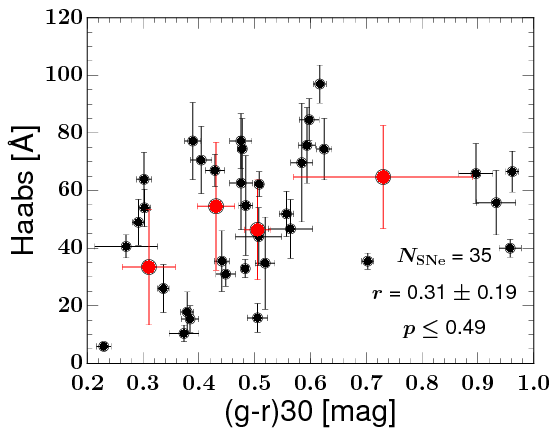}\includegraphics[width=4.8cm]{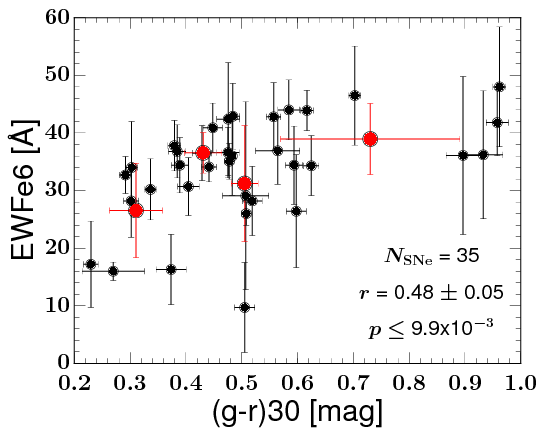}\\
\includegraphics[width=4.6cm]{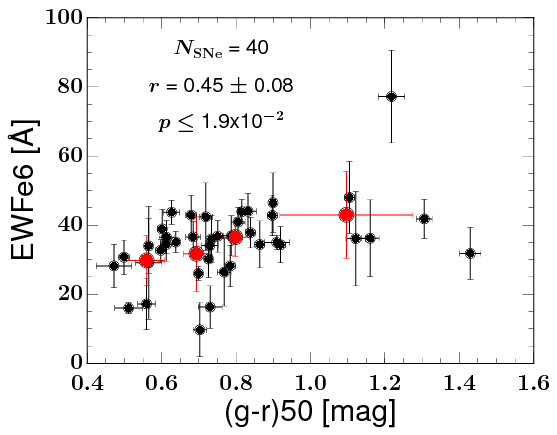}\includegraphics[width=4.6cm]{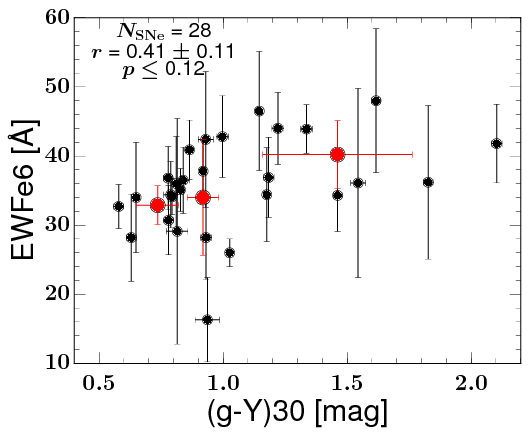}
\caption{Correlation between the absorption-line equivalent widths and the colours. The results of Monte Carlo simulations on the statistics of these two variables are noted, similarly to Figure \ref{fig:BV_correlations}.}
\label{fig_appen:EW_redder}
\end{figure*}

All correlations between the colour and the equivalent widths of lines of different elements are showed in Figure \ref{fig_appen:EW_redder}. 

\subsection{Correlation between Haabs and the Colour Slope.}\label{additional_figures}

\begin{figure*}
\includegraphics[width=8cm]{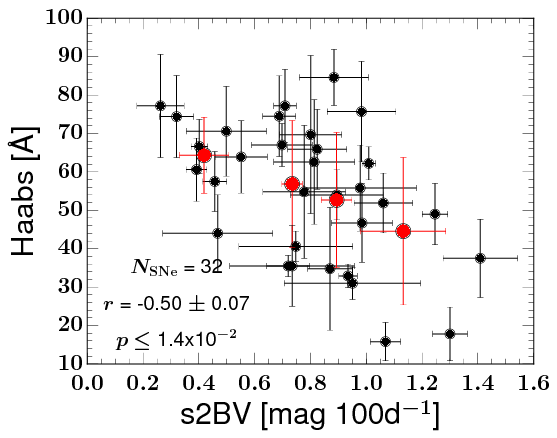}\includegraphics[width=8cm]{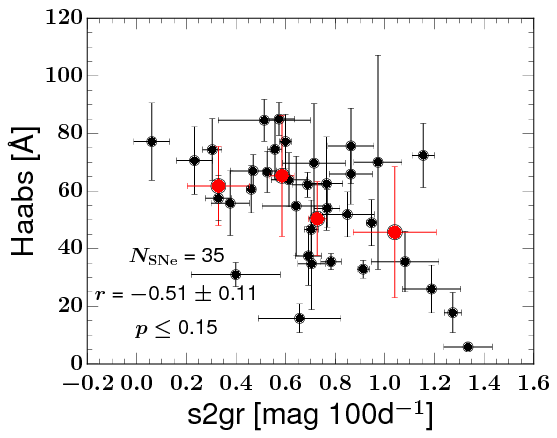}\\
\caption{Inverse correlations between $s_{2,(B-V)}$ and the strength of the H$\alpha$ absorption line (Haabs), in the sense that SNe~II with a fast cooling after transition show weaker H$\alpha$ absorption lines. The results of Monte Carlo simulations on the statistics of these two variables are noted at the bottom, similarly to Figure \ref{fig:BV_correlations}.}
\label{fig:Haabs_s2colour}
\end{figure*}

All correlations between the colour slope after transition and the strength of the H$\alpha$ absorption line are shown in Figure \ref{fig:Haabs_s2colour}.

\subsection{Are SNe~II with Faster Ejecta Velocities Redder?}\label{velo_velocity}

\begin{figure*}

\includegraphics[width=5.0cm]{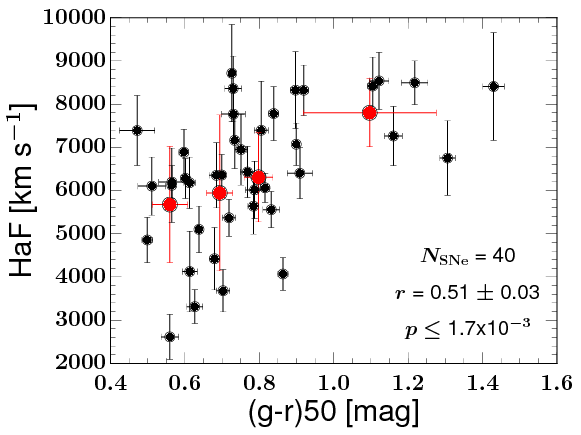}\includegraphics[width=5.0cm]{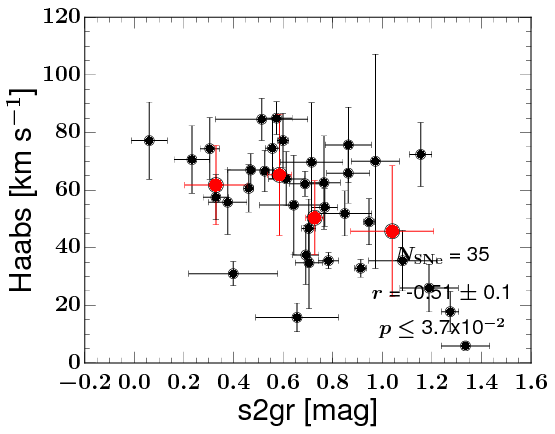}\includegraphics[width=5.0cm]{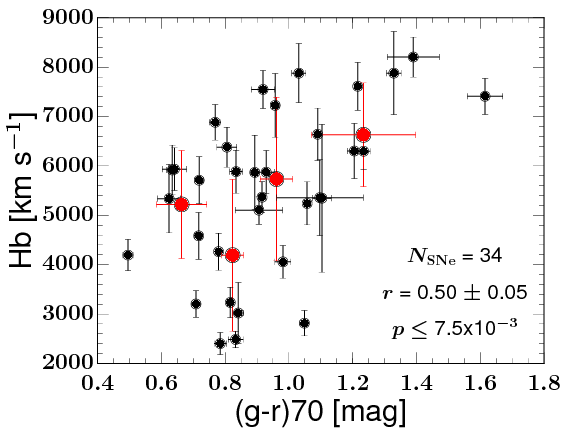}\\
\includegraphics[width=5.0cm]{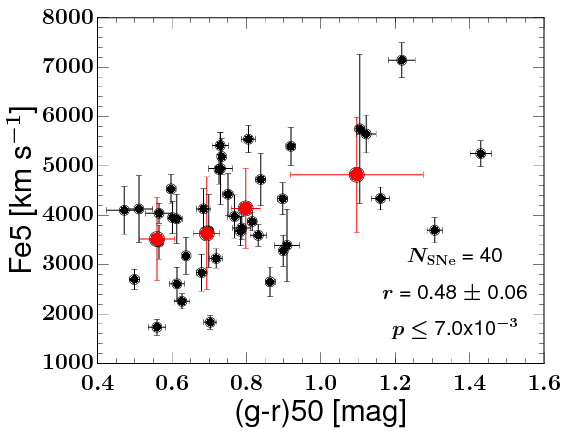}\includegraphics[width=5.0cm]{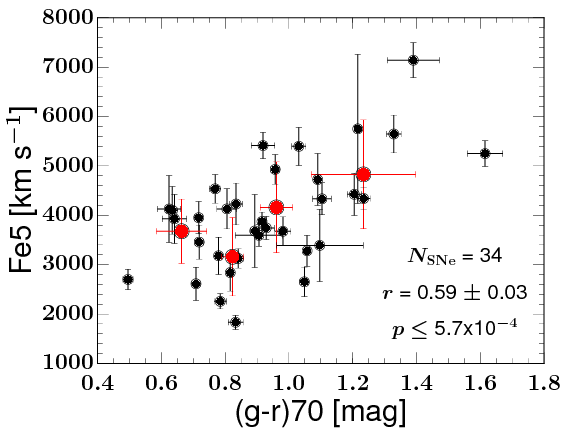}\includegraphics[width=5.0cm]{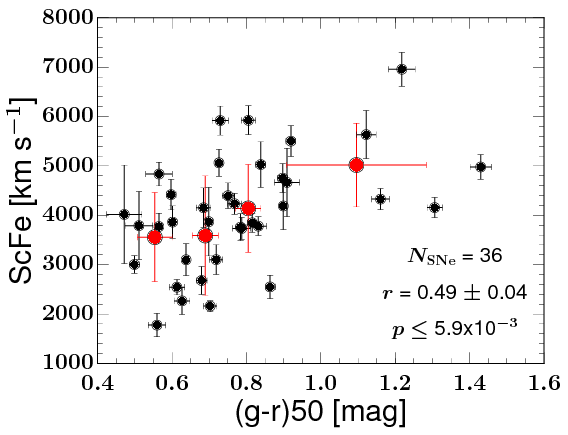}
\includegraphics[width=5.0cm]{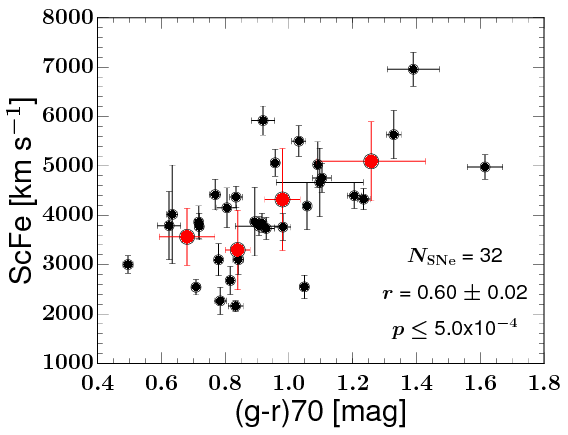}\includegraphics[width=5.0cm]{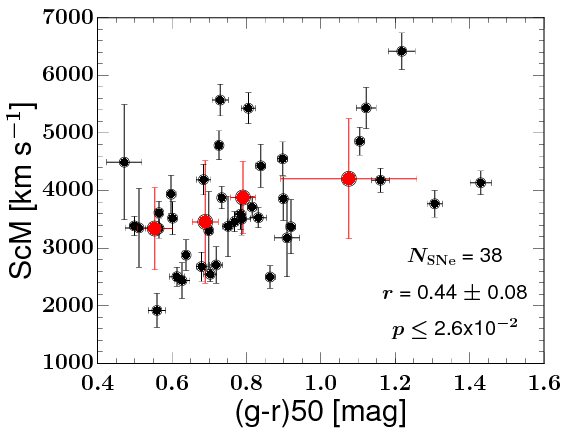}\includegraphics[width=5.0cm]{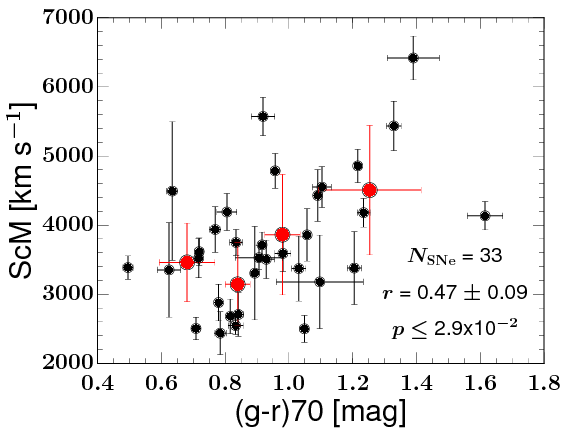}
\includegraphics[width=5.0cm]{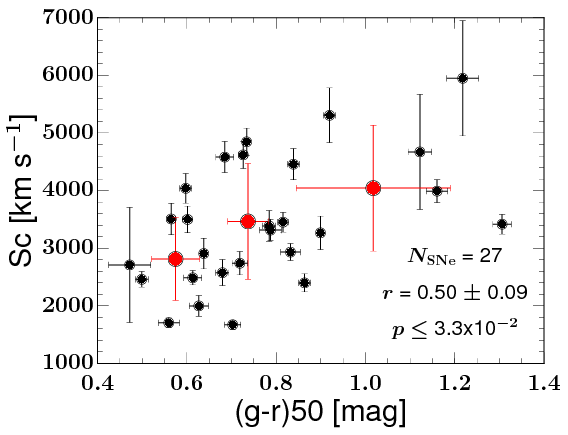}\includegraphics[width=5.0cm]{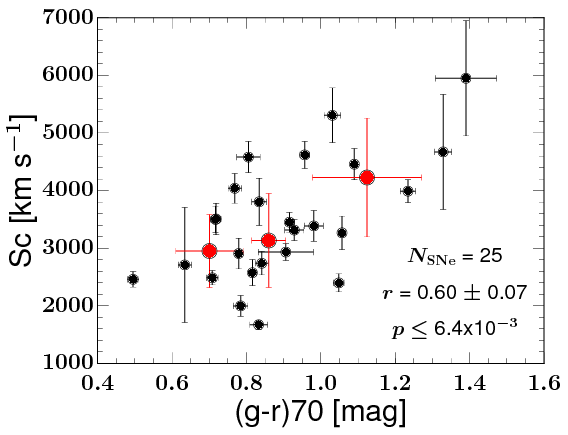}\includegraphics[width=5.0cm]{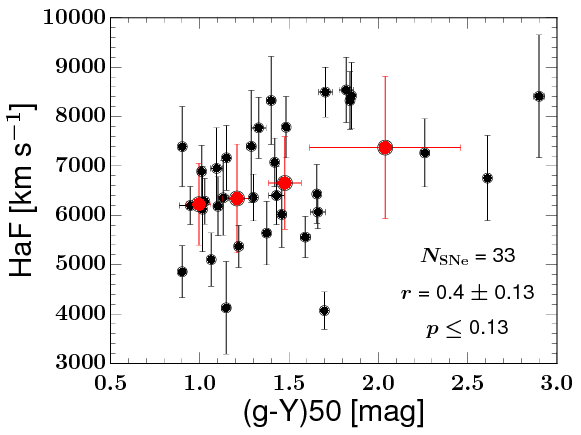}
\includegraphics[width=5.0cm]{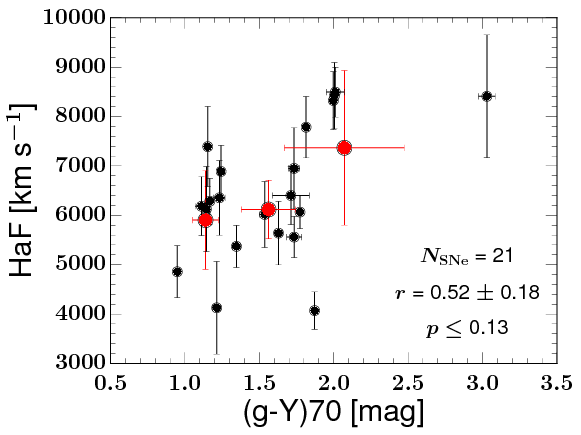}\includegraphics[width=5.0cm]{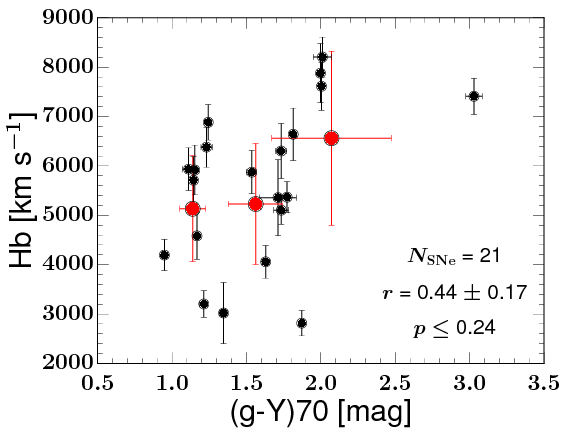}\includegraphics[width=5.0cm]{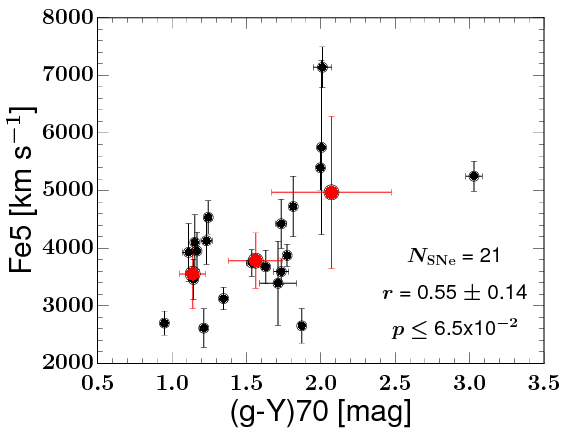}
\caption{Correlation between the ejecta velocities and the colours. The results of Monte Carlo simulations on the statistics of these two variables are noted, similarly to Figure \ref{fig:BV_correlations}.}
\label{fig_appen:velo_colour}
\end{figure*}

All correlations between the colour and the velocities of different elements are showed in Figure \ref{fig_appen:velo_colour}. We also display a trend between the colour $(g-Y)$ at epoch 15\,d and the velocities of HaF and ScM. For these correlations, we have only 8 and 6 objects, respectively.

\section{}\label{AppendixB}

\begin{figure*}
\includegraphics[width=6.0cm]{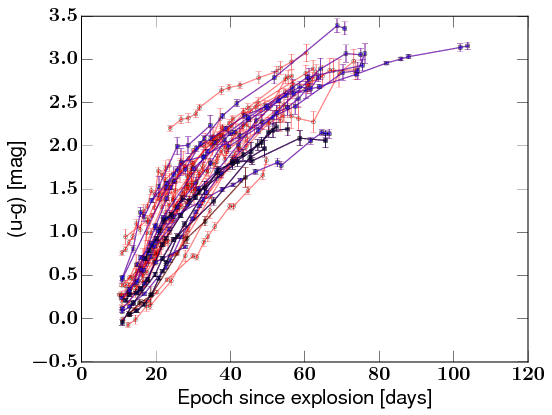}\includegraphics[width=6.00cm]{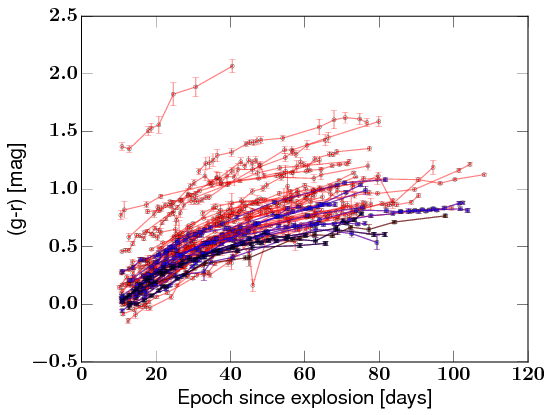}\includegraphics[width=5.9cm]{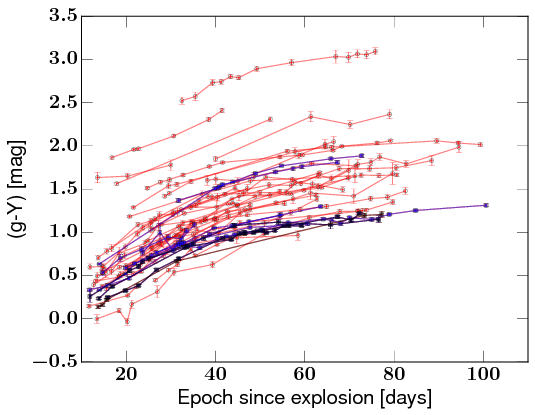}
\caption{Colour curves for the whole sample are shown in red (57 SNe~II). Blue light curves represent our unreddened subsample (19 SNe~II), while in black the bluest SNe~II from the unreddened subsample are shown (SN~2004fx, SN~2005dz, SN~2008M, and SN~2009bz). The colours $(u-g)$, $(g-r)$, and $(g-Y)$ are displayed in the left, middle, and right panels (respectively).}
\label{fig_appen:intrin_colour_evolution}
\end{figure*}

In this Appendix, the colour evolution for the entire sample and the three other colours used in this work are displayed in Figure \ref{fig_appen:intrin_colour_evolution}

%%%%%%%%%%%%%%%%%%%%%%%%%%%%%%%%%%%%%%%%%%%%%%%%%%
% Don't change these lines
\bsp	% typesetting comment
\label{lastpage}
\end{document}